\documentclass[reprint,
nofootinbib,
amsmath,amssymb,
aps,
prb,
]{revtex4-2}

\usepackage{amsfonts}
\usepackage{aas_macros}
\usepackage[english]{babel}
\usepackage{amsmath,amssymb,bm,epsfig,color,graphicx}
\usepackage{booktabs}
\usepackage[mathscr]{eucal}
\usepackage{slashed}
\usepackage{cancel}
\usepackage{yhmath}
\usepackage[colorlinks=true,linkcolor=blue,citecolor=blue,urlcolor=blue]{hyperref}
\usepackage{hyperref}
\usepackage{tikz,xcolor}
\usepackage{caption}
\usepackage{subcaption}
\usepackage{amsmath}
\usepackage{slashed}
\usepackage{comment}
\usepackage{cancel}
\usepackage{multirow}
\usepackage{tikz-feynman}
\tikzfeynmanset{compat=1.1.0}
\usepackage{ragged2e}

\usepackage[font=footnotesize,labelfont=bf]{caption}

\newcommand{\bea}{\begin{array}}
\newcommand{\eea}{\end{array}}
\newcommand{\beq}{\begin{eqnarray}}
\newcommand{\eeq}{\end{eqnarray}}

\begin{document}
\title{
Postsphaleron darkogenesis
\\
}

\author{Sudhakantha Girmohanta}
\email{sgirmohanta@sjtu.edu.cn}
\affiliation{Tsung-Dao Lee Institute, Shanghai Jiao Tong University,
No.~1 Lisuo Road, Pudong New Area, Shanghai 201210, China}
\affiliation{School of Physics and Astronomy, Shanghai Jiao Tong University,
800 Dongchuan Road, Shanghai 200240, China}

\author{Yuichiro Nakai}
\email{ynakai@sjtu.edu.cn}
\affiliation{Tsung-Dao Lee Institute, Shanghai Jiao Tong University,
No.~1 Lisuo Road, Pudong New Area, Shanghai 201210, China}
\affiliation{School of Physics and Astronomy, Shanghai Jiao Tong University,
800 Dongchuan Road, Shanghai 200240, China}

\author{Zhihao Zhang}
\email{zhangzhh@sjtu.edu.cn}
\affiliation{Tsung-Dao Lee Institute, Shanghai Jiao Tong University,
No.~1 Lisuo Road, Pudong New Area, Shanghai 201210, China}
\affiliation{School of Physics and Astronomy, Shanghai Jiao Tong University,
800 Dongchuan Road, Shanghai 200240, China}

\begin{abstract}
    A supercooled phase transition in a nearly conformal dark sector can provide a natural setting for darkogenesis via its out-of-equilibrium dynamics, where a particle-antiparticle number asymmetry in the dark sector can be reprocessed into the visible sector, yielding the observed baryon asymmetry and an asymmetric dark matter. We consider a scenario where the number asymmetry is generated from the decay of a  mother particle produced via parametric resonance during the phase transition induced due to its coupling to the dilaton associated with spontaneous breaking of scale invariance. It is shown that the correct baryon asymmetry and dark matter abundance can be realized for a dark phase transition at $\mathcal{O}(1) \, \rm GeV$, which can also explain the nano-Hz gravitational wave signal reported by pulsar timing array experiments. The scenario will be tested further in neutron-antineutron oscillation experiments.
\end{abstract}

\maketitle


\section{Introduction}
\label{sec:introduction}

The Standard Model (SM) is highly successful in accounting for the current experimental data, while it still leaves the question of dark matter (DM) and the origin of baryon asymmetry in the Universe (BAU) unanswered.
Usual baryogenesis scenarios take place at a high temperature where the sphaleron process is active
(however, see, e.g., Refs.~\cite{Babu:2006xc, Babu:2008rq, Bell:2018mgg} for post-sphaleron baryogenesis),
making it difficult to test them even in future experimental investigations.

Recently, the evidence of nano-Hz gravitational waves (GWs) has been observed by pulsar timing array (PTA) collaborations
\cite{NANOGrav:2023gor,NANOGrav:2023hvm, EPTA:2023fyk,Reardon:2023gzh,Xu:2023wog}.
A promising origin of such a GW background is
a supercooled phase transition (PT) in a dark sector (DS) at the ${\cal O}$(GeV) scale~\cite{Nakai:2020oit,Fujikura:2023lkn, Madge:2023dxc, Megias:2023kiy, Salvio:2023ynn, Salvio:2023blb, Gouttenoire:2023bqy, Addazi:2023jvg, Li:2023bxy, Ghosh:2023aum, Jiang:2023qbm, Wang:2023bbc, Li:2025nja,Fujikura:2025iam}. It is partly motivated by the fact that the PT interpretation provides a better fit to the observed GW spectra than the baseline expectation from the supermassive black hole binary (SMBHB) evolution~\cite{NANOGrav:2023hvm, Ellis:2023oxs, Winkler:2024olr}. This may have an interesting connection with the question of DM and BAU in the following way. The strong supercooling, which is essential for explaining the PTA signal amplitude, also dilutes away any baryon asymmetry and DM existing before the PT. Therefore, unless there was a huge pre-existing baryon asymmetry and DM abundance,
BAU and DM must be created after the dark supercooled PT at ${\cal O} (1)$ GeV to be consistent with the current cosmological observations, leading to an exciting consequence that a mechanism to produce BAU and DM is likely to be probed by ongoing and future experiments.

It is an intriguing possibility that a dark PT itself provides a mechanism for the creation of BAU and DM. Ref.~\cite{Fujikura:2024jto} has utilized the winding number changing dynamics of a dark $SU(2)_{\rm D}$ Higgs field and a chiral anomaly in the DS to generate a number asymmetry in the DS which is later reprocessed into the visible sector BAU through portal operators, leaving behind an asymmetric self-interacting DM~\cite{Shelton:2010ta, Garcia-Bellido:1999xos, Konstandin:2011ds, Hall:2019ank, Hall:2019rld, Girmohanta:2022dog}. Since there are a lot of ongoing and projected searches for DS particles, exploring the horizon of possibilities for darkogenesis~\cite{Shelton:2010ta} driven by a dark PT is of significant interest.

It is natural to ask if the decay of some GeV-scale Majorana fermion in a DS can account for BAU and DM after a dark supercooled PT. For this question, in the absence of resonance enhancement and tuning, to yield a significant asymmetry, a Majorana fermion usually needs to be very heavy, as in the standard thermal leptogenesis scenario~\cite{Fukugita:1986hr},
which does not fit the present exploration. However, if a dark supercooled PT provides an environment for the out-of-equilibrium dynamics to abundantly produce a GeV-scale Majorana fermion, the conclusion can be changed. 
In the present paper, we will consider such a possibility and show that a mother GeV-scale Majorana fermion $\psi$ is non-thermally generated through parametric resonance induced by post-transition oscillations of a CP-even scalar called the dilaton in a nearly conformal DS. The dilaton is the pseudo-Nambu Goldstone boson (pNGB) of the spontaneously broken scale invariance, and can be identified with the radion degree of freedom in a dual weakly coupled 5-dimensional description utilizing AdS/CFT~\cite{Maldacena:1997re,Gubser:1998bc,Witten:1998qj,Arkani-Hamed:2000ijo,Rattazzi:2000hs}.

We derive the dilaton/radion interaction with the Majorana fermion $\psi$ without performing any linear expansion as relevant for the induced parametric resonance phase and show explicitly the particle production during the PT via numerically solving the equation of motion with a time-dependent effective mass term caused by the dilaton/radion oscillations. The decay of $\psi$ into dark fermions $\chi_{1,2}$ and a scalar $\eta$ generates a DS asymmetry: the lighter $\chi_1$ becomes asymmetric DM, while the heavier $\chi_2$ transfers its asymmetry to the visible sector via a baryon- and dark-number-violating neutron portal operator. The scenario requires a Higgs-portal coupling of the scalar $\eta$, which keeps the DS and visible sector in thermal contact, and also helps transferring the false vacuum energy to the visible sector after the PT to be consistent with the standard Big Bang Nucleosynthesis (BBN). We show that this scenario can yield the observed BAU and DM abundance while also predicting neutron-antineutron oscillations potentially observable in upcoming experiments as a result of the Majorana nature of $\psi$ and the neutron portal operator.

The rest of the paper is organized as follows. In section~\ref{sec:setup}, we formulate the DS model. Section~\ref{sec:decay} discusses the non-thermal production of mother particles and a comparison of numerical results and analytical estimates for the particle production. Then, the abundance of BAU and DM is estimated and related phenomenological constraints are presented in section~\ref{sec:DM}. Section~\ref{sec:discuss} is devoted to conclusions and discussions. Appendices~\ref{sec:radionInt},~\ref{appendix:dilaton},~\ref{appendix:general},~\ref{appendix:5d} summarize relevant dilaton/radion interactions, general formulae for asymmetry generated by a particle decay, and a possible 5D setup.

\section{Setup}
\label{sec:setup}

\begin{figure}[t!]
     \centering
    \includegraphics[width=0.95\linewidth]{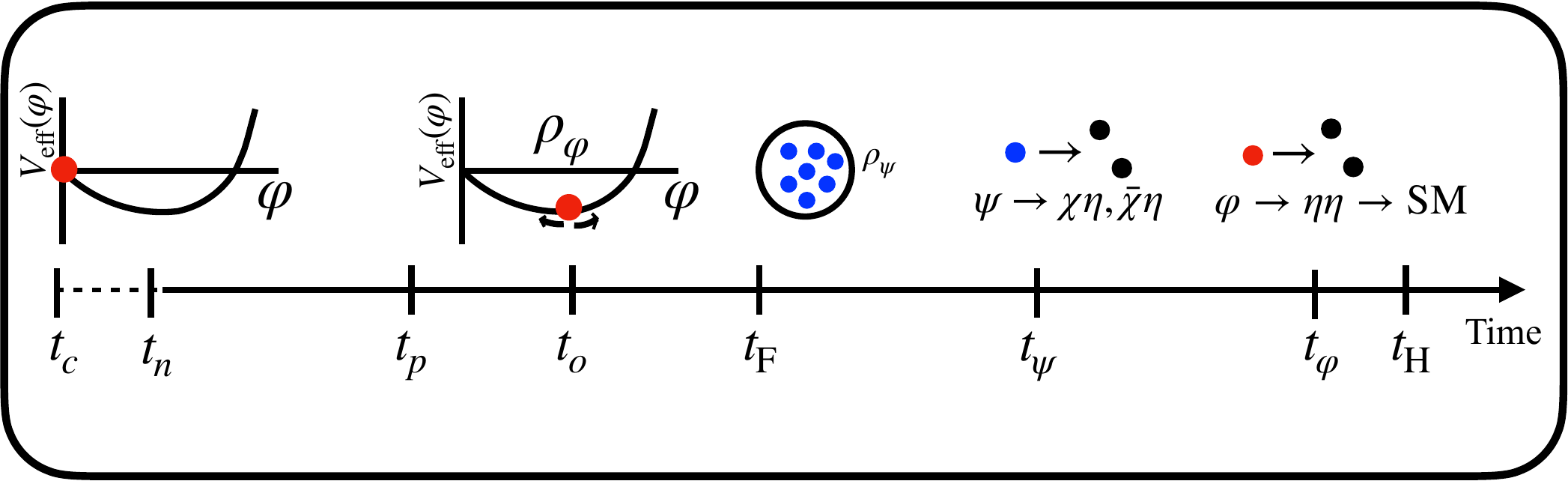}
     \captionsetup{justification=centerlast, singlelinecheck=false}
    \caption{Schematic and timeline of the scenario: $t_c$ corresponds to the critical time when the true minimum emerges. Strong supercooling and substantial dilution of preexisting baryon asymmetry and dark matter number density persist until nucleation at $t_n$. At $t_p$, bubbles of the true vacuum collide, reflect, and arrange the dilaton VEV near the basin of attraction of the true minimum. As the system approaches the minimum at $t_o$ while oscillating around it, $\psi$ fields are produced via parametric resonance. At $t_{\rm F}$, the large occupation of $\psi$ back-reacts, causing the production to freeze. At $t_\psi$, the produced $\psi$ decays into $\chi_{1,2}$ and $\eta$. $\chi_2$ transfers its asymmetry to the visible sector via a neutron portal coupling, while $\chi_1$ acts as the asymmetric DM. Thereafter, at $t_\varphi$, the remnant coherent dilaton field decays perturbatively due to the Higgs-portal coupling of $\eta$, completing the reheating process. All the time scales relevant are smaller than the Hubble time $t_{\rm H}$ for the GeV scale PT.}
    \label{fig:schematic}
\end{figure}

We present here our framework where a strong first-order PT in a nearly conformal DS generates both BAU and DM. The transition involves a CP-even dilaton field, denoted $\phi$, which arises from the spontaneous breaking of scale invariance. The oscillations of the dilaton just after the PT play a central role in producing out-of-equilibrium DS particles.
In Fig.~\ref{fig:schematic} we illustrate the schematic for the timescale of the PT and production of the mother particle $\psi$. Before the PT, the dilaton is stuck at the false vacuum with vacuum expectation value (VEV) $\langle \phi \rangle =0$. As the temperature drops sufficiently to allow for the tunneling rate to overcome the Hubble expansion rate, bubbles of the true vacuum with $\langle \phi \rangle \neq 0$ form that percolate and complete the PT. When the bubbles of the true vacuum collide during this stage, they bounce and the dilaton gets displaced from the true minimum, and starts to oscillate around it~\cite{Konstandin:2011ds}. As the bubbles sweep through the space, colliding and reflecting with each other, the space is filled with the oscillating dilaton field, which to a first approximation resembles a low-scale hybrid inflation scenario~\cite{Giudice:1999fb}.

The spontaneous breaking of scale invariance is a consequence of some nearly marginal operator ${\cal O}$ coupled to a conformal theory and its evolution to the low-energy. At the deep ultraviolet (UV), the effect of ${\cal O}$ on the conformal sector is negligible; however, it grows as one evolves towards the infrared (IR), and at some point triggers the breakdown of conformal invariance. A CP-even pseudo-Nambu-Goldstone boson emerges as a result of the spontaneously broken scale invariance, called the dilaton, whose mass is dictated by the scaling dimension of the perturbing operator at the breaking point of scale invariance~\cite{Chacko:2012sy, Girmohanta:2023tdr, Csaki:2000zn, Goldberger:1999uk, Bellazzini:2012vz}. We will remain agnostic about the nature of the operator $\cal{O}$, \textit{i.e.}, regarding the details of how the dilaton gets a mass, but consider an effective Lagrangian,
\begin{align}
    S_{\rm \varphi, eff} &= \int d^4 x \left[ \frac{1}{2} (\partial_\mu \varphi) (\partial^\mu \varphi) - V_{\rm eff}(\phi) + \frac{\varphi}{\langle \phi \rangle} T^\mu_\mu  \right] \ ,
    \label{Eq:dilaton_eff}
\end{align}
where $\phi \equiv \langle \phi \rangle + \varphi$, $\varphi$ denotes the canonically normalized dilaton, $V_{\rm eff} (\phi)$ encodes the effects of the breaking operator $\cal{O}$ in generating the dilaton potential, $\langle \phi \rangle$ is the dilaton VEV at the potential minimum, and its mass is given by $m_\varphi^2 = V''_{\rm eff}(\phi)|_{\phi=\langle \phi \rangle}$. The dilaton couplings to the matter are determined by the breaking of the dilatation current due to the trace of the energy-momentum tensor $T^\mu_\mu$. Given the details of $V_{\rm eff}(\phi)$, the phase transition dynamics and associated gravitational wave spectra can be analyzed~\cite{Fujikura:2024jto, Fujikura:2019oyi}. It is important to note that Eq.~\eqref{Eq:dilaton_eff} is based on a linear expansion in the dilaton field, which is valid when the dilaton has settled into its global minimum after the PT. However, immediately following bubble percolation, the dilaton oscillation amplitude remains large, and a more careful treatment of the dilaton couplings is required, which does not rely on a linear expansion. For further details, see appendices~\ref{sec:radionInt},~\ref{appendix:dilaton}.

\begin{table}[!t]
    \centering
    \setlength{\tabcolsep}{45pt} 
    \renewcommand{\arraystretch}{1.5} 
    \begin{tabular}{c|c}
        \hline
        \hline
        Fields & $U(1)_{\rm D}$ \\
        \hline
        \hline
        $\psi_i$ & $-1$ \\
        $\chi_j$ & $-1$ \\
        $\eta$   & 0 \\
        \hline
    \end{tabular}
    \captionsetup{justification=raggedright, singlelinecheck=false}
    \caption{Field content in the DS and their global $U(1)_{\rm D}$ charge assignments. Here, $i,j$ represent generational indices. $\psi$ decay produces asymmetry in $\chi_j$, and $\chi_1$ is the DM candidate. $\eta$ also acts as a Higgs portal scalar.}
    \label{tab:content}
\end{table}

Let us now specify the matter content of the DS. As shown in Tab.~\ref{tab:content}, in addition to the CFT and dynamics that give a mass to the dilaton, we introduce Majorana fermions $\psi_i$, Dirac fermions $\chi_j$, and a real scalar  field $\eta$, where $i$, $j$ represent generational indices. A notion of a dark number is represented by a global $U(1)_{\rm D}$ under which both $\psi$ and $\chi$ carry a charge $-1$. All the DS particles are singlets under the SM. We consider the following interactions for the DS:
\begin{align}
    \nonumber
    {V}_{\rm DS} &= \bar \chi \mu \chi \eta  + m_\chi \bar \chi \chi + \frac{1}{2} m_\psi \bar \psi^c_{\mathrm{R}} \psi_{\mathrm{R}}   + \frac{1}{2} m_\eta^2 \eta^2 \\ &  + \bar{\psi}_{\mathrm{R}} y \eta \chi_{\mathrm{L}} + {\rm h.c.} \ ,
    \label{Eq:VDS}
\end{align}
where generational indices are suppressed, $\mu$ and $y$ are real and complex coupling matrices, respectively, and the only violation of $U(1)_{\rm D}$ comes from the Majorana mass term of $\psi$, which violates it by 2 units. We assume that $\eta$ does not get any VEV. We choose a basis where $m_\chi$ and $m_\psi$ are diagonal. If $\psi$, $\chi$ have $N_{\psi}$, $N_\chi$ generations, respectively, $y_{ij}$ contains $N_\chi (N_\psi-1)$ physical CP-violating phases. We assume $N_\psi = N_\chi=2$, while the lightest $\chi$ will play the role of DM. We label the generational indices according to ascending mass, i.e., $m_{\chi_2} > m_{\chi_1}$, and so on.

If after the PT, all the energy remains within the DS, for instance, in a massless dark radiation component, the PT alone cannot account for the PTA signal without violating a constraint on the extra number of relativistic degrees of freedom, $\Delta N_{\rm eff}$~\cite{Nakai:2020oit, Bringmann:2023opz, Fujikura:2023lkn}. To address this, a portal operator is typically introduced to couple the DS with the visible sector, ensuring thermal equilibrium and facilitating energy transfer from the DS to the visible sector. To this end, we identify the field $\eta$ as the Higgs portal scalar. The effective Higgs portal and dilaton interactions for $\eta$ are given by
\begin{equation}
    {\cal L}_{\eta} \supset -g_{\eta} \, \eta  \left( \mathbf{H}^\dagger \mathbf{H} - \frac{v^2}{2}\right) -\frac{m_\eta^2}{2} \eta^2 \left( \frac{2 \varphi}{\langle \phi \rangle} + \frac{\varphi^2}{\langle \phi \rangle^2} \right) \ ,
    \label{Eq:HiggsPortal}
\end{equation}
where $\mathbf{H}$ denotes the SM Higgs doublet, and $v$ denotes its VEV.
Eq.~\eqref{Eq:HiggsPortal} induces a mixing angle between the physical Higgs and $\eta$ mass eigenstates, and is given by
\begin{align}
   \tan (2\theta_{h\eta}) = \frac{2 g_{\eta} v}{m_h^2-m_\eta^2} \ ,
    \label{Eq:thetaetah}
\end{align}
where $m_h$ is the physical Higgs mass. By virtue of the interaction in Eq.~\eqref{Eq:HiggsPortal}, $\varphi$ can decay to SM particles perturbatively.

If $\psi_1$ is thermally produced, the most efficient production of asymmetry happens when 
\begin{equation}
    \Gamma_{\psi_1} \simeq \frac{(y y^\dagger )_{11}}{8 \pi} m_{\psi_1} \lesssim H(T=m_{\psi_1}) = 0.3 \sqrt{g_*} \frac{m_{\psi_1}^2}{M_{\rm pl}} \ .
    \label{Eq:yEqu}
\end{equation}
Here, $g_*$ denotes the effective number of degrees of freedom, and $M_{\rm pl} = 2.4 \times 10^{18}$ GeV is the reduced Planck mass. For the relevant scale of PTA, {i.e.}, $m_{\psi_1} \sim$ GeV, assuming similar magnitude elements in the $y_{ij}$ matrix which we refer to as $|y|$, Eq.~\eqref{Eq:yEqu} implies $|y| \lesssim 5 \times 10^{-9}$. As a result, the asymmetry generated due to thermal $\psi$ decay $\simeq |y|^2/(4 \pi g_{*})$ would be negligibly small. On the other hand, when the condition in Eq.~\eqref{Eq:yEqu} is not satisfied, then the generated asymmetry suffers from a suppression $\sim 1/(K \ln K)$, where $K = \Gamma_{\psi_1}/H(T=m_{\psi_1})$. For $m_{\psi_1} \sim$ GeV, and $|y| \gtrsim 10^{-4}$, $(K \ln K)^{-1} \sim 10^{-10}$. Therefore, the decay of thermally produced GeV scale particles, as relevant for the PT explaining PTA, can not explain the observed BAU. We will circumvent this via a non-thermal production of $\psi_1$ that utilizes its coupling with the dilaton to induce parametric resonance during the PT.

\section{Mother particle production via dilaton oscillation}
\label{sec:decay}

Coherent oscillations of the dilaton field after the bubble percolation can cause the effective mass of a fermion to vary,
leading to parametric excitation of a fermion if its effective mass vanishes during the oscillation. This is contrary to the expectation that the fermion production suffers from Pauli blocking and is suppressed
compared to the case of the boson production. The coherently oscillating nature alters the naive expectation dramatically~\cite{Greene:1998nh}.
As evaluated in appendix~\ref{sec:radionInt},
the effective mass term during the dilaton oscillation in our setup is
\begin{align}
    m^{\rm eff}_\psi(t) = m_\psi \left[\frac{\{1+\xi(t) \cos(m_\varphi t)\}^{2b}-r}{1-r} \right] \ ,
    \label{Eq:meff2}
\end{align}
where in terms of the 5D dual description, $b$ corresponds to the 5D bulk mass parameter for $\psi$, $r$ denotes the ratio of its UV and IR localized mass terms,  $\xi(t)$ is the amplitude of the dilaton oscillation in units of $\langle \phi \rangle$, and $m_\psi$ is the physical $\psi$ mass today when the dilaton oscillation is frozen out. The initial value of $\xi(t=0)$ is denoted as $\xi_0$. $\xi_0$ is also the maximum amplitude of oscillation, and $\xi(t) < \xi_0$ at later times due to backreaction effects of the produced $\psi$. In terms of the 4D description, these are translated into how the field $\psi$ mixes with composite states of the CFT (see appendices~\ref{sec:radionInt},~\ref{appendix:dilaton} for further details).
As the effective mass of the Majorana fermion $\psi$ varies according to Eq.~\eqref{Eq:meff2}, the equation of motion appears analogous to that of a harmonic oscillator with a time-varying frequency. Efficient production of particles can take place when the adiabatic evolution is violated, in particular, at instants when $m_\psi^{\rm eff}(t)$ crosses zero. From Eq.~\eqref{Eq:meff2} we note that $m_\psi^{\rm eff}(t)$ never vanishes when $r=0$.
It is important not to expand the dilaton/radion interaction in a linear approximation, as it may appear to vanish at the level of a linear approximation for a certain range of parameters, while from the full expression, it is clear that the exponent never crosses zero when $r=0$. A necessary condition for $m_\psi^{\rm eff}(t)$ to vanish is 
\begin{align}
    \nonumber
   & (1-\xi_0)^{2b} \leq r \leq (1+\xi_0)^{2b} \ ; \ {\rm for \ } 0 < b < 1/2 \ , \\[1ex]
   & (1-\xi_0)^{2b} \geq r \geq (1+\xi_0)^{2b} \ ; \ {\rm for \ } b < 0 \ ,
   \label{Eq:neccCondn}
\end{align}
where we have used the fact that $\xi_0<1$, and $r \neq 0$.
As Eq.~\eqref{Eq:neccCondn} is a necessary condition for the particle production to occur, it can be ensured that the other fields coupled to $\varphi$ do not get produced similarly by appropriately choosing their $b,r$ parameters. Notice from Eq.~\eqref{Eq:neccCondn} that for a massive particle production to occur, $b \neq 0$. It is sufficient that only $\psi_1$, the lightest mother particle, is produced by this mechanism, as a virtual $\psi_2$ loop is adequate for the asymmetry generation. Only $\psi_1$ production is also desirable, as otherwise $\psi_2$ decay involving $\psi_1$ may wash out the asymmetry.

We now formulate the particle production, following Ref.~\cite{Giudice:1999fb}.
The effective equation of motion of $\psi$ in the background of oscillating $\varphi$ takes the form,
\begin{align}
    \left( i\gamma^\mu \partial_\mu -m^{\rm eff}_\psi (t) \right) \psi = 0 \ .
    \label{Eq:psiEOM}
\end{align}
To solve Eq.~\eqref{Eq:psiEOM} with the time-varying effective mass in Eq.~\eqref{Eq:meff2}, we decompose
\begin{align}
    \psi(x)=\int \frac{d^3p}{(2 \pi)^{3/2}} \frac{e^{i \vec{p}\cdot \vec{x}}}{\sqrt{2}} \sum_s \left[u_s(p,t)a_s(p)+v_s(p,t)a^\dagger_s(-p) \right] ,
\end{align}
where $a_s, a^\dagger_s$ denote annihilation and creation operators, respectively, the summation is over spin $s$, and the condition $v_s(p)=C \bar{u}_s^T(-p)$ is imposed to preserve the Majorana condition $\psi^c=\psi$.
The creation and annihilation operators satisfy the anticommutation relations,
\begin{align}
\nonumber
& \bigl\{a_s (p), a_{s'}^\dagger(p') \bigr\} = \delta_{ss'} \delta(\vec{p}-\vec{p}\,') \ , \\
&\bigl\{a_s (p), a_{s'} (p') \bigr\} = \bigl\{a_s^\dagger (p), a_{s'}^\dagger(p') \bigr\} = 0 \ .
\end{align}
The normalization of spinors is
$u_s^\dagger (p,t) u_{s'} (p,t) = 2 \delta_{ss'}$.
Defining $u \equiv \begin{pmatrix}
    u_+ \\
    u_-
\end{pmatrix}$, the Dirac equation \eqref{Eq:psiEOM} is reduced to an uncoupled set of differential equations,
\begin{align}
    \left [ \frac{d^2}{d t^2}+\omega^2 \pm i \dot{m}_\psi^{\rm eff} \right]u_\pm (p)=0 \ ,
    \label{uEoM}
\end{align}
where the dot denotes derivative with respect to time, $\omega^2(t)=p^2+(m_\psi^{\rm eff}(t))^2$, and $\gamma^0=\begin{pmatrix}
    \mathbf{I} & 0 \\
    0 & -\mathbf{I}
\end{pmatrix}$. Note that for a GeV scale PT, the Hubble rate is negligible in comparison with particle creation and dilaton oscillation scales. Therefore, we can effectively neglect the expansion of the Universe in the particle production stage and work in an effective Minkowski background. The effective Hamiltonian ${\cal H}$ is defined to be 
\begin{align}
    \nonumber
    {\cal H}(t) &=\frac{i}{2}\int d^3 x \psi^\dagger  \dot{\psi}
    \nonumber
    \\ & =\frac{1}{2}\int d^3p \sum_s \bigg\{  E_p(t) \left[a^\dagger_s(p)a_s(p)-a_s(-p)a^\dagger_s(-p) \right] 
    \nonumber
    \\ &+ F_p(t)a_s(-p)a_s(p)+F^*_p(t)a^\dagger_s(p)a^\dagger_s(-p) \bigg\} \ ,
\end{align}
where one obtains
\begin{align}
    & E_p  =p \Re(u^*_+ u_-)-m_\psi^{\rm eff}(1-u^*_+u_+) \ , \\[1ex]
    & F_p =\frac{p}{2}(u^2_+-u^2_-)-m_\psi^{\rm eff}u_+u_- \ , \\[1ex]
    & E^2_p+|F_p|^2  =\omega^2 \ .
\end{align}
The Hamiltonian ${\cal H}$ is diagonalized with the following time-dependent Bogolyubov transformation:
\begin{align}
\hat{a}(p,t)&=\alpha(p,t) a(p)+\beta(p,t)a^\dagger(-p) \ , \\[1ex]
\hat{a}^\dagger (-p,t)&=-\beta^*(p,t)a(p)+\alpha^*(p,t)a^\dagger(-p) \ ,
\end{align}
where $|\alpha|^2+|\beta|^2=1$, $\beta(-p)=-\beta(p)$ and $\alpha(-p)=\alpha(p)$. It can be shown that
\begin{align}
    \frac{\alpha}{\beta}&=\frac{E_p+\omega}{F_p^*} \ , \\[1ex] |\beta|^2&=\frac{|F_p|^2}{2\omega(\omega+E_p)}=\frac{\omega-E_p}{2\omega} \ .
\end{align}
The Hamiltonian is then reduced to 
\begin{align}
    {\cal H}(t)&=\frac{1}{2}\int d^3p \sum_s \omega(t)\left [\hat{a}^\dagger_s(p)\hat{a}_s(p)+\hat{a}^\dagger_s(-p)\hat{a}_s(-p)\right ] \\
    &=\int d^3p \sum_s \omega(t)\left [\hat{a}^\dagger_s(p)\hat{a}_s(p)\right ] .
\end{align}
The number density of produced $\psi$ is given by
\begin{align}
    n(t) =\frac{1}{2\pi^2}\int^\infty_0 dp \, p^2 |\beta|^2 \ .
    \label{ONC}
\end{align}
One can obtain $u_\pm(t)$ by numerically solving Eq.~\eqref{uEoM} with the following initial conditions at $t=0$:
\begin{align}
    u_\pm(0)&=\sqrt{1\pm\frac{m_\psi^{\rm eff}}{\omega}} \ , \\[1ex]
    \dot{u}_\pm(0)&=-ipu_\mp(0)\mp im_\psi^{\rm eff}u_\pm(0) \ .
\end{align}
Due to the initial conditions chosen, $E_p=\omega, F_p=0$ at $t=0$, \textit{i.e.,} one starts with vanishing particle number density.

\begin{figure*}[!t]
    \centering
    \includegraphics[width=0.46\linewidth]{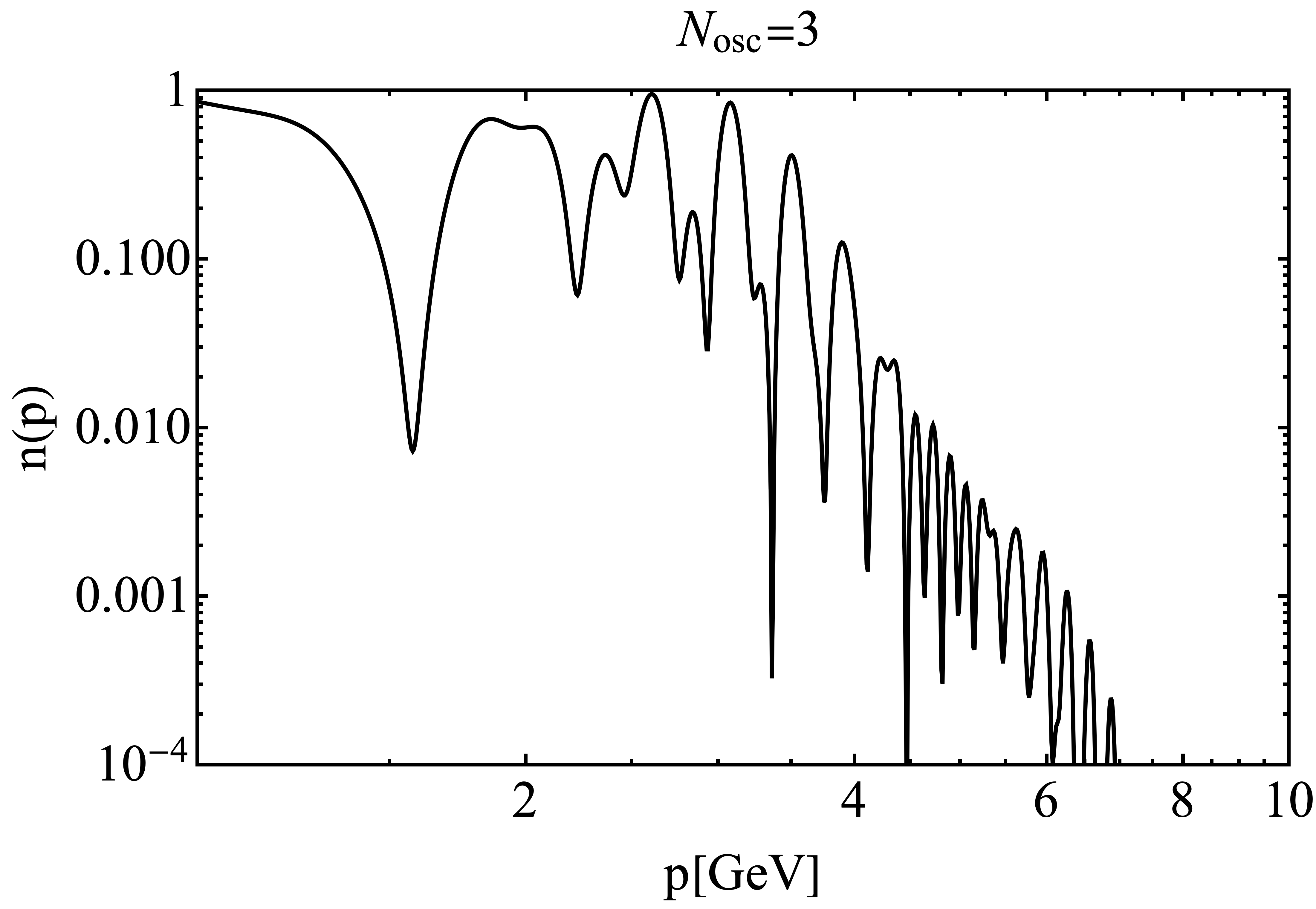}
    \hspace{0.5cm}
    \includegraphics[width=0.46\linewidth]{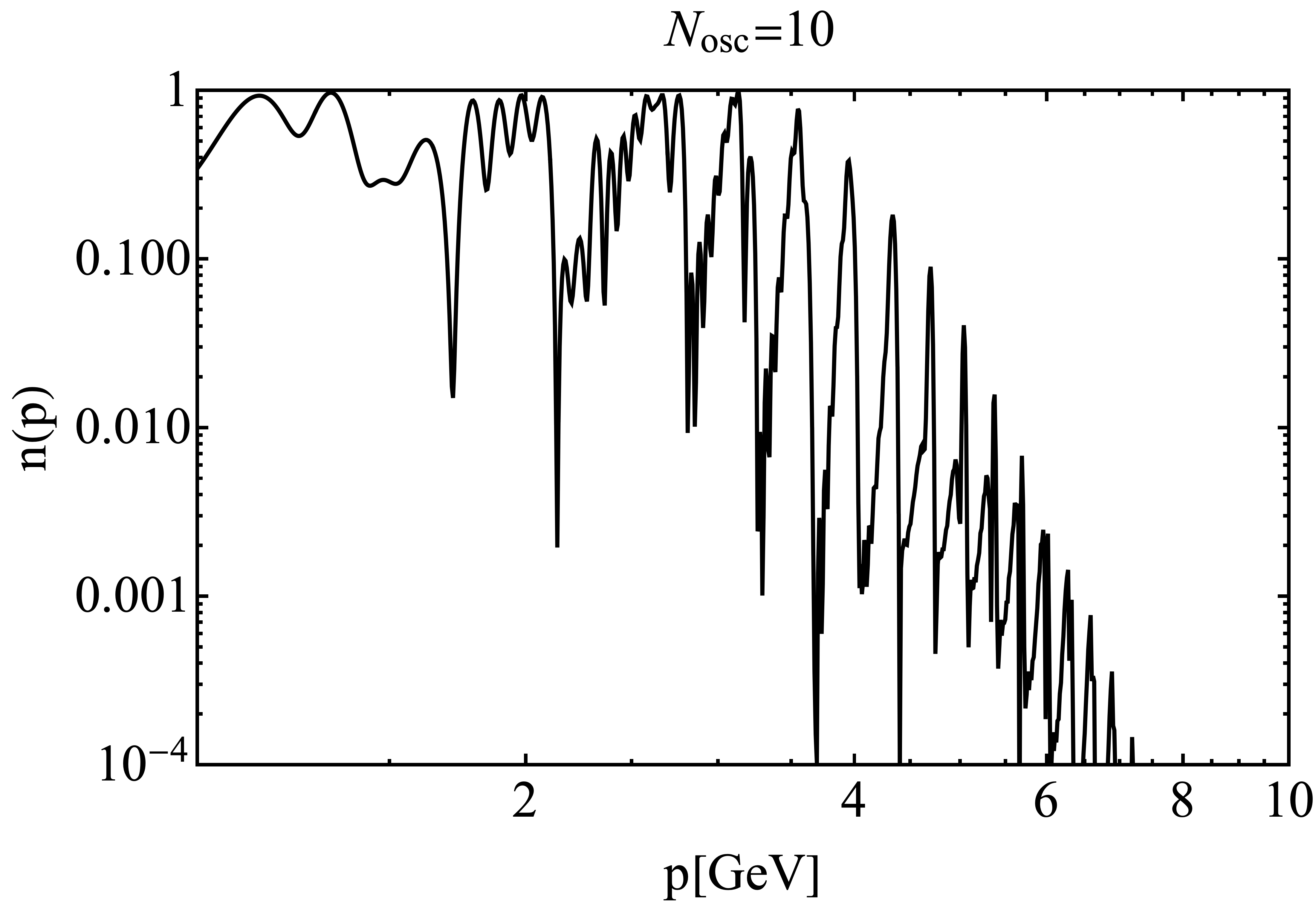}
    \captionsetup{justification=raggedright, singlelinecheck=false}
    \caption{Two different snapshots of the occupation number distribution for the Majorana fermion $\psi_1$, namely when the number of dilaton oscillation $N_{\rm osc}=3$ (left) and at the time of freeze-out, when $N_{\rm osc}=N_{\rm FO}=10$ (right). We have taken $m_{\psi_1}=5 \, \rm GeV$, $m_\varphi=0.5 \, \rm GeV$, $b=-2$, $\xi_0=0.8$, and $r=300$.}
    \label{fig:occupation_number}
\end{figure*}

Once the energy density produced in $\psi$ becomes comparable to the energy density stored in the oscillating dilaton field, the backreaction stops the parametric excitation.
The back reaction will effectively work as a friction term that will decrease the $\varphi$ oscillation amplitude over time, that is, $\xi(t)$ would decrease over time. When $\xi(t)$ cannot satisfy the condition~\eqref{Eq:neccCondn}, the $\psi$ particle density is frozen out.
The final freeze-out number density for $\psi$ is described as 
\begin{align}
    n_\psi^{\rm FO}= \frac{p^3_{\rm max}}{3\pi^2} \ ,
    \label{kmax}
\end{align}
 where the final density approximates a Fermi sphere up to the cut-off momentum $p_{\rm max}$. $p_{\rm max}$ essentially represents the maximum value of the physical momentum for $\psi$ below which the evolution can be non-adiabatic and hence particle production can take place. A trial mode function can be chosen as
\begin{align}
    u_+(t)=\sqrt{1+\frac{m_\psi^{\rm eff}}{\omega}}\times \exp \left(-i\int^t \omega dt' \right),
    \label{Eq:upTrial}
\end{align}
which results into $\beta(p)=0$. The trial mode in Eq.~\eqref{Eq:upTrial} satisfies the equation of motion in Eq.~\eqref{uEoM} approximately when the following condition holds:
\begin{align}
    \frac{1}{2} \left(1+\frac{m_\psi^{\rm eff}}{\omega} \right)\left| \frac{\ddot{m}_\psi^{\rm eff}}{\omega}+\frac{1}{2}\left( \frac{\dot{m}_\psi^{\rm eff}}{\omega}\right)^2\left( 1-\frac{5m_\psi^{\rm eff}}{\omega} \right)\right|\ll \omega^2 \ .
    \label{Eq:AdiaTrial}
\end{align}
Therefore, particle production takes place only when the condition in Eq.~\eqref{Eq:AdiaTrial} is violated. Hence, $p_{\rm max}$ can be determined from the following equation,
\begin{align}
    \frac{1}{2p_{\rm max}^2}\left | \frac{\ddot{m}_\psi^{\rm eff}}{p_{\rm max}}+\frac{1}{2}\left( \frac{\dot{m}_\psi^{\rm eff}}{p_{\rm max}}\right)^2 \right |_{t=\bar t}=1 \ ,
    \label{Eq:pmaxEv1}
\end{align}
where we use $m_\psi^{\rm eff}\simeq 0$ since particle production takes place at the instant of vanishing effective mass, denoted as $\bar t$, which is evaluated from Eq.~\eqref{Eq:meff2} to be
\begin{align}
    \cos(m_\varphi \bar t) = \frac{1}{\xi (\bar{t})} \left( r^{\frac{1}{2b}}-1\right) \ .
    \label{Eq:tbar1}
\end{align}
Finally, using Eqs.~\eqref{Eq:pmaxEv1},~\eqref{Eq:tbar1}, the analytical form of particle density is given by
\begin{align}
\label{NEK}
    \nonumber
    & n_\psi^{\rm FO}  \simeq \mathcal{M}(r,b)\left ( \frac{m_\psi m_\varphi^2}{6 \pi^2} \right ) ,
    \\
    & \mathcal{M}(r,b)  \equiv \left [ \frac{2br}{1-r}\left( r^{-\frac{1}{2b}}-1 \right) \right] ,
\end{align}
where we take $\xi=\xi_c=|r^{\frac{1}{2b}}-1|$, that is, the critical amplitude for vanishing effective mass.
Now the freeze-out $\varphi$ density stored in the coherent oscillation of the dilaton is evaluated from the condition when the effective mass in Eq.~\eqref{Eq:meff2} can no longer cross zero during its oscillation. This results in
\begin{equation}
    \rho_{\varphi}^{\rm FO} \simeq \frac{1}{2} m_\varphi^2 \langle \phi \rangle^2 \left[1-r^{\frac{1}{2b}} \right]^2 \ .
    \label{Eq:rhoPhiFO}
\end{equation}
Hence, one obtains
\begin{equation}
    \frac{\rho_\psi^{\rm FO}}{\rho_\varphi^{\rm FO}} \simeq \frac{m_\psi^2}{3 \pi^2 \langle \phi \rangle^2} \frac{{\cal M}(r,b)}{\left(1-r^{\frac{1}{2b}} \right)^2} \ .
    \label{Eq:rhoRatio}
\end{equation}

Realistically, the amplitude of dilaton oscillation would decrease due to backreaction from the $\psi$ occupation number, and therefore, both the dilaton and $\psi$ equations of motion should be solved simultaneously by taking the occupation of $\psi$ into account. Here we use a linearly decreasing $\xi(t)$ to simulate the backreaction effect, where particle production stops after a given number of dilaton oscillations, denoted as $N_{\rm FO}$, namely
\begin{align}
    \xi(t) = \xi_0 -(\xi_0-\xi_c) \left( \frac{m_\varphi t}{2 \pi N_{\rm FO}} \right) \ .
    \label{Eq:xiT}
\end{align}
We have checked that the final abundance is largely insensitive to different modeling of the backreaction effects by changing the form of Eq.~\eqref{Eq:xiT}. Fig.~\ref{fig:occupation_number} shows the occupation number distribution $n(p)$ against the physical momentum of $\psi_1$, denoted as $p$, for two different snapshots of time, namely after a different number of dilaton oscillations, $N_{\rm osc}=3$ (left), and $N_{\rm osc}=10$ (right). Here we have chosen $m_\psi=5$ GeV, $m_\varphi=0.5$ GeV, $b=-2$, $r=300$, $N_{\rm FO}=10$, and $\xi_0=0.8$. 
We numerically evaluate the total occupation number density from Eq.~\eqref{ONC} for the distribution shown in Fig.~\ref{fig:occupation_number} and compare it to the total density as estimated from Eq.~\eqref{kmax}. To illustrate, in Fig.~\ref{fig:occupation_number} the effective $p_{\rm max}=2.0\, \rm GeV$ for $N_{\rm osc}=3$ and $p_{\rm max}=2.2\, \rm GeV$ for $N_{\rm osc}=N_{\rm FO}=10$, while the analytical estimation from Eq.~\eqref{NEK} is $2.0 \, \rm GeV$. The occupation number of $\psi$ increases rapidly during the first few oscillations, and soon approaches the frozen-out number density, which approximates a Fermi sphere with the given $p_{\rm max}$. To conclude, we obtain good agreement between the particle number density obtained numerically and analytically as estimated by Eq.~\eqref{NEK}. We now use the results of this section to evaluate the DM abundance and BAU.


\section{DM and baryon asymmetry}
\label{sec:DM}

$\psi_1$ fermions produced by the parametric excitation are in out-of-thermal equilibrium, and through their decays to $\chi$, a non-zero asymmetry in the $\chi$ number density can emerge if $y_{ij}$ contains complex phases.
The Hubble scale during the PT is essentially set by the false vacuum energy, $\rho_{\rm vac} \sim c_{V} \langle \phi \rangle^4$, where $c_V$ is a model-dependent constant. For $\langle \phi \rangle \sim $ GeV, $H \sim \sqrt{\rho_{\rm vac}}/M_{\rm pl} \sim 10^{-10}$ eV. Hence the Hubble time is $t_{\rm H} \sim 10^{-6}$s. Now, the $\psi_1$ decay rate from Eq.~\eqref{Eq:yEqu} implies that its lifetime satisfies $t_{\psi_1} \ll t_{\rm H}$ for $|y| \gg 10^{-9}$. To justify the oscillation picture, $t_{\psi_1}$ should be larger than the timescale of dilaton oscillation $m_\varphi^{-1}$,
\begin{equation}
    |y| \lesssim \left(\frac{ 8 \pi m_\varphi}{m_{\psi_1}}\right)^{1/2} \ ,
    \label{Eq:yMin}
\end{equation}
which is easily satisfied for $|y| \lesssim 10^{-2}$ and $m_\varphi \gtrsim {\rm keV}$. Now, assuming anarchic $|y_{ij}|$ elements, the asymmetry in $\chi_\alpha$ denoted as $\delta {n_\chi}_\alpha = {n_\chi}_\alpha- {n_{\bar \chi}}_\alpha$ ($\alpha = 1,2$) is
\begin{align}
    \delta {n_\chi}_\alpha &\simeq \frac{|y|^2}{4 \pi} \frac{m_{\psi_1}}{m_{\psi_2}} n_{\psi_1}  \theta_{\rm CP}  \ ,
    \label{Eq:deltanChi}
\end{align}
where $\theta_{\rm CP} = {\cal O}(1)$ encodes the complex physical phase in $y_{ij}$. For a more precise estimation, including the effect of the $\mu$-coupling in Eq.~\eqref{Eq:VDS}, see appendix~\ref{appendix:general}.

The inverse decay $\chi \eta \to \psi_1$ tries to bring the system towards thermal equilibrium. Hence, one should ensure that the effective rate of this inverse decay is smaller than the Hubble rate at the maximum temperature achieved after the PT, namely at $T_{\rm RH}$. Using the principle of detailed balance, we can evaluate the ratio of the effective inverse decay rate to the Hubble rate as follows:
\begin{align}
    \nonumber
    &\frac{n_\eta^{\rm eq}  \langle \sigma v \rangle_{\chi \eta \to \psi_1}}{H(T_{\rm RH})} = \frac{\Gamma_{\psi_1 \to \chi \eta}}{H(T_{\rm RH})} \frac{n_{\psi_1}^{\rm eq}}{n_{\chi}^{\rm eq}}
    \\
    &  \simeq \frac{\Gamma_{\psi_1 \to \chi \eta}}{H(T_{\rm RH})} \left[ \frac{\sqrt{\pi/2}}{3  \zeta(3) } e^{-m_{\psi_1}/T_{\rm RH}} \left(\frac{m_{\psi_1}}{T_{\rm RH}} \right)^{3/2} \right] \ ,
    \label{Eq:inverse}
\end{align}
where we have assumed that the $\chi$, $\eta$ are in thermal equilibrium and $m_{\psi_1} > T_{\rm RH}$. For typical parameters, for \textit{e.g.}, $T_{\rm RH} \simeq 0.2$ GeV, which is required to explain the nano-Hz stochastic gravitational wave signal~\cite{Fujikura:2024jto}, and $m_{\psi_1} \simeq 5$ GeV, the term in the square bracket in Eq.~\eqref{Eq:inverse} obtains a strong exponential suppression, and for typical values of the coupling $|y| \simeq 10^{-4}$ as obtained later to explain the BAU, the ratio in Eq.~\eqref{Eq:inverse} is smaller than unity. Therefore, it can be ensured conservatively that the inverse decay does not bring the system into thermal equilibrium.

One also has to ensure that the asymmetry-destroying processes are out-of-equilibrium, otherwise, the generated asymmetry would be erased. To suppress $\chi \eta \to \bar\chi \eta$ through s-channel $\psi$ mediation, one needs to demand that the interaction rate for this process is less than the Hubble rate. Due to the portal interaction, the generated $\chi, \eta$ obtain thermal abundance, and it is sufficient to ensure that the rate is smaller than the Hubble rate at the reheating temperature $T_{\rm RH}$,
\begin{align}
   \left(\frac{ \zeta(3)}{ \pi^2} T_{\rm RH}^3 \right) \left(\frac{|y|^4}{8 \pi m_\psi^2} \right) &\lesssim 0.3 \sqrt{g_*} \frac{T_{\rm RH}^2}{M_{\rm pl}} \ ,
   \label{Eq:ymax1}
\end{align}
while the reheating temperature can be estimated using the false vacuum energy $\rho_{\rm vac}$ as 
\begin{equation}
    T_{\rm RH} \simeq \left( \frac{30 c_V}{\pi^2 g_*} \right)^{1/4} \langle \phi \rangle \ .
    \label{Eq:TRH}
\end{equation}
Eqs.~\eqref{Eq:ymax1},~\eqref{Eq:TRH} imply
\begin{align}
    |y| \lesssim 2 \times 10^{-4} \left( \frac{g_*}{100}\right)^{1/16} \left( \frac{m_\psi}{1 {\ \rm GeV}}\right)^{1/2} \left( \frac{\langle \phi \rangle}{1 {\ \rm GeV}}\right)^{-1/4} \ .
    \label{Eq:ymax2}
\end{align}
Furthermore, the Majorana mass mediated $U(1)_{\rm D}$ violating $\psi_1 \psi_1 \to \varphi \varphi$ annihilation rate should be smaller than the $\psi_1$ decay rate, which requires
\begin{align}
    \left( \frac{\rho_{\psi_1}^{\rm FO}}{m_{\psi_1}} \right) \left[ \frac{m_{\psi_1}^2 }{16 \pi^2 \langle \phi \rangle^4 } \left(\frac{2b}{1-r}\right)^4 \right] \lesssim \frac{|y|^2}{8 \pi} m_{\psi_1} \ .
    \label{Eq:psitophiWashout}
\end{align}
Together with the condition $t_{\psi_1} > m_\varphi^{-1}$, it is ensured that the backreaction due to this annihilation on the dilaton oscillation is negligible.
Let us introduce two dimensionless scales $\varepsilon \equiv m_\varphi/\langle \phi \rangle $, $\delta \equiv m_{\psi_1}/\langle \phi \rangle$. Then, Eq.~\eqref{Eq:psitophiWashout} becomes 
\begin{align}
     \varepsilon \delta  \lesssim  |y| \  {\cal W}(r,b) \ ,
     \label{Eq:epd}
\end{align}
where the dimensionless wash-out factor is defined as 
\begin{align}
    {\cal W}(r,b) \equiv \left[\frac{8  b^5 r}{3 \pi^3 (1-r)^5} \left( r^{-\frac{1}{2b}}-1 \right)\right]^{-1/2} \ .
    \label{Eq:Fdef}
\end{align}
Eqs.~\eqref{Eq:ymax2},~\eqref{Eq:psitophiWashout} imply
\begin{align}
    \varepsilon \lesssim 0.2 \left(\frac{1}{\delta}\right)^{1/2} \left( \frac{g_*}{100}\right)^{1/16}  \left( \frac{\langle \phi \rangle}{1 {\ \rm GeV}}\right)^{1/4} \left[ \frac{{\cal W}(r,b)}{10^3} \right]\ .
    \label{Eq:epsilonDelta}
\end{align}
In addition, to treat the effective coupling of the dilaton perturbatively, one demands (see Eq.~\eqref{Eq:radionPerturbative})
\begin{align}
    \left|\frac{2b}{1-r}\right| \lesssim \frac{4 \pi}{\delta}\ .
    \label{Eq:gammaMax}
\end{align}

After the dilaton condensate decays perturbatively, the Universe is reheated to $T_{\rm RH}$ as given in Eq.~\eqref{Eq:TRH}. Therefore, the entropy density at the time of reheating is evaluated to be 
\begin{equation}
    s(T_{\rm RH}) = \frac{2 \pi^2}{45} g_{*s} T_{\rm RH}^3 \ , 
    \label{Eq:sRH}
\end{equation}
where $g_{*s}$ denotes the number of degrees of freedom for entropy density. Now, utilizing Eqs.~\eqref{Eq:deltanChi},~\eqref{Eq:TRH},~\eqref{Eq:sRH} the final dark asymmetry is given by
\begin{align}
    \nonumber
    \frac{\delta n_{\chi_\alpha}}{s(T_{\rm RH})} & \simeq \frac{|y|^2}{4 \pi} \frac{m_{\psi_1}}{m_{\psi_2}} \frac{ \rho_{\psi_1}^{\rm FO}}{m_{\psi_1} s(T_{\rm RH})} \theta_{\rm CP} \\[1ex]
    & \simeq (3.5 \times 10^{-4}) |y|^2 \varepsilon^2 \delta {\cal M}(r,b) \theta_{\rm CP}  \ .
    \label{Eq:darkAsymmetry}
\end{align}
Here, we have assumed $m_{\psi_1}/m_{\psi_2} = 1/2$, $g_* \simeq g_{*s} = 100$, $c_{V}=1/2$.

As the electroweak sphalerons are frozen out well before the relevant PT temperature, to share the asymmetry between the DS and visible sector, one needs a portal interaction that violates both the baryon number $B$ and dark number $D$. The lowest-dimensional effective operator satisfying the criterion is the neutron portal operator,
\begin{equation}
    {\cal O}_{n} = \frac{1}{\Lambda_n^2} \chi_2 u_{R,i} d_{R, j} d_{R, k} \ ,
    \label{Eq:portalOp}
\end{equation}
where $i,j,k$ are generational indices. The stability of proton and bound neutrons is ensured by having $\chi_2$ heavier than the neutron. Note that the DM candidate $\chi_1$ should not appear in Eq.~\eqref{Eq:portalOp}, as it would cause the DM to decay with an unacceptable lifetime if $\chi_1$ is heavier than a neutron, or it would induce nucleon decay otherwise. This can be ensured, for example, in an extra-dimensional embedding of the model, where $\chi_2$ is localized towards the UV brane of a warped extra dimension, where the SM fields live, while $\chi_1$ is peaked towards the IR brane. The portal operator in Eq.~\eqref{Eq:portalOp} violates $B$, $D$ by 1 unit, but conserves $B+D$. DM stability is ensured by the wavefunction suppression, while the $\chi_2$-$\chi_1$ separation also suppresses off-diagonal $\mu$ terms in Eq.~\eqref{Eq:VDS}, hence, $\chi_2$ decay transfers asymmetry predominantly to the visible sector (see appendices~\ref{appendix:general},~\ref{appendix:5d} for more details). The neutron portal operator is probed by mono-jet searches in colliders, in processes such as $ud \to \bar d \bar \chi_2$, $dd \to \bar u \bar\chi_2$, and the current constraints from the LHC imply $\Lambda_n \gtrsim 2$ TeV~\cite{ATLAS:2021kxv, Ciscar-Monsalvatje:2023zkk}. As the decay rate of $\chi_2$ is 
\begin{equation}
    \Gamma_{\chi_2} \simeq \frac{m_{\chi_{2}}^5}{192 \pi^3 \Lambda_n^4} \ ,
    \label{Eq:GammaX2}
\end{equation}
$\Lambda_n \lesssim 200$ TeV is sufficient to ensure that $\chi_2$ decays to SM particles before the BBN for $m_{\chi_2} \gtrsim 1.5$ GeV. It is straightforward to show that the final dark number asymmetry $\delta n_{\rm ADM}$ and visible baryon asymmetry $\delta n_{\rm BAU}$ are given by the following equations for $N_\chi = 2$,
\begin{align}
    \nonumber
    \delta n_{\rm ADM} & \simeq  \delta n_{\chi_1} + \delta n_{\chi_2} \cdot  {\cal B}(\chi_2 \to \chi_1 + \eta) \ ,  \\[1ex]
    \delta n_{\rm BAU} &  \simeq \delta n_{\chi_2}  \cdot {\cal B}(\chi_2 \to udd)    \ ,
    \label{Eq:BAUfinal}
\end{align}
where $\delta n_{\chi_\alpha}$ has been obtained from Eq.~\eqref{Eq:darkAsymmetry}, and ${\cal B}$ denotes the branching ratio. We note that Eq.~\eqref{Eq:BAUfinal} is consistent with $B+D$ being a good symmetry, whose only source is the decay of Majorana $\psi$, as $\delta n_{\rm BAU}+ \delta n_{\rm ADM} = \sum_\alpha \delta n_{\chi_\alpha}$. Further, the ratio of the asymmetry in the DS and visible sector determines the DM mass, and can be controlled by the branching ratio in Eq.~\eqref{Eq:BAUfinal}, namely
\begin{align}
    \nonumber
    m_{\rm \chi_1} &\simeq 5 \left| \frac{\delta n_{\rm BAU}}{\delta n_{\rm ADM}} \right| {\ \rm GeV} \\
    & \simeq 5 \left[ \frac{1-{\cal B}(\chi_2 \to \chi_1+\eta)}{1+{\cal B}(\chi_2 \to \chi_1+\eta)} \right] {\ \rm GeV} \ .
    \label{Eq:DMmass}
\end{align}
Here, we have used the fact that for ${\cal B}(\chi_2 \to \chi_1+\eta) \lesssim 1$, $\delta n_{\chi_1} \simeq \delta n_{\chi_2}$. For instance, if one makes ${\cal B}(\chi_2 \to \chi_1 + \eta) = 2/3$, then $m_{\chi_1} \simeq 1$ GeV.

\begin{figure*}[t!]
	\centering
		\begin{subfigure}{0.45\textwidth}
		\centering
		\includegraphics[width=\textwidth]{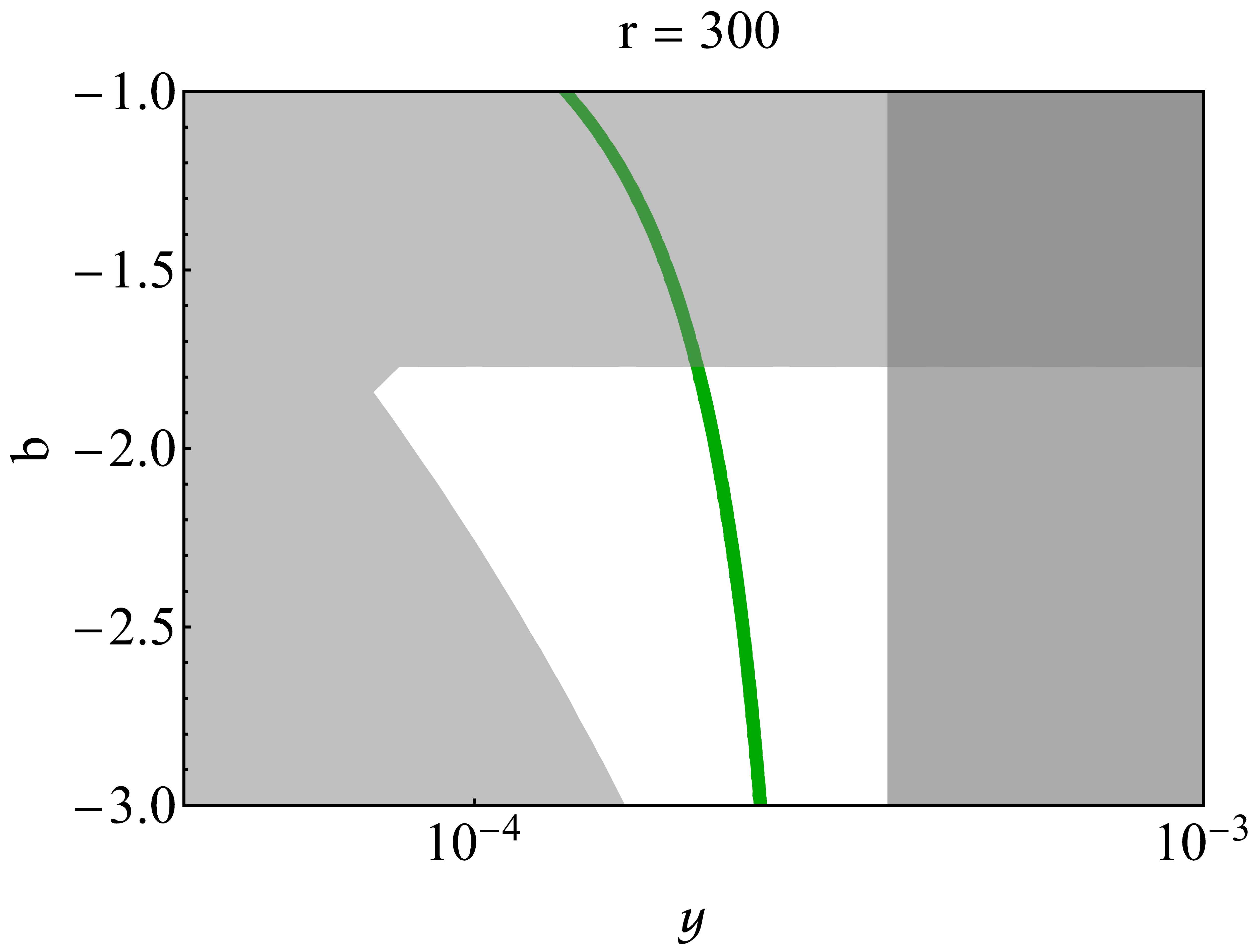}
		\subcaption{}
		\label{fig:pby}
		\end{subfigure}   
		\begin{subfigure}{0.45\textwidth}
		\centering
		\includegraphics[width=\textwidth]{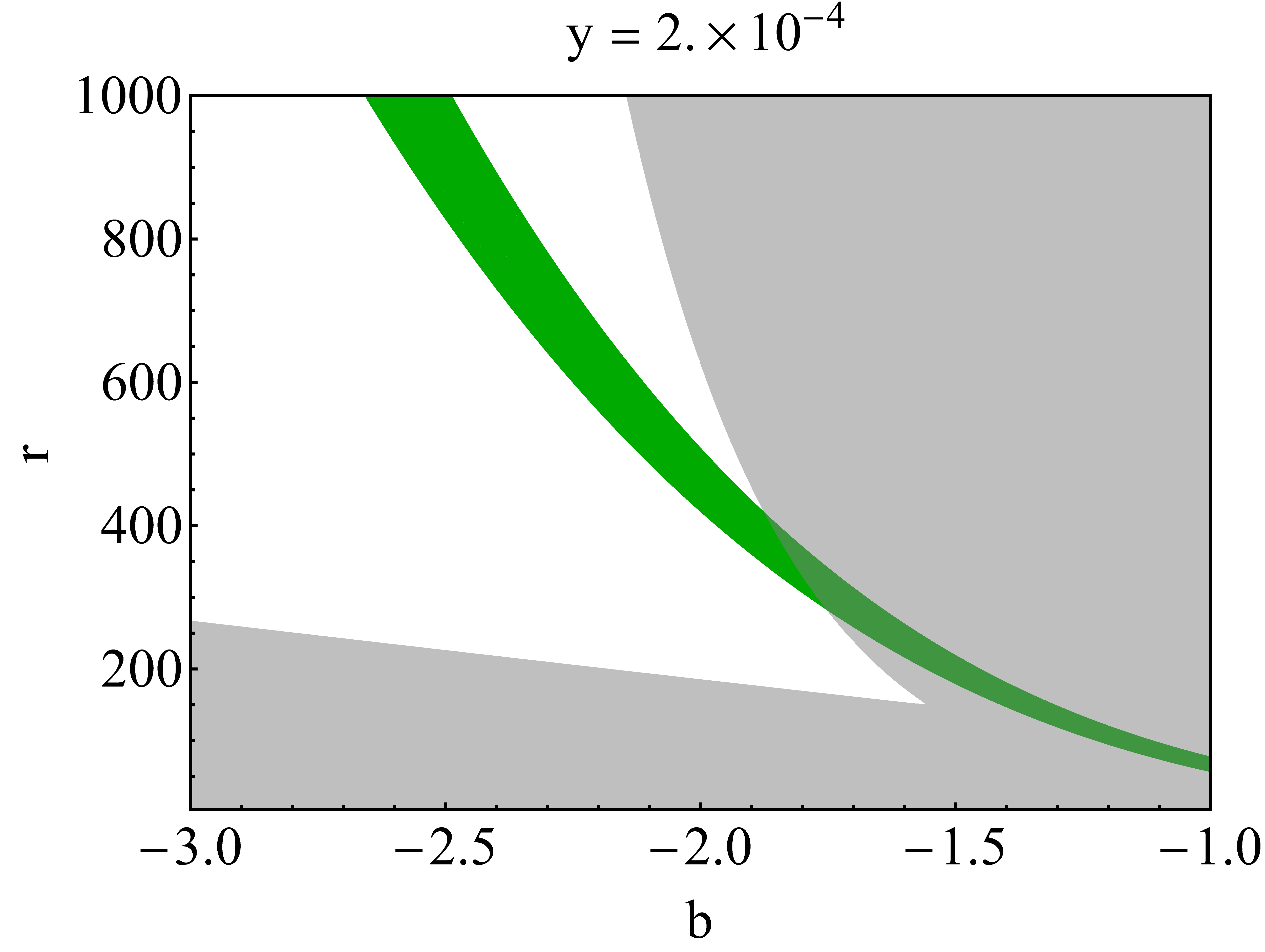}
		\subcaption{}
		\label{fig:pbr}
		\end{subfigure}
\hspace{0.4cm}
        \begin{subfigure}{0.45\textwidth}
		\centering
		\includegraphics[width=\textwidth]{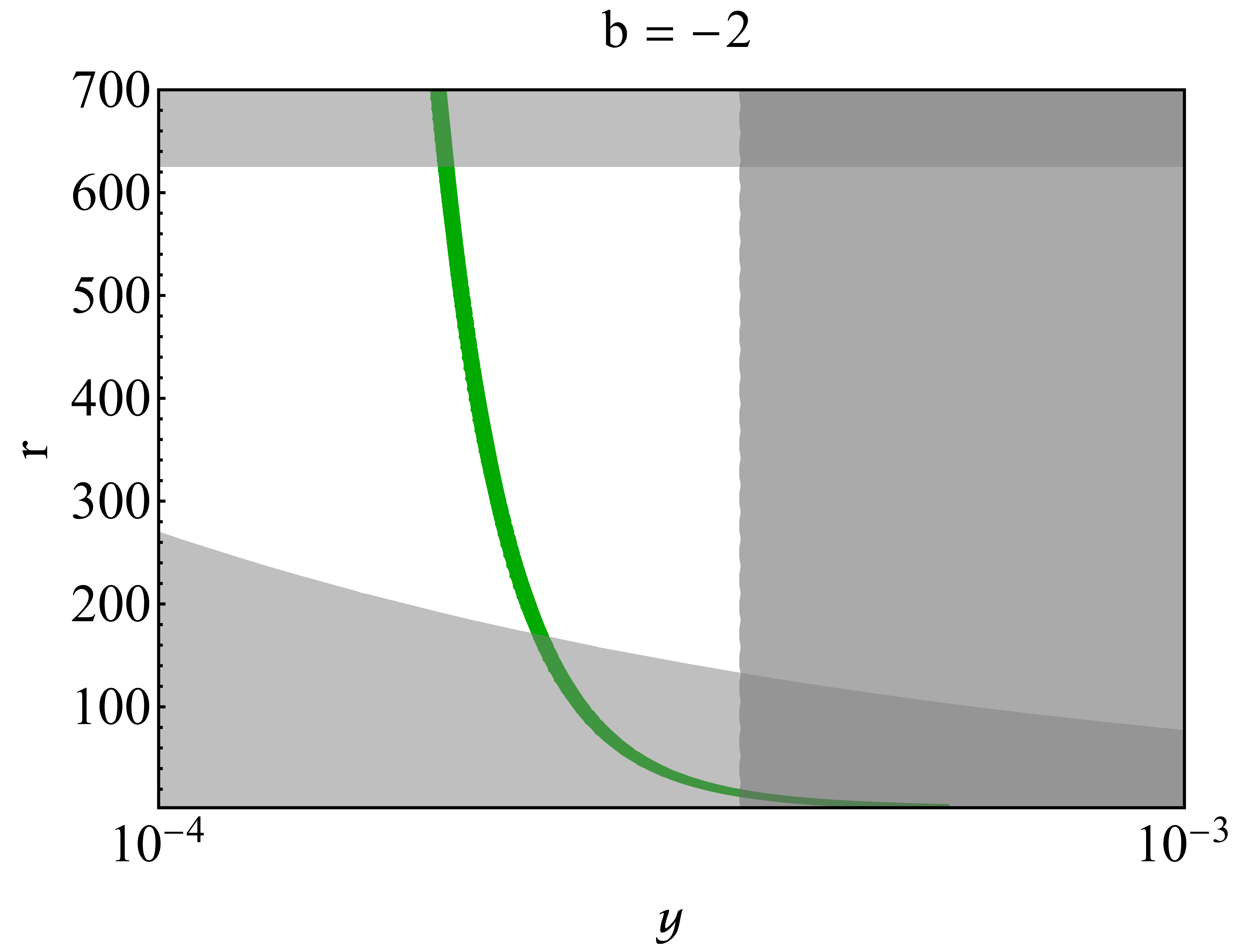}
		\subcaption{}
		\label{fig:pry}
		\end{subfigure}
    \captionsetup{justification=raggedright, singlelinecheck=false}
	\caption{
		(a) Contour in the parameter space of the bulk mass parameter for $\psi_1$, denoted as $b$ and coupling $|y|$, where the observed baryon asymmetry is reproduced (green), while in the dark shaded regions, washout processes destroy the asymmetry generated as embodied in Eqs.~\eqref{Eq:ymax2},~\eqref{Eq:epd}, or the particle production condition in Eq.~\eqref{Eq:neccCondn} is not satisfied. In (b), (c), similar contours are depicted in the $r$-vs-$b$ and $r$-vs-$y$ parameter space, where $r$ is the ratio of the effective UV and IR localized masses for $\psi_1$ as defined in Eq.~\eqref{Eq:rdef}. In all the subfigures, the chosen parameters are $\langle \phi \rangle=1 $ GeV, $m_{\psi_1}=5$ GeV, $m_\varphi = 0.5$ GeV, $\xi_0=0.8$, $g_*=100$, and $c_V = 0.5$, while the other fixed parameter for each subfigure is mentioned in the plot label.
	}
	\label{fig:paraSpace}
\end{figure*}

Fig.~\ref{fig:paraSpace} shows the viable parameter region as a green band where the observed BAU is reproduced, while in the gray shaded regions the asymmetry generated is washed out. Figs.~\ref{fig:pby},~\ref{fig:pbr},~\ref{fig:pry} depict the dependence of the BAU and the interplay of three possible combinations of the relevant parameters $b$, $y$, $r$, respectively. There we have chosen $\langle \phi \rangle = 1$ GeV, $m_{\psi_1}/m_{\psi_2} = 0.5$, $c_V = 0.5$, $\delta =5$, $\xi_0=0.8$ and $\varepsilon=0.5$. In the subfigures,  ${\cal B}(\chi_2 \to \chi_1 + \eta) = 2/3$. Hence, the DM mass of $m_{\chi_1} = 1$ GeV can account for the observed DM abundance. We also choose $m_\eta$ so that $m_{\chi_1}+ m_\eta<m_{\chi_2}<m_{\psi_1}-m_\eta$. The regions that do not satisfy the washout criteria in Eqs.~\eqref{Eq:ymax2},~\eqref{Eq:psitophiWashout}, or fail to satisfy the mother particle production criteria in Eq.~\eqref{Eq:neccCondn} are shaded in gray. The precise region and contours in Fig.~\ref{fig:paraSpace} can be varied somewhat by the different hierarchical structure in $y_{ij}$, and are illustrated in appendix~\ref{appendix:general}. Similarly, the ratio of the DS and visible asymmetries can be modified by order one factors utilizing the branching ratio of $\chi_2$, for example, with the variation of the wavefunction profile in the extra dimension, and therefore an appropriate ${\cal O}(\rm GeV)$ $\chi_1$ can explain the DM abundance today.

\begin{figure*}[t!]
	\centering
		\begin{subfigure}{0.44\textwidth}
		\centering
		\includegraphics[width=\textwidth]{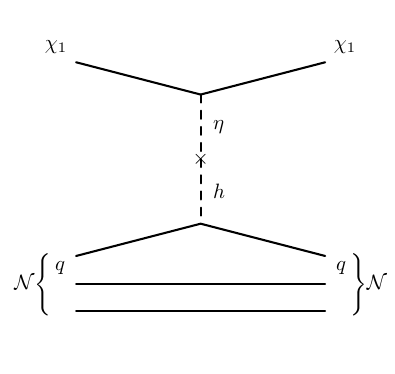}
		\subcaption{}
		\label{fig:DD}
		\end{subfigure}   
\hspace{0.4cm}
		\begin{subfigure}{0.48\textwidth}
		\centering
		\includegraphics[width=\textwidth]{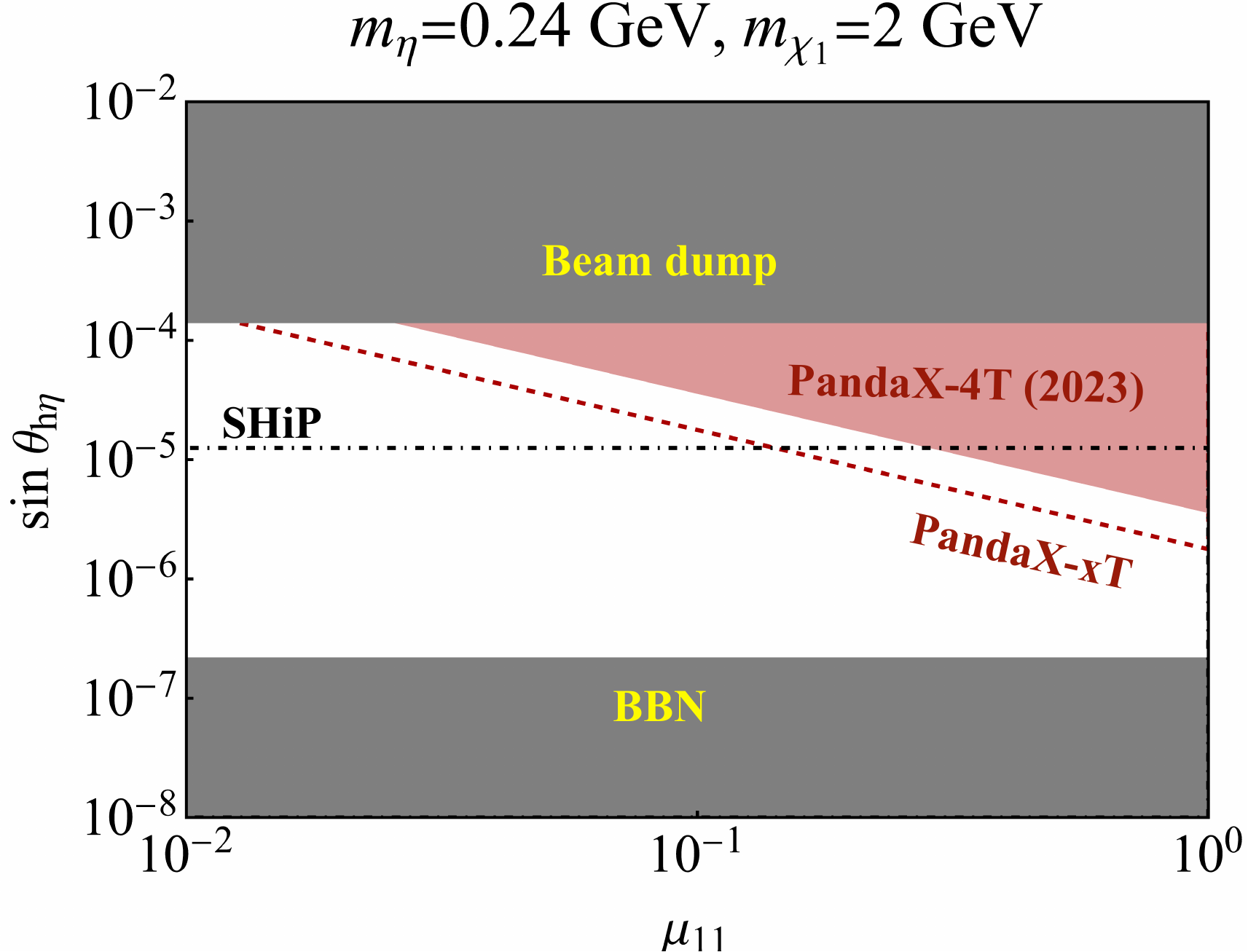}
		\subcaption{}
		\label{fig:pDDh}
		\end{subfigure}
    \captionsetup{justification=raggedright, singlelinecheck=false}
	\caption{(a) Feynman diagram for DM-nucleon interaction. (b) Constraints in the Higgs-portal coupling vs. $ \mu_{11}$ parameter space from DM direct detection experiment (red solid), beam dump experiments, and BBN (gray) for $m_{\chi_1}=2$ GeV, and $m_\eta=0.24$ GeV. Future projected bounds from SHiP and PandaX-xT are portrayed as black dot-dashed and red dashed lines.}
	\label{fig:pDDAll}
\end{figure*}

Let us now consider the lifetime of the dilaton and how the frozen out dilaton condensate decays perturbatively due to the couplings in Eq.~\eqref{Eq:HiggsPortal}. When $m_\varphi>2 m_\eta$, onshell $\varphi \to \eta \eta$ decay can proceed with a rate
\begin{align}
    \Gamma_{\varphi \to \eta \eta} \simeq \left( \frac{m_\eta^2}{\langle \phi \rangle}\right)^2 \frac{1}{8 \pi m_\varphi} \left(1-\frac{4 m_\eta^2}{m_\varphi^2} \right)^{1/2} \ .
    \label{Eq:GammaphiPerturbative}
\end{align}
For $m_\varphi=0.5$ GeV, and $\langle \phi \rangle = 1$ GeV, the perturbative lifetime of the dilaton condensate, denoted $t_\varphi$ is smaller than 1s for $2 {\, \rm keV} \lesssim m_\eta < m_\varphi/2$, which satisfies the BBN constraint. Furthermore, $t_\varphi \gg m_\varphi^{-1}$ for this whole parameter region as required by the particle production and dilaton oscillation picture to hold. Finally, the $\eta$ can decay to the SM, thanks to the Higgs portal coupling. The Higgs portal scalar has been studied extensively in the literature, and it is found that there exists a range of $\theta_{h\eta}$ and $m_\eta$ such that all laboratory and BBN constraints are satisfied (see for \textit{e.g.,} Ref.~\cite{Winkler:2018qyg}). To be more quantitative, let us take $2 m_\mu < m_\eta < m_\varphi/2$ such that $\eta$ can decay to on-shell $\mu^+ \mu^-$ pair. The corresponding decay rate is given by
\begin{align}
    \Gamma_{\eta \to \mu^+ \mu^-} \simeq  \sin^2 \theta_{h\eta}\frac{ G_{\rm F} m_\eta m_\mu^2}{4 \sqrt{2} \pi} \left( 1-\frac{4 m_\mu^2}{m_\eta^2} \right)^{3/2} \ ,
    \label{Eq:etaRate}
\end{align}
where $G_{\rm F} = 1.166 \times 10^{-5}$ GeV$^{-2}$ is the Fermi constant. Therefore, $\eta$ can decay with a lifetime smaller than $1$s provided $|\theta_{h\eta}| \gtrsim 10^{-7}$. On the other hand, LHCb constraints on $B \to K^{(*)} \eta(\to \mu^+ \mu^-)$, and constraints from beam dump experiments can be satisfied if $|\theta_{h \eta}| \lesssim  10^{-4}$~\cite{Winkler:2018qyg}. Similarly, a range of parameter space exists for $|\theta_{h \eta}|$-vs-$m_\eta$, which is consistent with all current experiments and also fulfills the cosmological bounds. This will be probed further by the SHiP experiment~\cite{SHiP:2025ows}.

The symmetric component of the DM is annihilated into $\eta$ and $\varphi$, when $m_{\chi_1} > m_\varphi$. The total annihilation cross-section is given by
\begin{equation}
  \langle  \sigma_{\rm ann} v \rangle_{\bar \chi_1 \chi_1 } \simeq \frac{ \mu_{11}^4}{8 \pi m_{\chi_1}^2 } + \frac{m_{\chi_1}^2}{8 \pi \langle \phi \rangle^4}  \ ,
  \label{Eq:annihilation}
\end{equation}
where $\mu_{11}$ is the coupling defined in Eq.~\eqref{Eq:VDS}. One usually requires $\langle  \sigma_{\rm ann} v  \rangle_{\bar \chi_1 \chi_1} \gtrsim 0.6 \times 10^{-25}$ cm$^{3}/s$ to ensure that the symmetric component annihilates well before structure formation. This is easily satisfied when $m_{\chi_1} \lesssim 5$ GeV.\footnote{
The mediation of $\eta$ and $\varphi$ creates an attractive force and causes self-interaction between $\chi_1$ that can be relevant for small-scale structure issues~\cite{Tulin:2017ara, Girmohanta:2022dog}.} Note that the process $\bar \chi_{1} \chi_1 \to \eta \eta $ via the $\mu_{11}$ coupling does not violate the $U(1)_{\rm D}$ number, and therefore, $\mu_{11}$ is not subjected to the washout constraints.

\begin{figure*}[t!]
	\centering
		\begin{subfigure}{0.47\textwidth}
		\centering
		\includegraphics[width=\textwidth]{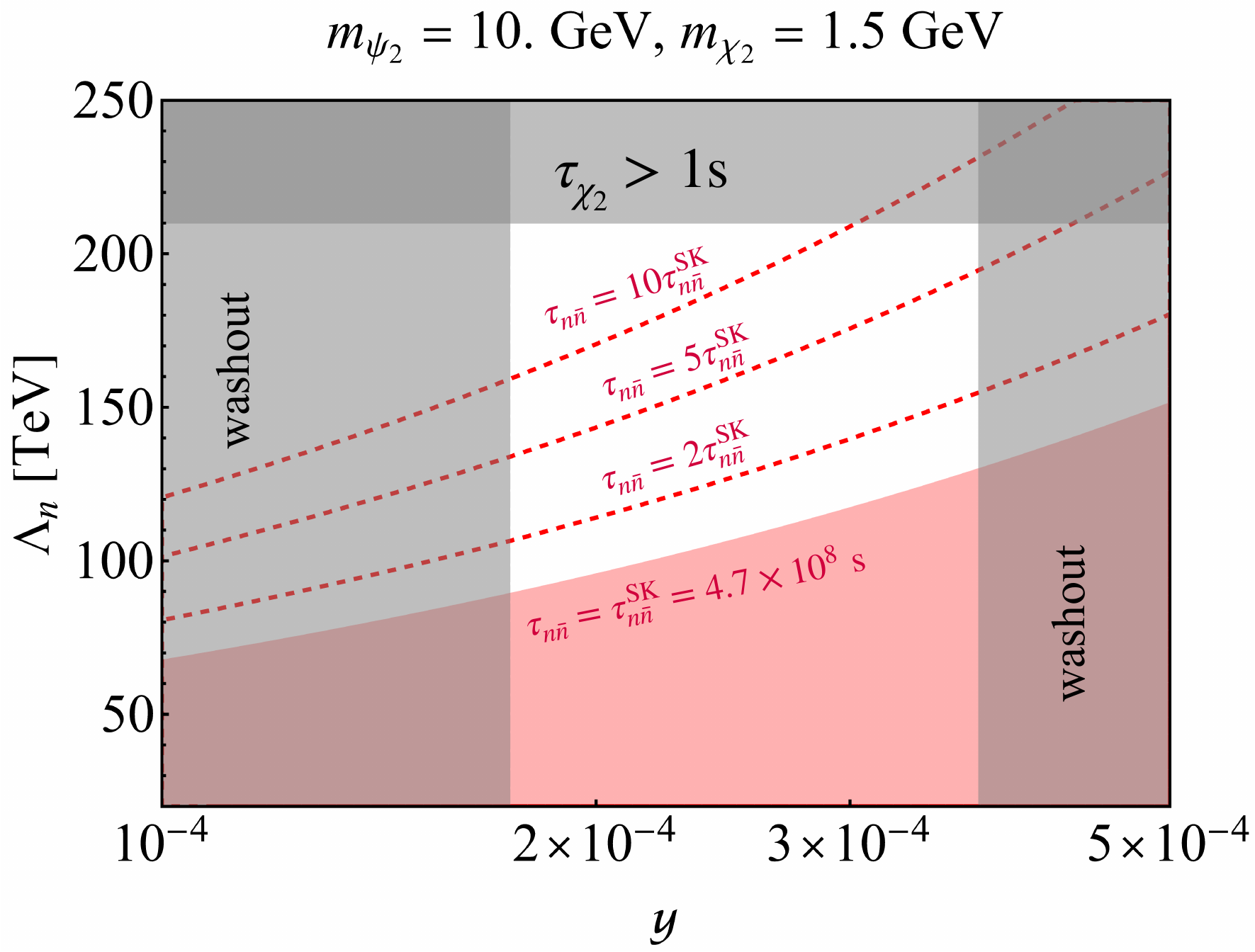}
		\subcaption{}
		\label{fig:nnby}
		\end{subfigure}   
\hspace{0.4cm}
		\begin{subfigure}{0.44\textwidth}
		\centering
		\includegraphics[width=\textwidth]{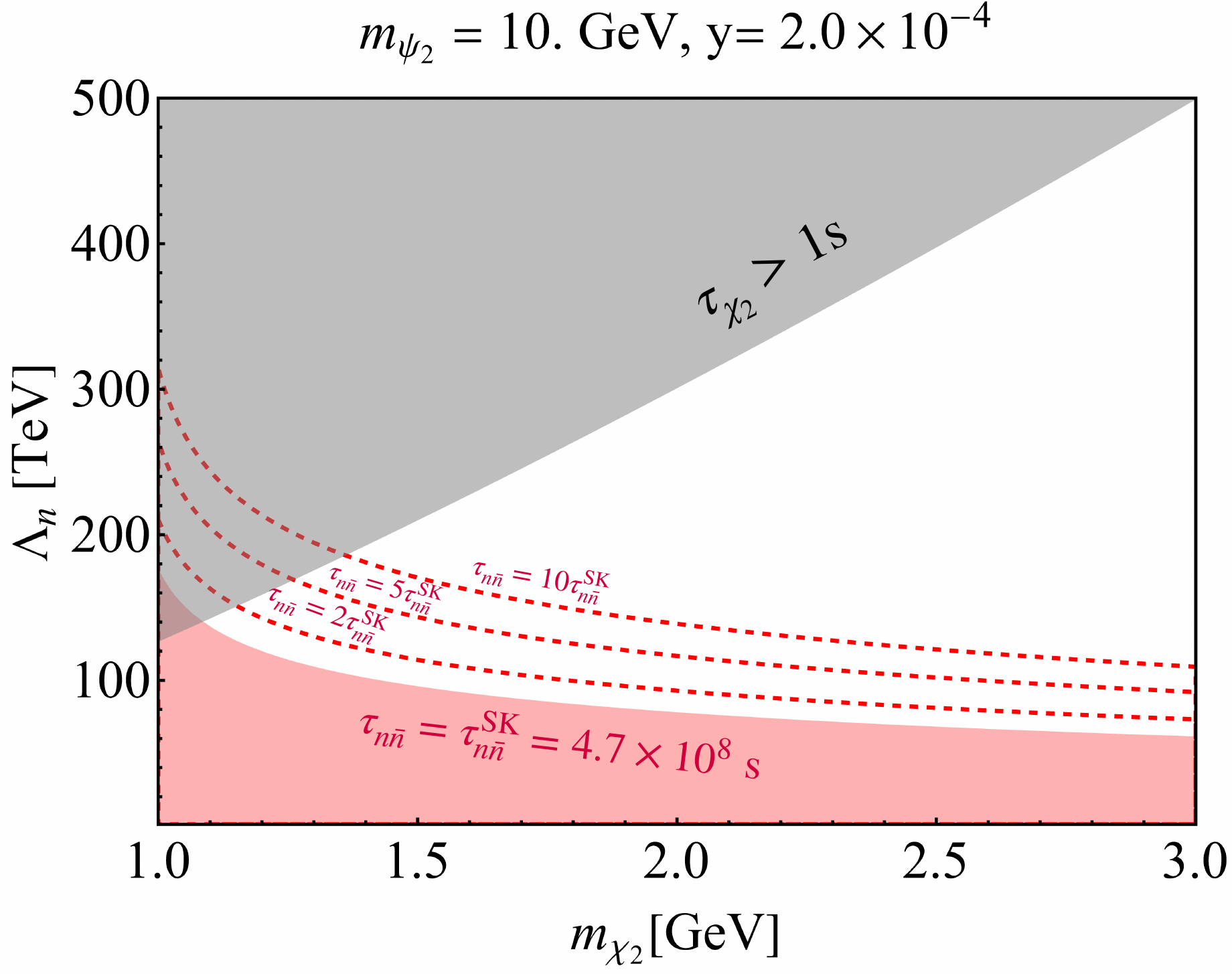}
		\subcaption{}
		\label{fig:nnbmx}
		\end{subfigure}
\hspace{0.4cm}
         \begin{subfigure}{0.60\textwidth}
        \centering
		\includegraphics[width=\textwidth]{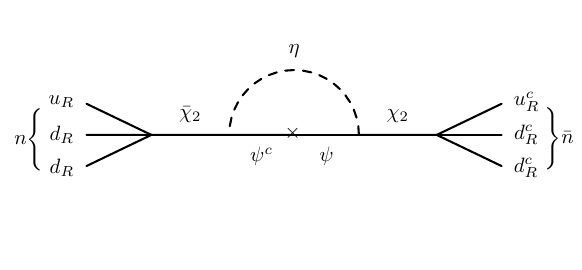}
		\subcaption{}
        \label{fig:nnbFeyn}    
        \end{subfigure}
    \captionsetup{justification=raggedright, singlelinecheck=false}
	\caption{(a, b) Constraint from $n$-$\bar n$ oscillations (red), and future projections by a factor of $2, 5,$ and $10$ above the current constraint (red dashed). Gray regions are constraints from successful baryogenesis and BBN, where we have chosen $\langle \phi \rangle=1$ GeV, $m_{\psi_1}/m_{\psi_2}=0.5$, $m_\varphi=0.5$ GeV, $b=-2$, $r=200$, $\xi_0=0.8$, and $c_V=0.5$ to determine the washout constraints for the left subfigure. All other relevant parameters are mentioned on the plot labels. (c) Feynman diagram for $n$-$\bar n$ oscillations generated by the Majorana mass for $\psi$.}
	\label{fig:nnb}
\end{figure*}

Due to the Higgs-portal interaction, the DM also has a non-vanishing scattering cross-section with the SM nucleons, which are probed by direct detection experiments. The dominant DM-nucleon interaction is generated due to $\eta$ mediation, as shown in Fig.~\ref{fig:DD}, and the spin-independent direct detection cross-section is evaluated to be
\begin{align}
    \nonumber
    \sigma_{\chi_1 {\cal N}}^{\rm SI} & \simeq \frac{4 \mu_{\chi_1 \cal N}^2}{\pi} \left( \frac{\mu_{11} \sin \theta_{h \eta}}{v} \frac{m_{\cal N}}{m_\eta^2 + m_{\chi_1}^2 v_{\rm DM}^2}\right)^2 \\
    & \times \left( \sum_{Q=u,d,s} f_{T_{Q}}^{(\cal N)} + \frac{2}{9} f_{T_{ G}}^{(\cal N)} \right)^2 \ ,
\end{align}
where $f_{T_{Q,G}}^{(\cal N)}$ encodes the nuclear matrix elements~\cite{Jungman:1995df}, ${\cal N}$ denotes SM nucleon, $\mu_{\chi_1 {\cal N}}$ is the DM-nucleon reduced mass, and $v_{\rm DM} \simeq 10^{-3}c$ is the typical DM velocity in the halo. In Fig.~\ref{fig:pDDh} we depict the constraints from DM direct detection experiments from PandaX-4T~\cite{PandaX:2022aac} (solid red) on the parameter space of $\sin \theta_{h \eta}$ and $\mu_{11}$ for a chosen DM mass of $2$ GeV and $m_\eta=0.24$ GeV. Existing laboratory constraints from the beam dump experiment and cosmological bounds from the success of BBN are portrayed in solid gray~\cite{Winkler:2018qyg}, while the white region can satisfy all the constraints. Future projections from PandaX-xT~\cite{PandaX:2024oxq} and SHiP~\cite{SHiP:2025ows} experiments are also overlaid as red dashed and black dot-dashed lines, respectively.

There is a crucial difference regarding the prediction for neutron-antineutron ($n$-$\bar n$) oscillations in the current setup from Ref.~\cite{Fujikura:2024jto}, where the source of $U(1)_{\rm D}$ violation is the $SU(2)_{\rm D}$ sphaleron-like process. In contrast, the source of $U(1)_{\rm D}$ violation is the Majorana mass of $\psi$ in the current scenario. Therefore, $n$-$\bar n$ oscillations can proceed via the Feynman diagram in Fig.~\ref{fig:nnbFeyn}. This generates the following effective operators:
\begin{align}
    \nonumber
    {\cal O}_{n \bar n}^{(1)} &= c_{n \bar n}^{(1)} \left[ u_{R}^{\alpha T} C u_{R}^\beta\right] \left[ d_{R}^{\gamma T} C d_{R}^\delta\right] \left[ d_{R}^{\rho T} C d_{R}^\sigma\right] (T_s)_{\alpha \beta \gamma \delta \rho \sigma} \ , \\[1ex]
    {\cal O}_{n \bar n}^{(2)} &= c_{n \bar n}^{(2)} \left[ u_{R}^{\alpha T} C d_{R}^\beta\right] \left[ u_{R}^{\gamma T} C d_{R}^\delta\right] \left[ d_{R}^{\rho T} C d_{R}^\sigma\right] (T_s)_{\alpha \beta \gamma \delta \rho \sigma} \ ,
    \label{Eq:Onnb}
\end{align}
where $C$ is the Dirac charge conjugation matrix, and the symmetric color contraction tensor is expressed by
\begin{equation}
    (T_s)_{\alpha \beta \gamma \delta \rho \sigma}  = \epsilon_{\rho \alpha \gamma } \epsilon_{\sigma \beta \delta} + \epsilon_{\sigma \alpha \gamma } \epsilon_{\rho \beta \delta} + \epsilon_{\sigma \beta \gamma } \epsilon_{\rho \alpha \delta} \ .
    \label{Eq:Ts}
\end{equation}
The current experimental limit on the $n$-$\bar n$ oscillation time $\tau_{n \bar n} \equiv |\delta m|^{-1} > 4.7 \times 10^{8} $s~\cite{Super-Kamiokande:2020bov}, where
\begin{align}
    |\delta m| = \sum_{i=1,2} \langle \bar n | {\cal O}_{n \bar n}^{(i)}|n \rangle \simeq \frac{|y|^2}{\Lambda_n^4} \frac{m_{\psi_2}}{16 \pi^2 (m_{\chi_2}^2-m_n^2)} \Lambda_{\rm QCD}^6 \ .
    \label{Eq:Tnnb}
\end{align}
In Figs.~\ref{fig:nnby},~\ref{fig:nnbmx} we show the constraint from $n$-$\bar n$ oscillation in red in the $\Lambda_n$-$y$ and $\Lambda_n$-$m_{\chi_2}$ parameter space, respectively, where we have chosen $m_{\psi_2} = 2 m_{\psi_1} = 10$ GeV, $\langle \phi \rangle = 1$ GeV, $m_\varphi = 0.5$ GeV, $g_* = 100$, $r=200$, $b=-2$ and $c_V = 0.5$. In Fig.~\ref{fig:nnby} $m_{\chi_2} = 1.5$ GeV is fixed, and three gray exclusion regions from theory are depicted, namely regions where $\tau_{\chi_2} > 1$s, and the washout conditions in Eqs.~\eqref{Eq:ymax2},~\eqref{Eq:epd} are not satisfied. In Fig.~\ref{fig:nnbmx}, $y=2 \times 10^{-4}$ is kept fixed instead, letting $m_{\chi_2}$ to vary, while the exclusion region $\tau_{\chi_2} > 1$s is portrayed in gray. We have also overlaid the future projections on $\tau_{n \bar n}$ that will probe our model further. In particular, a factor of $2,5,10$ improvement over the existing limit set by Super-Kamiokande~\cite{Super-Kamiokande:2020bov} is overlaid as red dashed curves. In the future, NNBAR at the ESS plans to improve the limit on $\tau_{n \bar n}$ to $2.8 \times 10^9$s (for 2MW ESS operation), $4.4 \times 10^9$s (for 5MW ESS operation)~\cite{Santoro:2024lvc}. Competitive bounds will also be obtained by DUNE~\cite{DUNE:2024wvj}, Hyper-Kamiokande~\cite{Hyper-Kamiokande:2018ofw}, and JUNO~\cite{JUNO:2021vlw}.

\section{Conclusions}
\label{sec:discuss}

We have realized a post-sphaleron darkogenesis scenario that simultaneously explains the observed baryon asymmetry, dark matter, and the nano-Hz stochastic gravitational wave signal observed by the PTA. A first-order supercooled phase transition in a nearly conformal dark sector generates the mother particle via parametric resonance, satisfying the out-of-equilibrium condition at $\sim$ GeV temperatures. The model predicts potentially observable neutron-antineutron oscillations, testable in NNBAR at ESS~\cite{Santoro:2024lvc}, DUNE~\cite{DUNE:2024wvj}, Hyper-K~\cite{Hyper-Kamiokande:2018ofw}, and JUNO~\cite{JUNO:2021vlw}.

Finally, we note that it is straightforward to extend this framework of baryon asymmetry generation to phase transition at higher temperature, where more parameter space opens up, which corresponds to gravitational waves having peak frequency between the nano-Hz, and milli-Hz range, and could be probed by astrometric measurements of Gaia and THEIA~\cite{Garcia-Bellido:2021zgu}.

\section*{Acknowledgements}

We are grateful to Kohei Fujikura for useful discussions.
YN is supported by Natural Science Foundation of Shanghai.

\appendix

\section{Radion interaction with a bulk Majorana fermion}
\label{sec:radionInt}

Utilizing the holographic AdS/CFT correspondence, a weakly coupled dual description of the phase transition can be given in terms of the radion, which is the stabilized modulus field in a Randall-Sundrum (RS) type warped extra-dimensional model
\cite{Randall:1999ee}.
We derive the radion interaction with a fermion in the bulk of the RS spacetime
having boundary localized mass terms without performing any linear expansion for the radion. This formalism is appropriate for treating the effective mass term for a fermion just after the phase transition, when the radion field is oscillating around its minima, and the oscillation amplitude is not small. We utilize this effective mass as an input to the evolution of the fermion modes and the possible parametric excitation of the fermion.

The RS model contains a ${\mathbf{R}_4 \otimes \mathbf{S}^{1}/{\mathbf{Z}_2}}$ compactified slice of AdS$_5$ with the metric,
\begin{equation}
    ds^2 = g_{M N} dx^M dx^N = e^{-2 k T(x) |\theta|} \eta_{\mu \nu} dx^\mu dx^\nu + T^2(x) d\theta^2 \  ,
    \label{Eq:metric}
\end{equation}
where $k$ is the AdS curvature scale, $\theta \in [-\pi, \pi]$ denotes the extra-dimensional coordinate, $M, N$ represent generic 5D coordinates, while $\mu, \nu=0,\cdots,3$ are reserved for coordinates along the non-compact directions. Two 3-branes are placed at the fixed points $\theta=0, \pi$, referred to as the UV and IR branes, respectively, where $T(x)$ is the modulus field. After the Kaluza-Klein (KK) reduction of the 5D Einstein-Hilbert action constructed from the 5D Ricci scalar, and performing integration over the extra dimension, one obtains a 4D effective action for the radion~\cite{Goldberger:1999un},
\begin{equation}
    S_\varphi \supset \int d^4 x \ \left[ \frac{1}{2}\partial_\mu \varphi \partial^\mu \varphi - V_{\rm eff}(\phi) \right] \ ,
\end{equation}
where the radion $\phi(x) = Z e^{-\pi k T(x)}$ with $Z=\sqrt{24 M_*^3/k}$, and $M_*$ is the 5D Planck constant.
The radion $\phi$ is expanded around its VEV, $\phi = \langle \phi \rangle + \varphi$ where $\langle \phi \rangle = Z e^{-\pi k T_0}$, with $T_0$ being the modulus VEV.
We use the same notation for the radion and dilaton, as they are dual to each other.
The potential $V_{\rm eff}(\phi)$ arises as a result of introducing a bulk scalar with brane localized potentials~\cite{Goldberger:1999uk} or some bulk confining gauge field that breaks the conformal invariance~\cite{Fujikura:2019oyi}.
The PT in the 5D description is given as follows~\cite{Creminelli:2001th}: at a temperature much higher than the IR brane mass scale, the RS geometry becomes unstable, and a black hole horizon replaces the IR brane, while the Hawking radiation keeps the bulk in a thermal bath. Only when the temperature drops much below the IR mass scale, the RS geometry with a stabilized IR brane configuration becomes thermodynamically favorable, and a PT takes place via the nucleation of IR brane bubbles. The associated gravitational wave production is evaluated by analyzing the radion potential~\cite{Fujikura:2019oyi, Fujikura:2023lkn, Fujikura:2024jto}. 

A 5D Dirac fermion $\Psi_{\rm D}(x^\mu, \theta)$ is introduced in the bulk, and its action reads
\begin{align}
    S_{\rm \Psi_{\rm D}} &=  \int d^5 x \sqrt{|g|} k  \bigg[ \bar{\Psi}_{\rm D}i \gamma^N e_{N}^{M} (\partial_M + \omega_M) \Psi_{\rm D}  - m_{\Psi} \bar{\Psi}_{\rm D} \Psi_{\rm D} \bigg] \ ,
    \label{Eq:SPsiD}
\end{align}
where $g$ denotes the metric determinant, $\gamma^N$ satisfies the 5D Dirac algebra in Minkowski space-time, $e_{N}^{M}$ is the vielbein, $\omega_M$ is the spin connection, and $m_\Psi = b k$ is the 5D bulk mass parameter. Decomposing $\Psi_{\rm D} = \Psi_{+} + \Psi_{-}$, such that $\Psi_{\pm} = \mp \gamma^5 \Psi_{\pm}$, and performing a KK decomposition,
\begin{equation}
    \Psi_{\pm}(x^\mu, \theta) = \sum_{n=0}^{\infty} {\cal X}^{(n)}_{\pm} (x^\mu) f^{(n)}_{\pm}(\theta) \ ,
    \label{Eq:KKcom}
\end{equation}
one can solve the equation of motion for the zero-mode $n=0$. It can be shown that one of the chiral modes is always removed by the boundary condition, and the bulk profiles for the zero modes are
\begin{equation}
    f_{\pm}^{(0)} (\theta) = {\cal N}_{\pm} e^{(2 \mp b)k T_0 |\theta| } \ ,
    \label{Eq:fpm}
\end{equation}
where the normalization constant ${\cal N}_{\pm}$ is obtained by canonically normalizing the zero-mode kinetic term at the radion VEV, namely
\begin{equation}
    {\cal N}_{+} (\phi=\langle \phi \rangle) = \left[ \frac{b-1/2}{ 1-e^{(1-2b)\pi k T_0} }\right]^{1/2} \ .
    \label{Eq:Npm}
\end{equation}
Without loss of generality, we can remove the right-handed zero-mode ${\cal X}_{-}$, as similar results can be obtained for $b \to -b$ for ${\cal X}_{-}$. For $b>1/2 \  (b<1/2)$ the zero-mode is localized towards the UV (IR) from the perspective of a 5D flat metric.

A 4D Majorana fermion can be obtained from a 5D Dirac fermion if the chiral zero-mode gets Majorana mass terms at the UV and IR branes. We will consider such a scenario. We package the chiral zero-mode ${\cal X}_{+}^{(0)}$ into a bulk Majorana fermion denoted $\Psi_{\rm M}$:
\begin{equation}
  \Psi_{\rm M} (x^\mu, \theta) = \psi (x^\mu) f_\psi(\theta)= {\cal N}_{+} e^{(2-b)k T_0 |\theta|} \begin{pmatrix}
        {\cal X}_{+}^{(0)} (x^\mu)  \\
        {\cal X}_{+}^{(0) \dagger} (x^\mu) 
    \end{pmatrix} \ .
    \label{Eq:bulkMajorana}
\end{equation}
We consider $b<1/2$, such that $\psi$ is localized towards the IR. Let us write down the kinetic term for $\psi$ including a large radion fluctuation. For $b<1/2$, one can neglect the $1$ in the denominator of Eq.~\eqref{Eq:Npm} in comparison with the exponential, and the normalization factor becomes
\begin{equation}
    {\cal N}_{+} (\phi) = {\cal N}_{+} (\phi = \langle \phi \rangle) \left( \frac{\phi}{\langle \phi \rangle} \right)^{\frac{1-2b}{2}} \ .
    \label{Eq:Npgeneral}
\end{equation}
Hence, the effective 4D kinetic term, including the radion fluctuation, is 
\begin{equation}
    S_{\rm kin, eff}^{\psi} = \frac{1}{2} \int d^4 x  \left( \frac{\phi}{\langle \phi \rangle} \right)^{{1-2b}} \bar{\psi}(x) i \gamma^\mu \partial_\mu \psi(x) \ .
    \label{Eq:kinMaj}
\end{equation}
Consider the following IR localized mass term for $\psi$,
\begin{equation}
    S_\psi^{\rm IR} = -\int d^5 x  \ k  \sqrt{|g|} \ \frac{1}{2} \widetilde{m}_{\rm IR} \bar{\Psi}^c_{\rm M} \Psi_{\rm M} \  \delta(\theta-\pi) \ ,
    \label{Eq:IRbare}
\end{equation}
where $\widetilde{m}_{\rm IR} \sim {\cal O}(k)$ is the bare IR mass term, and $c$ denotes charge conjugation.  Using Eq.~\eqref{Eq:bulkMajorana}, one arrives at the 4D effective mass,
\begin{equation}
    S_{\psi, \rm{eff}}^{\rm IR} = -\int d^4 x \frac{1}{2} m_{\rm IR} \left( \frac{\phi}{\langle \phi \rangle} \right) \bar{\psi}^c(x) \psi(x) \ ,
    \label{Eq:mIRReduced}
\end{equation}
where the warped down IR mass is
\begin{equation}
    m_{\rm IR} \equiv \left(\frac{1}{2}-b \right)  \frac{\langle \phi \rangle}{Z} \widetilde{m}_{\rm IR} \ .
    \label{Eq:mIRWarpped}
\end{equation}
Further, we can consider a UV localized mass term as 
\begin{equation}
    S_\psi^{\rm UV} = -\int d^5 x  \ k  \sqrt{|g|} \ \frac{1}{2} \widetilde{m}_{\rm UV} \bar{\Psi}^c_{\rm M} \Psi_{\rm M} \  \delta(\theta) \ ,
    \label{Eq:UVbare}
\end{equation}
with bare mass term $\widetilde{m}_{\rm UV}$. After the integration, this reduces to the following effective mass: 
\begin{equation}
    S_{\psi, \rm{eff}}^{\rm UV} = -\int d^4 x \frac{1}{2} m_{\rm UV} \left( \frac{\phi}{\langle \phi \rangle} \right)^{1-2b} \bar{\psi}^c(x) \psi(x) \ ,
    \label{Eq:mUVReduced}
\end{equation}
where
\begin{equation}
    m_{\rm UV} \equiv \left(\frac{1}{2}-b \right)  \left( \frac{\langle \phi \rangle}{Z} \right)^{1-2b} \widetilde{m}_{\rm UV} \ .
    \label{Eq:mUVwaveFunction}
\end{equation}
Here, the suppression $\left( {\langle \phi \rangle}/{Z} \right)^{1-2b}$ is due to the wavefunction overlap at the UV brane as can be seen from Eq.~\eqref{Eq:Npm}. Therefore, including the radion fluctuations, the 4D effective Majorana action is given by
\begin{align}
    \nonumber
    S_{\psi}^{\rm eff} &= \frac{1}{2}\int d^4 x \left( \frac{\phi}{\langle \phi \rangle} \right)^{{1-2b}} \bigg[  \bar{\psi}(x) i \gamma^\mu \partial_\mu \psi(x) \\
    & - \left\{ m_{\rm IR} \left( \frac{\phi}{\langle \phi \rangle} \right)^{2b} + m_{\rm UV} \right\} \bar{\psi}^c(x) \psi(x)  \bigg] \  ; \ (b<1/2) \ ,
    \label{Eq:MajoranaTotal}
\end{align}
and is applicable for $b < 1/2$. The equation of motion for $\psi$ from Eq.~\eqref{Eq:MajoranaTotal} is evaluated to be
\begin{align}
    \left( i\gamma^\mu \partial_\mu -m^{\rm eff}_\psi \right) \psi = 0 \ ,
    \label{Eq:psiEOMAppendix}
\end{align}
where the effective mass $m_{\rm eff}$, including the radion fluctuations as a function of time $\phi(t) = \langle \phi \rangle + \varphi (t)$, is given by
\begin{equation}
    m^{\rm eff}_\psi(t) = m_{\rm UV} + m_{\rm IR} \left(1+\frac{\varphi(t)}{\langle \phi \rangle} \right)^{2b} \ .
    \label{Eq:meff1}
\end{equation}
We can parametrize the radion oscillation as $\varphi(t) = \varphi_A(t) \cos(m_\varphi t)$, where $m_\varphi$ denotes the radion mass, and the amplitude of oscillation $\varphi_A (t) = \xi (t) \langle \phi \rangle$.\footnote{
As the dilaton gets rearranged close to the symmetric phase after the bubble collision and reflection $\varphi_{\rm A} \sim \langle \phi \rangle$ during the first few oscillations~\cite{Konstandin:2011ds}, and therefore the dilaton mass near the true vacuum $m_\varphi$ might not appropriately describe the oscillation frequency, although here we treat it as a phenomenological parameter and assume that it is not too different from the true dilaton mass.} Clearly, $\xi <1$. $\xi$  itself depends on $t$ in general. Further, we introduce the ratio of the UV to the IR mass as 
\begin{equation}
    r \equiv -\frac{m_{\rm UV}}{m_{\rm IR}} = -\left(\frac{\widetilde{m}_{\rm UV}}{\widetilde{m}_{\rm IR}} \right) \left( \frac{\langle \phi \rangle}{Z} \right)^{-2b} \ ,
    \label{Eq:rdef}
\end{equation}
where a relative sign between the UV and IR mass terms is assumed. Hence, the effective mass can be expressed as 
\begin{align}
    m^{\rm eff}_\psi(t) = m_\psi \left[\frac{\{1+\xi(t) \cos(m_\varphi t)\}^{2b}-r}{1-r} \right] \ ,
    \label{Eq:meff2Appendix}
\end{align}
where the present-day $\psi$ mass, \textit{i.e.,} at $\phi = \langle \phi \rangle$ is denoted as $m_\psi = m_{\rm IR}(1-r)$. Notice $r \neq 0$ for $m_\psi \neq 0$. Also, as only the relative sign between the UV and IR mass terms is physical, we can take $r>0$ without loss of generality. 

A similar result is obtained for a 5D fermion peaked towards the UV, \textit{i.e.,} for $b>1/2$. The difference from the IR peaked case is because the exponential term in the denominator of Eq.~\eqref{Eq:Npm} can be neglected in comparison with 1. The 4D effective action, including the brane-localized IR and UV mass terms, is given by
\begin{align}
        S_{\psi}^{\rm eff} &= \frac{1}{2}\int d^4 x  \bigg[  \bar{\psi}(x) i \gamma^\mu \partial_\mu \psi(x) \\
    & - \left\{ m_{\rm IR} \left( \frac{\phi}{\langle \phi \rangle} \right)^{2b} + m_{\rm UV} \right\} \bar{\psi}^c(x) \psi(x)  \bigg] \  ; \ (b>1/2) \ ,
\end{align}
and the effective 4D mass terms $m_{\rm IR}$, $m_{\rm UV}$ are related to the corresponding bare quantities as
\begin{align}
    \nonumber
    m_{\rm IR} &= \left(\frac{1}{2}-b \right) \left( \frac{\langle \phi \rangle}{Z} \right)^{2b} \widetilde{m}_{\rm IR} \\[1ex]
    m_{\rm UV} &= \left(\frac{1}{2}-b \right)  \widetilde{m}_{\rm UV} \ \quad ; \quad \quad \ (b>1/2) \ .
    \label{Eq:mUVIRbg}
\end{align}
The effective mass from the equation of motion remains in the same form as Eqs.~\eqref{Eq:meff1}-\eqref{Eq:meff2Appendix} with the modified definitions in Eq.~\eqref{Eq:mUVIRbg} for individual $m_{\rm UV, IR}$. Finally, if the zero-mode is flat, \textit{i.e.,} $b=1/2$, then after a similar calculation, one arrives at Eq.~\eqref{Eq:meff1} with $b=1/2$.  Hence, one can utilize Eqs.~\eqref{Eq:meff1}-\eqref{Eq:meff2Appendix} to evaluate the effective mass for a general fermion $\psi$ during radion oscillations after the phase transition with any set of parameters $b,r$. After the radion has settled down to its minimum, the perturbative $\varphi-\psi$ Feynman vertex can be evaluated using Eq.~\eqref{Eq:MajoranaTotal}, and is given by
\begin{align}
    {\cal L}_{\varphi \psi} = -\frac{1}{2} m_\psi \left( \frac{2b}{1-r}\right) \frac{\varphi}{\langle \phi \rangle} \bar \psi^c \psi  \ .
    \label{Eq:radionPerturbative}
\end{align}

\section{Dilaton effective field theory}
\label{appendix:dilaton}

Let us recall the expectation of dilaton couplings from an effective field theory (EFT) point of view~\cite{Goldberger:2007zk, Chacko:2012sy}. We consider a Lagrangian in the basis of anomalous dimension eigen-operators,
\begin{equation}
    L = \sum_i g_i (\mu) {\cal{O}}_i (x) \ ,
    \label{Eq:GenLagrangian}
\end{equation}
where $\mu$ is the renormalization energy scale, ${\cal{O}}_i$ denotes an operator with scaling dimension $d_i$
and $g_i (\mu)$ represents its coupling coefficient.
The dilatation current $S^\mu = T^{\mu}_{ \ \nu} x^\nu$, where $T^{\mu \nu}$ is the symmetric energy-momentum tensor, can be obtained from the scale transformation defined by $x \to e^\lambda x$, $\mu \to e^{-\lambda} \mu$, and ${\cal O}_i (x) \to e^{\lambda d_i} {\cal O}_i (e^\lambda x)$. The divergence of the dilatation current is given by
\begin{equation}
    \partial_\mu S^\mu = T^{\mu}_{ \ \mu} = \sum_i \left[ (d_i-4) g_i (\mu) -\beta (g_i)  \right] {\cal O}_i (x) \ .
    \label{Eq:Tmumu}
\end{equation}
Here, $\beta (g_i) \equiv \partial g_i/ \partial(\ln \mu)$ is the beta function. The scale invariance can be restored with the proper insertion of $\phi$, following the prescription,
\begin{equation}
   g_{i} (\mu) \to g_{i} \left(\mu \left(1+ \frac{\varphi}{\langle \phi \rangle}\right)\right) \left(1+ \frac{\varphi}{\langle \phi \rangle}\right)^{4-d_i} \ ,
   \label{Eq:conformalCompensator}
\end{equation}
where ${\phi} (x) \to e^\lambda  \phi (e^\lambda x)$ under the scale transformation. From Eqs.~\eqref{Eq:GenLagrangian}-\eqref{Eq:conformalCompensator}, the dilaton interaction is determined as $(\varphi/\langle \phi \rangle)  T^{\mu}_{ \ \mu}$. For illustration, let us consider a fermion $\psi$ whose dynamically generated Majorana mass term is given by $(1/2) m_\psi \bar\psi^c \psi$, with $c$ denoting the charge conjugation. Then, the dilaton interaction is 
\begin{equation}
    {\cal L}_{\psi \varphi} = - \frac{1}{2} m_{\psi} (1+ \gamma_\psi) \frac{\varphi}{ \langle \phi \rangle} \bar\psi^c \psi \ .
    \label{Eq:MajoranaDilaton}
\end{equation}
Here, $\gamma_\psi \equiv \partial(\ln m_\psi)/\partial(\ln \mu)$ is the associated anomalous dimension.
The anomalous dimension is determined from the partial compositeness of the fermion, i.e., how it mixes with the CFT sector. When the radion interaction in Eq.~\eqref{Eq:MajoranaTotal} is expanded linearly, one can see the correspondence between the bulk mass parameter $b$, the ratio of the UV and IR mass $r$ and the anomalous dimension $\gamma_\psi$ by comparing with Eq.~\eqref{Eq:radionPerturbative}.

\section{General formulae for asymmetry generated}
\label{appendix:general}

We begin with the Lagrangian density in Eq.~\eqref{Eq:VDS}, namely
\begin{align}
\nonumber
& \mathcal{L}_{\rm DS} \supset  \ \frac{i}{2}\bar \psi_i \partial_\mu \gamma^\mu \psi_i + i\bar \chi_\alpha \partial_\mu \gamma^\mu \chi_\alpha + \frac{1}{2} (\partial \eta)^2 - m_{\chi_\alpha} \bar \chi_\alpha \chi_\alpha \\ & - \frac{1}{2} m_\eta^2 \eta^2   -  \mu_{\alpha \beta} \bar \chi_\alpha \chi_\beta \eta  -\frac{1}{2} m_{\psi_i} \bar \psi_{i\mathrm{R}}^c \psi_{i\mathrm{R}} -( y_{i \alpha}\bar{\psi}_{i \mathrm{R}} \chi_{\alpha \mathrm{L}} \eta + {\rm h.c.})   \ ,
\end{align}
where $i, \alpha = 1,2$ are the $\psi$, $\chi$ generational indices, respectively. 

The asymmetric number density of  $\chi_{\alpha}$ i.e. $\delta n_{\chi_{\alpha}}=n_{\chi_{\alpha}}-n_{\Bar{\chi}_{\alpha}}$ is produced due to the difference between the decay rates $\Gamma_{\alpha}(\psi_1 \rightarrow \chi_{\alpha} \eta)$ and $\Bar{\Gamma}_{\alpha}(\psi_1 \rightarrow \Bar{\chi}_{\alpha} \eta)$
\begin{align}
\delta n_{\chi_{\alpha}}=n_{\psi_1}\cdot \frac{\Gamma_{\alpha}-\bar{\Gamma}_{\alpha}}{\Gamma_{\rm tot}} \ ,
\end{align}
where $\Gamma_{\rm tot}$ is the total decay width of $\psi_1$. The channels corresponding to the decay processes at tree and one-loop level are the following:

\begin{center}
    \noindent\includegraphics[width=0.95\linewidth]{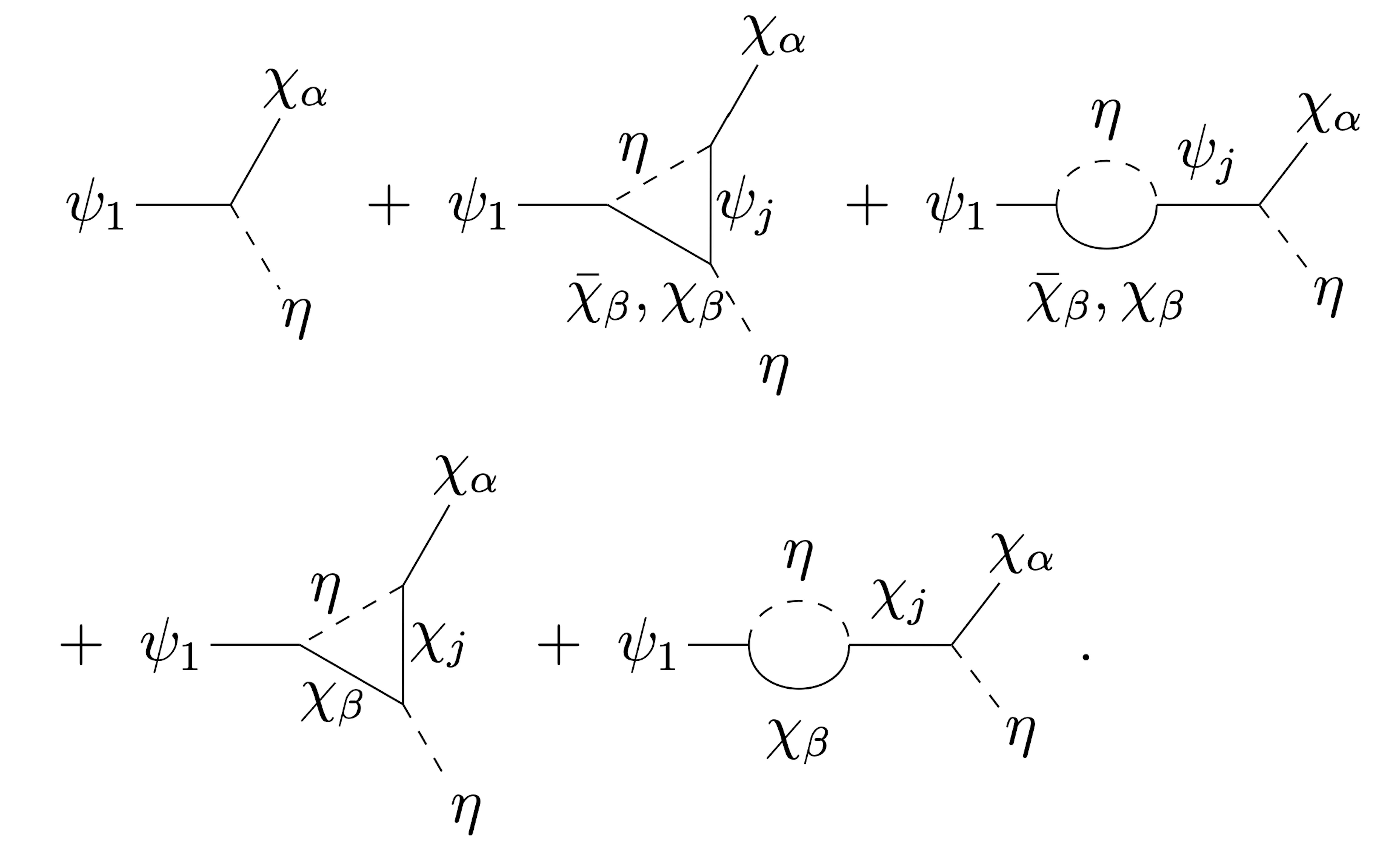}
\end{center}
Note that the four diagrams involving $\chi_\beta$ rather than $\bar\chi_\beta$ inside the loop do not contribute to the total asymmetry, but contribute to asymmetry in the individual generation of $\chi_\alpha$. The asymmetry arises from the interference between the tree level and the one-loop level. We use $\delta n_{\chi_{\alpha}}^{(y)}$ and $\delta n_{\chi_{\alpha}}^{(\mu)}$ to represent the asymmetry generated due to $y\bar{\psi}\chi \eta$ and $\mu \bar{\chi} \chi \eta$ couplings, respectively. We obtain

\begin{align}
    & \delta n_{\chi_{\alpha}}^{(y)}\approx n_{\psi_1}\cdot 4 \sum_j \Im(L_{\psi_j}) \frac{\Re(\sum_{\beta}y_{1\beta}y_{j\beta}^*)}{\sum_{\gamma}|y_{1\gamma}|^2} \Im(y_{1\alpha}y_{j\alpha}^*) \ , \\[1ex]
    & \delta n_{\chi_{\alpha}}^{(\mu)} \approx n_{\psi_1}\cdot 2 \sum_j \Im(L_{\chi_j}) \frac{\Im(y_{1\alpha}\mu_{j\alpha}^* \cdot \sum_{\beta}y^*_{1\beta}\mu_{j\beta})}{\sum_{\gamma}|y_{1\gamma}|^2} \ , \\[1ex]
    &     \delta n_{\chi_\alpha} =   \delta n_{\chi_{\alpha}}^{(y)}+ \delta n_{\chi_{\alpha}}^{(\mu)} \ ,
\end{align}
where $L_{\psi_j}=L(m_{\psi_1},m_{\psi_j}), L_{\chi_j}=L(m_{\psi_1},m_{\chi_j})$ are the corresponding loop functions (see for \textit{e.g.,} Ref.~\cite{Pilaftsis:1998pd}).

After $\chi_2$ decays due to the neutron portal coupling $\frac{1}{\Lambda_n^2} \chi_2 u_{R} d_{R} d_{R}$, it transfers the asymmetry to the visible baryons. Therefore, the final asymmetry for the DM, denoted as $\delta n_{\rm ADM}$, is stored in $\chi_1$, while the visible asymmetry is represented as $\delta n_{\rm BAU}$. They are given by
\begin{align}
    & \delta n_{\rm ADM} = \delta n_{\chi_1}+\delta n_{\chi_2}\cdot \frac{\Gamma_{\chi_2 \rightarrow \chi_1+\eta}}{\Gamma_{\chi_2}} \ , \label{Eq:ADM}\\[1ex]
    & \delta n_{\rm BAU} =\delta n_{\chi_2}\cdot \frac{\Gamma_{\chi_2 \rightarrow udd}}{\Gamma_{\chi_2}} \ ,
    \label{Eq:BAU}
\end{align}
where $\Gamma_{\chi_2}=\Gamma_{\chi_2 \rightarrow \chi_1+\eta}+\Gamma_{\chi_2 \rightarrow udd}$, and 
\begin{align}
    \nonumber
     \Gamma_{\chi_2 \rightarrow \chi_1+\eta} \approx & \, \frac{\mu_{12}^2}{8 \pi}\frac{(m_{\chi_1}+m_{\chi_2})^2-m_{\eta}^2}{m_{\chi_2}^3} \\ \nonumber & \times [(m_{\chi_2}^2-m_{\chi_1}^2-m_{\eta}^2)^2-4m_{\chi_1}^2m_{\eta}^2]^{\frac{1}{2}} \ , \\[1ex]
     \Gamma_{\chi_2 \rightarrow udd} \approx & \, \frac{1}{\Lambda_n^4}\frac{m_{\chi_2}^5}{192 \pi^3} \ .
\end{align}
Here, we assume $m_{\chi_2}\gg m_u,m_d$.

For simplicity, let us assume $|y_{ij}|=|y|, |\mu_{11}|=|\mu_{22}|$ and all phases are random. Then one obtains
\begin{align}
   & \delta n_{\chi_{\alpha}}^{(y)} \approx n_{\psi_1}\cdot 4   \Im(L_{\psi_2}) \cdot |y|^2 \ ,  \\[1ex]
    &\delta n_{\chi_{\alpha}}^{(\mu)} \approx n_{\psi_1}\cdot  (\Im(L_{\chi_1})+\Im(L_{\chi_2}))\cdot |\mu_{11}|\cdot |\mu_{12}| \ .
\end{align}
The loop functions can be estimated as (see $e.g.$ Ref.~\cite{Pilaftsis:1998pd})
\begin{align}
    \nonumber
    \Im \left(L(M_1,M_2) \right) & =\frac{1}{8\pi}\cdot \bigg [\frac{\Delta M^2 M_1 M_2}{(\Delta M^2)^2+M_1^2 M_2^2\cdot (\frac{|y|^2}{8\pi})^2} \\
   & + f \left(\frac{M_2^2}{M_1^2}\right)\bigg] \ ,
    \label{Eq:loopInt}
\end{align}
where $\Delta M^2=M_1^2-M_2^2$ and $f(x)=\sqrt{x}(1-(1+x) \ln(1+\frac{1}{x}))$. Here, the first term comes from the self-energy diagram, while the second term is due to the vertex diagram. In our model, we typically have $m_{\psi_2}> m_{\psi_1}> m_{\chi_2}> m_{\chi_1}$. For simplicity, we assume $\frac{m_{\psi_2}}{m_{\psi_1}}\sim \frac{m_{\psi_1}}{m_{\chi_{1,2}}}> 1$.
Finally, the total asymmetry in each generation can be expressed as: 
\begin{align}
    & \delta n_{\chi_{1}} \simeq n_{\psi_1}\cdot \frac{m_{\psi_1}}{m_{\psi_2}} (|y|^2+|\mu_{11}||\mu_{12}|) \ , \\[1ex]
    & \delta n_{\chi_{2}} \simeq n_{\psi_1}\cdot \frac{m_{\psi_1}}{m_{\psi_2}} (|y|^2-|\mu_{11}||\mu_{12}|) \ \label{Eq:asymmetry} .
\end{align}
 Together with Eqs.~\eqref{Eq:ADM}-\eqref{Eq:asymmetry}, we then find
\begin{align}
    \nonumber
    \frac{\delta n_{\rm BAU}}{s(T_{\rm RH})} & \approx   
    \frac{(|y|^2-|\mu_{11}||\mu_{12}|) n_{\psi_1}}{4 \pi} \\ 
    & \times \left[ 1+12 \pi^2 \mu_{12}^2\cdot \left({\Lambda_n}/{m_{\chi_2}}\right)^4 \right]^{-1} \ ,
    \label{Eq:BAUAs}
\end{align}
which implies for $m_{\chi_2} \simeq $ GeV and $\Lambda_n \simeq 100$ TeV, the visible baryon asymmetry would be highly suppressed unless $\mu_{12} \lesssim m_{\chi_2}^2/(2 \sqrt3 \pi \Lambda_n^2)$. For $m_{\chi_2} = 3$ GeV, $\Lambda_n = 100$ TeV, this corresponds to $\mu_{12} \lesssim 10^{-10}$. One can arrange such a small $\mu_{12}$ via the wavefunction overlap in an extra dimension, as illustrated in appendix~\ref{appendix:5d}. Consequently, the contribution of $\mu$ terms are negligible for the resultant asymmetry in Eq.~\eqref{Eq:BAUAs}, and the final expression of visible asymmetry produced by heavy $\psi_1$ decay is
\begin{align}
    \delta n_{\chi_\alpha}\simeq n_{\psi_1}\frac{|y|^2}{4\pi}\cdot \frac{m_{\psi_1}}{m_{\psi_2}}\cdot \theta_{\rm CP} \ ,
\end{align}
where $\theta_{\rm CP}$ is the ${\cal O}(1)$ CP violating phase in the $y$-matrix.

\section{A possible 5D setup}
\label{appendix:5d}

\begin{figure}[!t]
    \centering
    \includegraphics[width=0.6\linewidth]{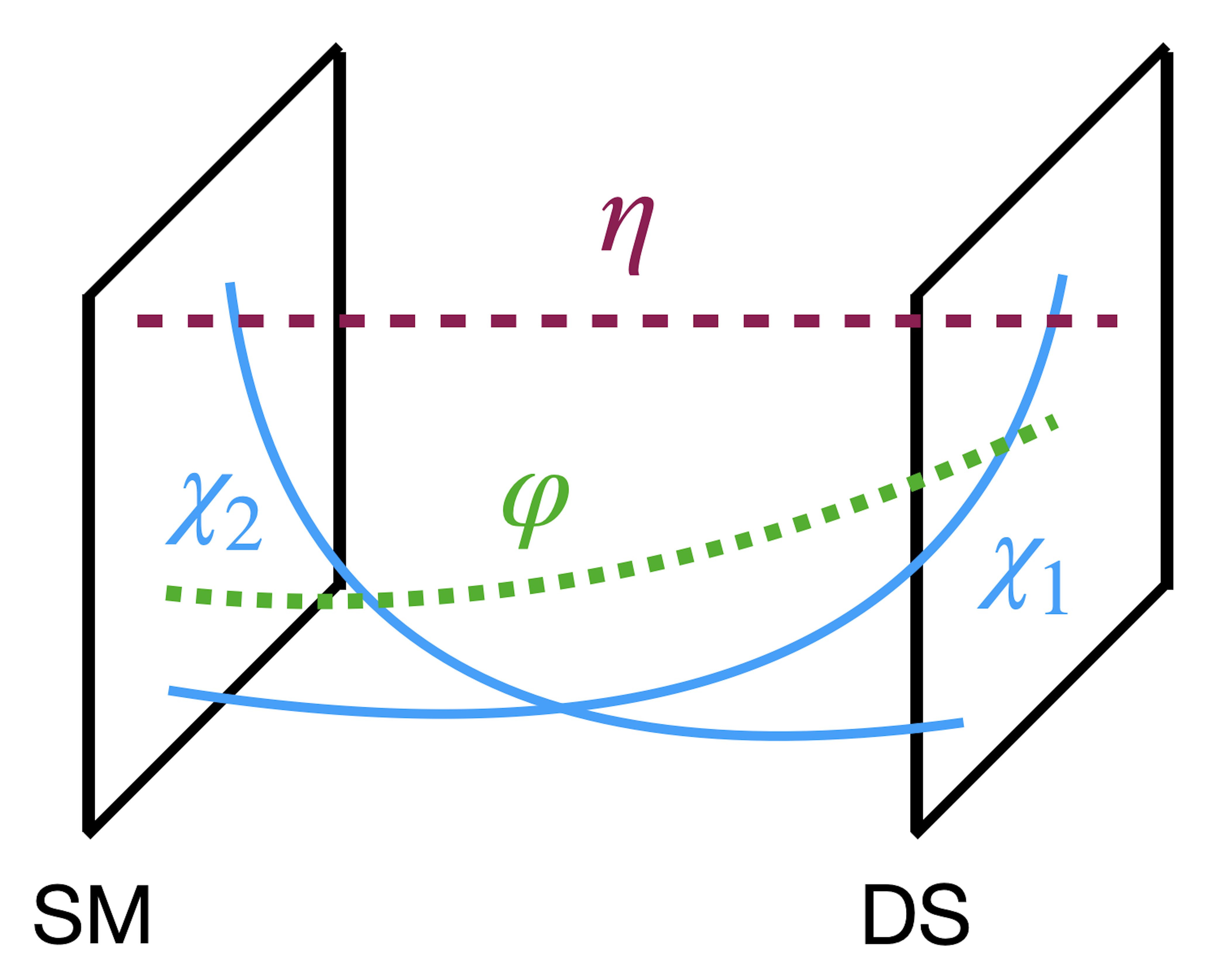}
    \captionsetup{justification=raggedright, singlelinecheck=false}
    \caption{A possible embedding of the dark sector in a 5D warped geometry, where the SM lives on the UV brane.}
    \label{fig:extradimension}
\end{figure}

 Let us illustrate how $\mu_{12} \lesssim 10^{-10}$ can be achieved in a holographic embedding with a warped extra dimension bounded by two 3-branes, where the SM is taken to live on the UV brane, as shown in Fig.~\ref{fig:extradimension}. The relevant $\mu_{12}$ term in 5D is described as
\begin{align}
    S\supset \int d^5x \, k \sqrt{|g|} \ \tilde{\mu}_{12} \bar{\chi}_2(x,\theta)\chi_1(x,\theta)\eta(x,\theta)  \ ,
\end{align}
where $\tilde{\mu}_{12}$ is the corresponding 5D coupling. Let us take the zero-mode of $\eta$ to have a flat profile, while the zero-modes of $\chi_{1,2}$ are peaked towards the IR and UV branes, respectively. Their 5D profiles are given by 
\begin{align}
    & f_{\chi_i}(\theta)\simeq \sqrt{\frac{(1/2-c_i)}{e^{(1-2c_i)\pi k T_0}-1}}e^{(2-c_i)kT_0 |\theta|} \ ,  \\[1ex]
    & f_{\eta}(\theta) = \frac{1}{\sqrt{2\pi k T_0}} \ ,
\end{align}
where $c_{1,2}$ are the 5D bulk mass parameters for $\chi_{1,2}$. By choosing $c_1<\frac{1}{2}<c_2$, we get a IR localized $\chi_1$, and a UV localized $\chi_2$. After integrating over the extra dimension, we obtain the 4D effective $\mu_{12}$ as
\begin{align}
    \nonumber
    \mu_{12} & \simeq \frac{\tilde{\mu}_{12}}{c_1+c_2} \cdot  \sqrt{\frac{{(2c_2-1)(1-2 c_1)}}{(1-e^{-(2c_2-1) \pi k T_0}) (e^{(1-2c_1) \pi k T_0}-1)}}\\
    & \times \frac{\left(1-e^{-(c_1+c_2) \pi k T_0 } \right)}{\sqrt{2 \pi k T_0}} \ .
\end{align}
The redshift factor $z_{\rm IR}$ from the UV mass scale $M_{\rm UV}$ to the IR mass scale $M_{\rm IR}$ is defined as ${M_{\rm IR}}= {M_{\rm UV}} \, e^{-\pi k T_0}= {M_{\rm UV}} \, z_{\rm IR}$. If $M_{\rm UV} = 10$ TeV,\footnote{See \textit{e.g.}, Ref.~\cite{Girmohanta:2025uue} for a consistent low-cutoff theory, and Refs.~\cite{Lee:2021wau,Girmohanta:2023sjv} for a three 3-brane setup and its associated cosmology, within which the present model can be embedded.} and $M_{\rm IR} = 1$ GeV, $z_{\rm IR} = 10^{-4}$. Taking $\tilde{\mu}_{12} \simeq {\cal O}(1)$, we can achieve $\mu_{12} \lesssim 10^{-10}$ for $c_1 \lesssim -2.1$, $c_2 \gtrsim 2.5$. The parameter space can be enlarged further if one assumes $\tilde{\mu}_{12} \ll 1$. As a result of being localized towards the IR brane, the DM candidate $\chi_1$ is stable against decay to lighter SM visible particles, while it is also the lightest DS fermion and does not decay to DS particles. This explains the stability of the DM.

\bibliography{Nanoref}

\begin{thebibliography}{68}%
\makeatletter
\providecommand \@ifxundefined [1]{%
 \@ifx{#1\undefined}
}%
\providecommand \@ifnum [1]{%
 \ifnum #1\expandafter \@firstoftwo
 \else \expandafter \@secondoftwo
 \fi
}%
\providecommand \@ifx [1]{%
 \ifx #1\expandafter \@firstoftwo
 \else \expandafter \@secondoftwo
 \fi
}%
\providecommand \natexlab [1]{#1}%
\providecommand \enquote  [1]{``#1''}%
\providecommand \bibnamefont  [1]{#1}%
\providecommand \bibfnamefont [1]{#1}%
\providecommand \citenamefont [1]{#1}%
\providecommand \href@noop [0]{\@secondoftwo}%
\providecommand \href [0]{\begingroup \@sanitize@url \@href}%
\providecommand \@href[1]{\@@startlink{#1}\@@href}%
\providecommand \@@href[1]{\endgroup#1\@@endlink}%
\providecommand \@sanitize@url [0]{\catcode `\\12\catcode `\$12\catcode
  `\&12\catcode `\#12\catcode `\^12\catcode `\_12\catcode `\%12\relax}%
\providecommand \@@startlink[1]{}%
\providecommand \@@endlink[0]{}%
\providecommand \url  [0]{\begingroup\@sanitize@url \@url }%
\providecommand \@url [1]{\endgroup\@href {#1}{\urlprefix }}%
\providecommand \urlprefix  [0]{URL }%
\providecommand \Eprint [0]{\href }%
\providecommand \doibase [0]{https://doi.org/}%
\providecommand \selectlanguage [0]{\@gobble}%
\providecommand \bibinfo  [0]{\@secondoftwo}%
\providecommand \bibfield  [0]{\@secondoftwo}%
\providecommand \translation [1]{[#1]}%
\providecommand \BibitemOpen [0]{}%
\providecommand \bibitemStop [0]{}%
\providecommand \bibitemNoStop [0]{.\EOS\space}%
\providecommand \EOS [0]{\spacefactor3000\relax}%
\providecommand \BibitemShut  [1]{\csname bibitem#1\endcsname}%
\let\auto@bib@innerbib\@empty
\bibitem [{\citenamefont {Babu}\ \emph {et~al.}(2006)\citenamefont {Babu},
  \citenamefont {Mohapatra},\ and\ \citenamefont {Nasri}}]{Babu:2006xc}%
  \BibitemOpen
  \bibfield  {author} {\bibinfo {author} {\bibfnamefont {K.~S.}\ \bibnamefont
  {Babu}}, \bibinfo {author} {\bibfnamefont {R.~N.}\ \bibnamefont
  {Mohapatra}},\ and\ \bibinfo {author} {\bibfnamefont {S.}~\bibnamefont
  {Nasri}},\ }\bibfield  {title} {\bibinfo {title} {{Post-Sphaleron
  Baryogenesis}},\ }\href {https://doi.org/10.1103/PhysRevLett.97.131301}
  {\bibfield  {journal} {\bibinfo  {journal} {Phys. Rev. Lett.}\ }\textbf
  {\bibinfo {volume} {97}},\ \bibinfo {pages} {131301} (\bibinfo {year}
  {2006})},\ \Eprint {https://arxiv.org/abs/hep-ph/0606144}
  {arXiv:hep-ph/0606144} \BibitemShut {NoStop}%
\bibitem [{\citenamefont {Babu}\ \emph {et~al.}(2009)\citenamefont {Babu},
  \citenamefont {Bhupal~Dev},\ and\ \citenamefont {Mohapatra}}]{Babu:2008rq}%
  \BibitemOpen
  \bibfield  {author} {\bibinfo {author} {\bibfnamefont {K.~S.}\ \bibnamefont
  {Babu}}, \bibinfo {author} {\bibfnamefont {P.~S.}\ \bibnamefont
  {Bhupal~Dev}},\ and\ \bibinfo {author} {\bibfnamefont {R.~N.}\ \bibnamefont
  {Mohapatra}},\ }\bibfield  {title} {\bibinfo {title} {{Neutrino mass
  hierarchy, neutron - anti-neutron oscillation from baryogenesis}},\ }\href
  {https://doi.org/10.1103/PhysRevD.79.015017} {\bibfield  {journal} {\bibinfo
  {journal} {Phys. Rev. D}\ }\textbf {\bibinfo {volume} {79}},\ \bibinfo
  {pages} {015017} (\bibinfo {year} {2009})},\ \Eprint
  {https://arxiv.org/abs/0811.3411} {arXiv:0811.3411 [hep-ph]} \BibitemShut
  {NoStop}%
\bibitem [{\citenamefont {Bell}\ \emph {et~al.}(2019)\citenamefont {Bell},
  \citenamefont {Corbett}, \citenamefont {Nee},\ and\ \citenamefont
  {Ramsey-Musolf}}]{Bell:2018mgg}%
  \BibitemOpen
  \bibfield  {author} {\bibinfo {author} {\bibfnamefont {N.~F.}\ \bibnamefont
  {Bell}}, \bibinfo {author} {\bibfnamefont {T.}~\bibnamefont {Corbett}},
  \bibinfo {author} {\bibfnamefont {M.}~\bibnamefont {Nee}},\ and\ \bibinfo
  {author} {\bibfnamefont {M.~J.}\ \bibnamefont {Ramsey-Musolf}},\ }\bibfield
  {title} {\bibinfo {title} {{Electric dipole moments from postsphaleron
  baryogenesis}},\ }\href {https://doi.org/10.1103/PhysRevD.99.015034}
  {\bibfield  {journal} {\bibinfo  {journal} {Phys. Rev. D}\ }\textbf {\bibinfo
  {volume} {99}},\ \bibinfo {pages} {015034} (\bibinfo {year} {2019})},\
  \Eprint {https://arxiv.org/abs/1808.10597} {arXiv:1808.10597 [hep-ph]}
  \BibitemShut {NoStop}%
\bibitem [{\citenamefont {Agazie}\ \emph {et~al.}(2023)\citenamefont {Agazie}
  \emph {et~al.}}]{NANOGrav:2023gor}%
  \BibitemOpen
  \bibfield  {author} {\bibinfo {author} {\bibfnamefont {G.}~\bibnamefont
  {Agazie}} \emph {et~al.} (\bibinfo {collaboration} {NANOGrav}),\ }\bibfield
  {title} {\bibinfo {title} {{The NANOGrav 15 yr Data Set: Evidence for a
  Gravitational-wave Background}},\ }\href
  {https://doi.org/10.3847/2041-8213/acdac6} {\bibfield  {journal} {\bibinfo
  {journal} {Astrophys. J. Lett.}\ }\textbf {\bibinfo {volume} {951}},\
  \bibinfo {pages} {L8} (\bibinfo {year} {2023})},\ \Eprint
  {https://arxiv.org/abs/2306.16213} {arXiv:2306.16213 [astro-ph.HE]}
  \BibitemShut {NoStop}%
\bibitem [{\citenamefont {Afzal}\ \emph {et~al.}(2023)\citenamefont {Afzal}
  \emph {et~al.}}]{NANOGrav:2023hvm}%
  \BibitemOpen
  \bibfield  {author} {\bibinfo {author} {\bibfnamefont {A.}~\bibnamefont
  {Afzal}} \emph {et~al.} (\bibinfo {collaboration} {NANOGrav}),\ }\bibfield
  {title} {\bibinfo {title} {{The NANOGrav 15 yr Data Set: Search for Signals
  from New Physics}},\ }\href {https://doi.org/10.3847/2041-8213/acdc91}
  {\bibfield  {journal} {\bibinfo  {journal} {Astrophys. J. Lett.}\ }\textbf
  {\bibinfo {volume} {951}},\ \bibinfo {pages} {L11} (\bibinfo {year}
  {2023})},\ \Eprint {https://arxiv.org/abs/2306.16219} {arXiv:2306.16219
  [astro-ph.HE]} \BibitemShut {NoStop}%
\bibitem [{\citenamefont {Antoniadis}\ \emph {et~al.}(2023)\citenamefont
  {Antoniadis} \emph {et~al.}}]{EPTA:2023fyk}%
  \BibitemOpen
  \bibfield  {author} {\bibinfo {author} {\bibfnamefont {J.}~\bibnamefont
  {Antoniadis}} \emph {et~al.} (\bibinfo {collaboration} {EPTA, InPTA:}),\
  }\bibfield  {title} {\bibinfo {title} {{The second data release from the
  European Pulsar Timing Array - III. Search for gravitational wave signals}},\
  }\href {https://doi.org/10.1051/0004-6361/202346844} {\bibfield  {journal}
  {\bibinfo  {journal} {Astron. Astrophys.}\ }\textbf {\bibinfo {volume}
  {678}},\ \bibinfo {pages} {A50} (\bibinfo {year} {2023})},\ \Eprint
  {https://arxiv.org/abs/2306.16214} {arXiv:2306.16214 [astro-ph.HE]}
  \BibitemShut {NoStop}%
\bibitem [{\citenamefont {Reardon}\ \emph {et~al.}(2023)\citenamefont {Reardon}
  \emph {et~al.}}]{Reardon:2023gzh}%
  \BibitemOpen
  \bibfield  {author} {\bibinfo {author} {\bibfnamefont {D.~J.}\ \bibnamefont
  {Reardon}} \emph {et~al.},\ }\bibfield  {title} {\bibinfo {title} {{Search
  for an Isotropic Gravitational-wave Background with the Parkes Pulsar Timing
  Array}},\ }\href {https://doi.org/10.3847/2041-8213/acdd02} {\bibfield
  {journal} {\bibinfo  {journal} {Astrophys. J. Lett.}\ }\textbf {\bibinfo
  {volume} {951}},\ \bibinfo {pages} {L6} (\bibinfo {year} {2023})},\ \Eprint
  {https://arxiv.org/abs/2306.16215} {arXiv:2306.16215 [astro-ph.HE]}
  \BibitemShut {NoStop}%
\bibitem [{\citenamefont {Xu}\ \emph {et~al.}(2023)\citenamefont {Xu} \emph
  {et~al.}}]{Xu:2023wog}%
  \BibitemOpen
  \bibfield  {author} {\bibinfo {author} {\bibfnamefont {H.}~\bibnamefont {Xu}}
  \emph {et~al.},\ }\bibfield  {title} {\bibinfo {title} {{Searching for the
  Nano-Hertz Stochastic Gravitational Wave Background with the Chinese Pulsar
  Timing Array Data Release I}},\ }\href
  {https://doi.org/10.1088/1674-4527/acdfa5} {\bibfield  {journal} {\bibinfo
  {journal} {Res. Astron. Astrophys.}\ }\textbf {\bibinfo {volume} {23}},\
  \bibinfo {pages} {075024} (\bibinfo {year} {2023})},\ \Eprint
  {https://arxiv.org/abs/2306.16216} {arXiv:2306.16216 [astro-ph.HE]}
  \BibitemShut {NoStop}%
\bibitem [{\citenamefont {Nakai}\ \emph {et~al.}(2021)\citenamefont {Nakai},
  \citenamefont {Suzuki}, \citenamefont {Takahashi},\ and\ \citenamefont
  {Yamada}}]{Nakai:2020oit}%
  \BibitemOpen
  \bibfield  {author} {\bibinfo {author} {\bibfnamefont {Y.}~\bibnamefont
  {Nakai}}, \bibinfo {author} {\bibfnamefont {M.}~\bibnamefont {Suzuki}},
  \bibinfo {author} {\bibfnamefont {F.}~\bibnamefont {Takahashi}},\ and\
  \bibinfo {author} {\bibfnamefont {M.}~\bibnamefont {Yamada}},\ }\bibfield
  {title} {\bibinfo {title} {{Gravitational Waves and Dark Radiation from Dark
  Phase Transition: Connecting NANOGrav Pulsar Timing Data and Hubble
  Tension}},\ }\href {https://doi.org/10.1016/j.physletb.2021.136238}
  {\bibfield  {journal} {\bibinfo  {journal} {Phys. Lett. B}\ }\textbf
  {\bibinfo {volume} {816}},\ \bibinfo {pages} {136238} (\bibinfo {year}
  {2021})},\ \Eprint {https://arxiv.org/abs/2009.09754} {arXiv:2009.09754
  [astro-ph.CO]} \BibitemShut {NoStop}%
\bibitem [{\citenamefont {Fujikura}\ \emph {et~al.}(2023)\citenamefont
  {Fujikura}, \citenamefont {Girmohanta}, \citenamefont {Nakai},\ and\
  \citenamefont {Suzuki}}]{Fujikura:2023lkn}%
  \BibitemOpen
  \bibfield  {author} {\bibinfo {author} {\bibfnamefont {K.}~\bibnamefont
  {Fujikura}}, \bibinfo {author} {\bibfnamefont {S.}~\bibnamefont
  {Girmohanta}}, \bibinfo {author} {\bibfnamefont {Y.}~\bibnamefont {Nakai}},\
  and\ \bibinfo {author} {\bibfnamefont {M.}~\bibnamefont {Suzuki}},\
  }\bibfield  {title} {\bibinfo {title} {{NANOGrav signal from a dark conformal
  phase transition}},\ }\href {https://doi.org/10.1016/j.physletb.2023.138203}
  {\bibfield  {journal} {\bibinfo  {journal} {Phys. Lett. B}\ }\textbf
  {\bibinfo {volume} {846}},\ \bibinfo {pages} {138203} (\bibinfo {year}
  {2023})},\ \Eprint {https://arxiv.org/abs/2306.17086} {arXiv:2306.17086
  [hep-ph]} \BibitemShut {NoStop}%
\bibitem [{\citenamefont {Madge}\ \emph {et~al.}(2023)\citenamefont {Madge},
  \citenamefont {Morgante}, \citenamefont {Puchades-Ib\'a\~nez}, \citenamefont
  {Ramberg}, \citenamefont {Ratzinger}, \citenamefont {Schenk},\ and\
  \citenamefont {Schwaller}}]{Madge:2023dxc}%
  \BibitemOpen
  \bibfield  {author} {\bibinfo {author} {\bibfnamefont {E.}~\bibnamefont
  {Madge}}, \bibinfo {author} {\bibfnamefont {E.}~\bibnamefont {Morgante}},
  \bibinfo {author} {\bibfnamefont {C.}~\bibnamefont {Puchades-Ib\'a\~nez}},
  \bibinfo {author} {\bibfnamefont {N.}~\bibnamefont {Ramberg}}, \bibinfo
  {author} {\bibfnamefont {W.}~\bibnamefont {Ratzinger}}, \bibinfo {author}
  {\bibfnamefont {S.}~\bibnamefont {Schenk}},\ and\ \bibinfo {author}
  {\bibfnamefont {P.}~\bibnamefont {Schwaller}},\ }\bibfield  {title} {\bibinfo
  {title} {{Primordial gravitational waves in the nano-Hertz regime and PTA
  data \textemdash{} towards solving the GW inverse problem}},\ }\href
  {https://doi.org/10.1007/JHEP10(2023)171} {\bibfield  {journal} {\bibinfo
  {journal} {JHEP}\ }\textbf {\bibinfo {volume} {10}},\ \bibinfo {pages}
  {171}},\ \Eprint {https://arxiv.org/abs/2306.14856} {arXiv:2306.14856
  [hep-ph]} \BibitemShut {NoStop}%
\bibitem [{\citenamefont {Megias}\ \emph {et~al.}(2023)\citenamefont {Megias},
  \citenamefont {Nardini},\ and\ \citenamefont {Quiros}}]{Megias:2023kiy}%
  \BibitemOpen
  \bibfield  {author} {\bibinfo {author} {\bibfnamefont {E.}~\bibnamefont
  {Megias}}, \bibinfo {author} {\bibfnamefont {G.}~\bibnamefont {Nardini}},\
  and\ \bibinfo {author} {\bibfnamefont {M.}~\bibnamefont {Quiros}},\
  }\bibfield  {title} {\bibinfo {title} {{Pulsar timing array stochastic
  background from light Kaluza-Klein resonances}},\ }\href
  {https://doi.org/10.1103/PhysRevD.108.095017} {\bibfield  {journal} {\bibinfo
   {journal} {Phys. Rev. D}\ }\textbf {\bibinfo {volume} {108}},\ \bibinfo
  {pages} {095017} (\bibinfo {year} {2023})},\ \Eprint
  {https://arxiv.org/abs/2306.17071} {arXiv:2306.17071 [hep-ph]} \BibitemShut
  {NoStop}%
\bibitem [{\citenamefont {Salvio}(2023)}]{Salvio:2023ynn}%
  \BibitemOpen
  \bibfield  {author} {\bibinfo {author} {\bibfnamefont {A.}~\bibnamefont
  {Salvio}},\ }\bibfield  {title} {\bibinfo {title} {{Supercooling in radiative
  symmetry breaking: theory extensions, gravitational wave detection and
  primordial black holes}},\ }\href
  {https://doi.org/10.1088/1475-7516/2023/12/046} {\bibfield  {journal}
  {\bibinfo  {journal} {JCAP}\ }\textbf {\bibinfo {volume} {12}},\ \bibinfo
  {pages} {046}},\ \Eprint {https://arxiv.org/abs/2307.04694} {arXiv:2307.04694
  [hep-ph]} \BibitemShut {NoStop}%
\bibitem [{\citenamefont {Salvio}(2024)}]{Salvio:2023blb}%
  \BibitemOpen
  \bibfield  {author} {\bibinfo {author} {\bibfnamefont {A.}~\bibnamefont
  {Salvio}},\ }\bibfield  {title} {\bibinfo {title} {{Pulsar timing arrays and
  primordial black holes from a supercooled phase transition}},\ }\href
  {https://doi.org/10.1016/j.physletb.2024.138639} {\bibfield  {journal}
  {\bibinfo  {journal} {Phys. Lett. B}\ }\textbf {\bibinfo {volume} {852}},\
  \bibinfo {pages} {138639} (\bibinfo {year} {2024})},\ \Eprint
  {https://arxiv.org/abs/2312.04628} {arXiv:2312.04628 [hep-ph]} \BibitemShut
  {NoStop}%
\bibitem [{\citenamefont {Gouttenoire}(2023)}]{Gouttenoire:2023bqy}%
  \BibitemOpen
  \bibfield  {author} {\bibinfo {author} {\bibfnamefont {Y.}~\bibnamefont
  {Gouttenoire}},\ }\bibfield  {title} {\bibinfo {title} {{First-Order Phase
  Transition Interpretation of Pulsar Timing Array Signal Is Consistent with
  Solar-Mass Black Holes}},\ }\href
  {https://doi.org/10.1103/PhysRevLett.131.171404} {\bibfield  {journal}
  {\bibinfo  {journal} {Phys. Rev. Lett.}\ }\textbf {\bibinfo {volume} {131}},\
  \bibinfo {pages} {171404} (\bibinfo {year} {2023})},\ \Eprint
  {https://arxiv.org/abs/2307.04239} {arXiv:2307.04239 [hep-ph]} \BibitemShut
  {NoStop}%
\bibitem [{\citenamefont {Addazi}\ \emph {et~al.}(2024)\citenamefont {Addazi},
  \citenamefont {Cai}, \citenamefont {Marciano},\ and\ \citenamefont
  {Visinelli}}]{Addazi:2023jvg}%
  \BibitemOpen
  \bibfield  {author} {\bibinfo {author} {\bibfnamefont {A.}~\bibnamefont
  {Addazi}}, \bibinfo {author} {\bibfnamefont {Y.-F.}\ \bibnamefont {Cai}},
  \bibinfo {author} {\bibfnamefont {A.}~\bibnamefont {Marciano}},\ and\
  \bibinfo {author} {\bibfnamefont {L.}~\bibnamefont {Visinelli}},\ }\bibfield
  {title} {\bibinfo {title} {{Have pulsar timing array methods detected a
  cosmological phase transition?}},\ }\href
  {https://doi.org/10.1103/PhysRevD.109.015028} {\bibfield  {journal} {\bibinfo
   {journal} {Phys. Rev. D}\ }\textbf {\bibinfo {volume} {109}},\ \bibinfo
  {pages} {015028} (\bibinfo {year} {2024})},\ \Eprint
  {https://arxiv.org/abs/2306.17205} {arXiv:2306.17205 [astro-ph.CO]}
  \BibitemShut {NoStop}%
\bibitem [{\citenamefont {Li}\ and\ \citenamefont {Xie}(2023)}]{Li:2023bxy}%
  \BibitemOpen
  \bibfield  {author} {\bibinfo {author} {\bibfnamefont {S.-P.}\ \bibnamefont
  {Li}}\ and\ \bibinfo {author} {\bibfnamefont {K.-P.}\ \bibnamefont {Xie}},\
  }\bibfield  {title} {\bibinfo {title} {{Collider test of nano-Hertz
  gravitational waves from pulsar timing arrays}},\ }\href
  {https://doi.org/10.1103/PhysRevD.108.055018} {\bibfield  {journal} {\bibinfo
   {journal} {Phys. Rev. D}\ }\textbf {\bibinfo {volume} {108}},\ \bibinfo
  {pages} {055018} (\bibinfo {year} {2023})},\ \Eprint
  {https://arxiv.org/abs/2307.01086} {arXiv:2307.01086 [hep-ph]} \BibitemShut
  {NoStop}%
\bibitem [{\citenamefont {Ghosh}\ \emph {et~al.}(2024)\citenamefont {Ghosh},
  \citenamefont {Ghoshal}, \citenamefont {Guo}, \citenamefont {Hajkarim},
  \citenamefont {King}, \citenamefont {Sinha}, \citenamefont {Wang},\ and\
  \citenamefont {White}}]{Ghosh:2023aum}%
  \BibitemOpen
  \bibfield  {author} {\bibinfo {author} {\bibfnamefont {T.}~\bibnamefont
  {Ghosh}}, \bibinfo {author} {\bibfnamefont {A.}~\bibnamefont {Ghoshal}},
  \bibinfo {author} {\bibfnamefont {H.-K.}\ \bibnamefont {Guo}}, \bibinfo
  {author} {\bibfnamefont {F.}~\bibnamefont {Hajkarim}}, \bibinfo {author}
  {\bibfnamefont {S.~F.}\ \bibnamefont {King}}, \bibinfo {author}
  {\bibfnamefont {K.}~\bibnamefont {Sinha}}, \bibinfo {author} {\bibfnamefont
  {X.}~\bibnamefont {Wang}},\ and\ \bibinfo {author} {\bibfnamefont
  {G.}~\bibnamefont {White}},\ }\bibfield  {title} {\bibinfo {title} {{Did we
  hear the sound of the Universe boiling? Analysis using the full fluid
  velocity profiles and NANOGrav 15-year data}},\ }\href
  {https://doi.org/10.1088/1475-7516/2024/05/100} {\bibfield  {journal}
  {\bibinfo  {journal} {JCAP}\ }\textbf {\bibinfo {volume} {05}},\ \bibinfo
  {pages} {100}},\ \Eprint {https://arxiv.org/abs/2307.02259} {arXiv:2307.02259
  [astro-ph.HE]} \BibitemShut {NoStop}%
\bibitem [{\citenamefont {Jiang}\ \emph {et~al.}(2024)\citenamefont {Jiang},
  \citenamefont {Yang}, \citenamefont {Ma},\ and\ \citenamefont
  {Huang}}]{Jiang:2023qbm}%
  \BibitemOpen
  \bibfield  {author} {\bibinfo {author} {\bibfnamefont {S.}~\bibnamefont
  {Jiang}}, \bibinfo {author} {\bibfnamefont {A.}~\bibnamefont {Yang}},
  \bibinfo {author} {\bibfnamefont {J.}~\bibnamefont {Ma}},\ and\ \bibinfo
  {author} {\bibfnamefont {F.~P.}\ \bibnamefont {Huang}},\ }\bibfield  {title}
  {\bibinfo {title} {{Implication of nano-Hertz stochastic gravitational wave
  on dynamical dark matter through a dark first-order phase transition}},\
  }\href {https://doi.org/10.1088/1361-6382/ad24c6} {\bibfield  {journal}
  {\bibinfo  {journal} {Class. Quant. Grav.}\ }\textbf {\bibinfo {volume}
  {41}},\ \bibinfo {pages} {065009} (\bibinfo {year} {2024})},\ \Eprint
  {https://arxiv.org/abs/2306.17827} {arXiv:2306.17827 [hep-ph]} \BibitemShut
  {NoStop}%
\bibitem [{\citenamefont {Wang}(2023)}]{Wang:2023bbc}%
  \BibitemOpen
  \bibfield  {author} {\bibinfo {author} {\bibfnamefont {D.}~\bibnamefont
  {Wang}},\ }\bibfield  {title} {\bibinfo {title} {{Constraining Cosmological
  Phase Transitions with Chinese Pulsar Timing Array Data Release 1}},\
  }\href@noop {} {\  (\bibinfo {year} {2023})},\ \Eprint
  {https://arxiv.org/abs/2307.15970} {arXiv:2307.15970 [astro-ph.CO]}
  \BibitemShut {NoStop}%
\bibitem [{\citenamefont {Li}\ and\ \citenamefont {Nath}(2025)}]{Li:2025nja}%
  \BibitemOpen
  \bibfield  {author} {\bibinfo {author} {\bibfnamefont {J.}~\bibnamefont
  {Li}}\ and\ \bibinfo {author} {\bibfnamefont {P.}~\bibnamefont {Nath}},\
  }\bibfield  {title} {\bibinfo {title} {{Supercooled phase transitions: Why
  thermal history of hidden sector matters in analysis of pulsar timing array
  signals}},\ }\href {https://doi.org/10.1103/79cb-rssl} {\bibfield  {journal}
  {\bibinfo  {journal} {Phys. Rev. D}\ }\textbf {\bibinfo {volume} {111}},\
  \bibinfo {pages} {123007} (\bibinfo {year} {2025})},\ \Eprint
  {https://arxiv.org/abs/2501.14986} {arXiv:2501.14986 [hep-ph]} \BibitemShut
  {NoStop}%
\bibitem [{\citenamefont {Fujikura}\ \emph {et~al.}(2025)\citenamefont
  {Fujikura}, \citenamefont {Nakagawa}, \citenamefont {Nakai}, \citenamefont
  {Sun},\ and\ \citenamefont {Zhang}}]{Fujikura:2025iam}%
  \BibitemOpen
  \bibfield  {author} {\bibinfo {author} {\bibfnamefont {K.}~\bibnamefont
  {Fujikura}}, \bibinfo {author} {\bibfnamefont {S.}~\bibnamefont {Nakagawa}},
  \bibinfo {author} {\bibfnamefont {Y.}~\bibnamefont {Nakai}}, \bibinfo
  {author} {\bibfnamefont {P.}~\bibnamefont {Sun}},\ and\ \bibinfo {author}
  {\bibfnamefont {Y.}~\bibnamefont {Zhang}},\ }\bibfield  {title} {\bibinfo
  {title} {{Conformal Phase Transition in Supersymmetric QCD}},\ }\href@noop {}
  {\  (\bibinfo {year} {2025})},\ \Eprint {https://arxiv.org/abs/2503.22293}
  {arXiv:2503.22293 [hep-ph]} \BibitemShut {NoStop}%
\bibitem [{\citenamefont {Ellis}\ \emph {et~al.}(2024)\citenamefont {Ellis},
  \citenamefont {Fairbairn}, \citenamefont {Franciolini}, \citenamefont
  {H\"utsi}, \citenamefont {Iovino}, \citenamefont {Lewicki}, \citenamefont
  {Raidal}, \citenamefont {Urrutia}, \citenamefont {Vaskonen},\ and\
  \citenamefont {Veerm\"ae}}]{Ellis:2023oxs}%
  \BibitemOpen
  \bibfield  {author} {\bibinfo {author} {\bibfnamefont {J.}~\bibnamefont
  {Ellis}}, \bibinfo {author} {\bibfnamefont {M.}~\bibnamefont {Fairbairn}},
  \bibinfo {author} {\bibfnamefont {G.}~\bibnamefont {Franciolini}}, \bibinfo
  {author} {\bibfnamefont {G.}~\bibnamefont {H\"utsi}}, \bibinfo {author}
  {\bibfnamefont {A.}~\bibnamefont {Iovino}}, \bibinfo {author} {\bibfnamefont
  {M.}~\bibnamefont {Lewicki}}, \bibinfo {author} {\bibfnamefont
  {M.}~\bibnamefont {Raidal}}, \bibinfo {author} {\bibfnamefont
  {J.}~\bibnamefont {Urrutia}}, \bibinfo {author} {\bibfnamefont
  {V.}~\bibnamefont {Vaskonen}},\ and\ \bibinfo {author} {\bibfnamefont
  {H.}~\bibnamefont {Veerm\"ae}},\ }\bibfield  {title} {\bibinfo {title} {{What
  is the source of the PTA GW signal?}},\ }\href
  {https://doi.org/10.1103/PhysRevD.109.023522} {\bibfield  {journal} {\bibinfo
   {journal} {Phys. Rev. D}\ }\textbf {\bibinfo {volume} {109}},\ \bibinfo
  {pages} {023522} (\bibinfo {year} {2024})},\ \Eprint
  {https://arxiv.org/abs/2308.08546} {arXiv:2308.08546 [astro-ph.CO]}
  \BibitemShut {NoStop}%
\bibitem [{\citenamefont {Winkler}\ and\ \citenamefont
  {Freese}(2025)}]{Winkler:2024olr}%
  \BibitemOpen
  \bibfield  {author} {\bibinfo {author} {\bibfnamefont {M.~W.}\ \bibnamefont
  {Winkler}}\ and\ \bibinfo {author} {\bibfnamefont {K.}~\bibnamefont
  {Freese}},\ }\bibfield  {title} {\bibinfo {title} {{Origin of the stochastic
  gravitational wave background: First-order phase transition versus black hole
  mergers}},\ }\href {https://doi.org/10.1103/PhysRevD.111.083509} {\bibfield
  {journal} {\bibinfo  {journal} {Phys. Rev. D}\ }\textbf {\bibinfo {volume}
  {111}},\ \bibinfo {pages} {083509} (\bibinfo {year} {2025})},\ \Eprint
  {https://arxiv.org/abs/2401.13729} {arXiv:2401.13729 [astro-ph.CO]}
  \BibitemShut {NoStop}%
\bibitem [{\citenamefont {Fujikura}\ \emph {et~al.}(2024)\citenamefont
  {Fujikura}, \citenamefont {Girmohanta}, \citenamefont {Nakai},\ and\
  \citenamefont {Zhang}}]{Fujikura:2024jto}%
  \BibitemOpen
  \bibfield  {author} {\bibinfo {author} {\bibfnamefont {K.}~\bibnamefont
  {Fujikura}}, \bibinfo {author} {\bibfnamefont {S.}~\bibnamefont
  {Girmohanta}}, \bibinfo {author} {\bibfnamefont {Y.}~\bibnamefont {Nakai}},\
  and\ \bibinfo {author} {\bibfnamefont {Z.}~\bibnamefont {Zhang}},\ }\bibfield
   {title} {\bibinfo {title} {{Cold darkogenesis: Dark matter and baryon
  asymmetry in light of the PTA signal}},\ }\href
  {https://doi.org/10.1016/j.physletb.2024.139045} {\bibfield  {journal}
  {\bibinfo  {journal} {Phys. Lett. B}\ }\textbf {\bibinfo {volume} {858}},\
  \bibinfo {pages} {139045} (\bibinfo {year} {2024})},\ \Eprint
  {https://arxiv.org/abs/2406.12956} {arXiv:2406.12956 [hep-ph]} \BibitemShut
  {NoStop}%
\bibitem [{\citenamefont {Shelton}\ and\ \citenamefont
  {Zurek}(2010)}]{Shelton:2010ta}%
  \BibitemOpen
  \bibfield  {author} {\bibinfo {author} {\bibfnamefont {J.}~\bibnamefont
  {Shelton}}\ and\ \bibinfo {author} {\bibfnamefont {K.~M.}\ \bibnamefont
  {Zurek}},\ }\bibfield  {title} {\bibinfo {title} {{Darkogenesis: A baryon
  asymmetry from the dark matter sector}},\ }\href
  {https://doi.org/10.1103/PhysRevD.82.123512} {\bibfield  {journal} {\bibinfo
  {journal} {Phys. Rev. D}\ }\textbf {\bibinfo {volume} {82}},\ \bibinfo
  {pages} {123512} (\bibinfo {year} {2010})},\ \Eprint
  {https://arxiv.org/abs/1008.1997} {arXiv:1008.1997 [hep-ph]} \BibitemShut
  {NoStop}%
\bibitem [{\citenamefont {Garcia-Bellido}\ \emph {et~al.}(1999)\citenamefont
  {Garcia-Bellido}, \citenamefont {Grigoriev}, \citenamefont {Kusenko},\ and\
  \citenamefont {Shaposhnikov}}]{Garcia-Bellido:1999xos}%
  \BibitemOpen
  \bibfield  {author} {\bibinfo {author} {\bibfnamefont {J.}~\bibnamefont
  {Garcia-Bellido}}, \bibinfo {author} {\bibfnamefont {D.~Y.}\ \bibnamefont
  {Grigoriev}}, \bibinfo {author} {\bibfnamefont {A.}~\bibnamefont {Kusenko}},\
  and\ \bibinfo {author} {\bibfnamefont {M.~E.}\ \bibnamefont {Shaposhnikov}},\
  }\bibfield  {title} {\bibinfo {title} {{Nonequilibrium electroweak
  baryogenesis from preheating after inflation}},\ }\href
  {https://doi.org/10.1103/PhysRevD.60.123504} {\bibfield  {journal} {\bibinfo
  {journal} {Phys. Rev. D}\ }\textbf {\bibinfo {volume} {60}},\ \bibinfo
  {pages} {123504} (\bibinfo {year} {1999})},\ \Eprint
  {https://arxiv.org/abs/hep-ph/9902449} {arXiv:hep-ph/9902449} \BibitemShut
  {NoStop}%
\bibitem [{\citenamefont {Konstandin}\ and\ \citenamefont
  {Servant}(2011)}]{Konstandin:2011ds}%
  \BibitemOpen
  \bibfield  {author} {\bibinfo {author} {\bibfnamefont {T.}~\bibnamefont
  {Konstandin}}\ and\ \bibinfo {author} {\bibfnamefont {G.}~\bibnamefont
  {Servant}},\ }\bibfield  {title} {\bibinfo {title} {{Natural Cold
  Baryogenesis from Strongly Interacting Electroweak Symmetry Breaking}},\
  }\href {https://doi.org/10.1088/1475-7516/2011/07/024} {\bibfield  {journal}
  {\bibinfo  {journal} {JCAP}\ }\textbf {\bibinfo {volume} {07}},\ \bibinfo
  {pages} {024}},\ \Eprint {https://arxiv.org/abs/1104.4793} {arXiv:1104.4793
  [hep-ph]} \BibitemShut {NoStop}%
\bibitem [{\citenamefont {Hall}\ \emph {et~al.}(2020)\citenamefont {Hall},
  \citenamefont {Konstandin}, \citenamefont {McGehee}, \citenamefont
  {Murayama},\ and\ \citenamefont {Servant}}]{Hall:2019ank}%
  \BibitemOpen
  \bibfield  {author} {\bibinfo {author} {\bibfnamefont {E.}~\bibnamefont
  {Hall}}, \bibinfo {author} {\bibfnamefont {T.}~\bibnamefont {Konstandin}},
  \bibinfo {author} {\bibfnamefont {R.}~\bibnamefont {McGehee}}, \bibinfo
  {author} {\bibfnamefont {H.}~\bibnamefont {Murayama}},\ and\ \bibinfo
  {author} {\bibfnamefont {G.}~\bibnamefont {Servant}},\ }\bibfield  {title}
  {\bibinfo {title} {{Baryogenesis From a Dark First-Order Phase Transition}},\
  }\href {https://doi.org/10.1007/JHEP04(2020)042} {\bibfield  {journal}
  {\bibinfo  {journal} {JHEP}\ }\textbf {\bibinfo {volume} {04}},\ \bibinfo
  {pages} {042}},\ \Eprint {https://arxiv.org/abs/1910.08068} {arXiv:1910.08068
  [hep-ph]} \BibitemShut {NoStop}%
\bibitem [{\citenamefont {Hall}\ \emph {et~al.}(2023)\citenamefont {Hall},
  \citenamefont {Konstandin}, \citenamefont {McGehee},\ and\ \citenamefont
  {Murayama}}]{Hall:2019rld}%
  \BibitemOpen
  \bibfield  {author} {\bibinfo {author} {\bibfnamefont {E.}~\bibnamefont
  {Hall}}, \bibinfo {author} {\bibfnamefont {T.}~\bibnamefont {Konstandin}},
  \bibinfo {author} {\bibfnamefont {R.}~\bibnamefont {McGehee}},\ and\ \bibinfo
  {author} {\bibfnamefont {H.}~\bibnamefont {Murayama}},\ }\bibfield  {title}
  {\bibinfo {title} {{Asymmetric matter from a dark first-order phase
  transition}},\ }\href {https://doi.org/10.1103/PhysRevD.107.055011}
  {\bibfield  {journal} {\bibinfo  {journal} {Phys. Rev. D}\ }\textbf {\bibinfo
  {volume} {107}},\ \bibinfo {pages} {055011} (\bibinfo {year} {2023})},\
  \Eprint {https://arxiv.org/abs/1911.12342} {arXiv:1911.12342 [hep-ph]}
  \BibitemShut {NoStop}%
\bibitem [{\citenamefont {Girmohanta}\ and\ \citenamefont
  {Shrock}(2022)}]{Girmohanta:2022dog}%
  \BibitemOpen
  \bibfield  {author} {\bibinfo {author} {\bibfnamefont {S.}~\bibnamefont
  {Girmohanta}}\ and\ \bibinfo {author} {\bibfnamefont {R.}~\bibnamefont
  {Shrock}},\ }\bibfield  {title} {\bibinfo {title} {{Cross section
  calculations in theories of self-interacting dark matter}},\ }\href
  {https://doi.org/10.1103/PhysRevD.106.063013} {\bibfield  {journal} {\bibinfo
   {journal} {Phys. Rev. D}\ }\textbf {\bibinfo {volume} {106}},\ \bibinfo
  {pages} {063013} (\bibinfo {year} {2022})},\ \Eprint
  {https://arxiv.org/abs/2206.14395} {arXiv:2206.14395 [hep-ph]} \BibitemShut
  {NoStop}%
\bibitem [{\citenamefont {Fukugita}\ and\ \citenamefont
  {Yanagida}(1986)}]{Fukugita:1986hr}%
  \BibitemOpen
  \bibfield  {author} {\bibinfo {author} {\bibfnamefont {M.}~\bibnamefont
  {Fukugita}}\ and\ \bibinfo {author} {\bibfnamefont {T.}~\bibnamefont
  {Yanagida}},\ }\bibfield  {title} {\bibinfo {title} {{Baryogenesis Without
  Grand Unification}},\ }\href {https://doi.org/10.1016/0370-2693(86)91126-3}
  {\bibfield  {journal} {\bibinfo  {journal} {Phys. Lett. B}\ }\textbf
  {\bibinfo {volume} {174}},\ \bibinfo {pages} {45} (\bibinfo {year}
  {1986})}\BibitemShut {NoStop}%
\bibitem [{\citenamefont {Maldacena}(1998)}]{Maldacena:1997re}%
  \BibitemOpen
  \bibfield  {author} {\bibinfo {author} {\bibfnamefont {J.~M.}\ \bibnamefont
  {Maldacena}},\ }\bibfield  {title} {\bibinfo {title} {{The Large N limit of
  superconformal field theories and supergravity}},\ }\href
  {https://doi.org/10.4310/ATMP.1998.v2.n2.a1} {\bibfield  {journal} {\bibinfo
  {journal} {Adv. Theor. Math. Phys.}\ }\textbf {\bibinfo {volume} {2}},\
  \bibinfo {pages} {231} (\bibinfo {year} {1998})},\ \Eprint
  {https://arxiv.org/abs/hep-th/9711200} {arXiv:hep-th/9711200} \BibitemShut
  {NoStop}%
\bibitem [{\citenamefont {Gubser}\ \emph {et~al.}(1998)\citenamefont {Gubser},
  \citenamefont {Klebanov},\ and\ \citenamefont {Polyakov}}]{Gubser:1998bc}%
  \BibitemOpen
  \bibfield  {author} {\bibinfo {author} {\bibfnamefont {S.~S.}\ \bibnamefont
  {Gubser}}, \bibinfo {author} {\bibfnamefont {I.~R.}\ \bibnamefont
  {Klebanov}},\ and\ \bibinfo {author} {\bibfnamefont {A.~M.}\ \bibnamefont
  {Polyakov}},\ }\bibfield  {title} {\bibinfo {title} {{Gauge theory
  correlators from noncritical string theory}},\ }\href
  {https://doi.org/10.1016/S0370-2693(98)00377-3} {\bibfield  {journal}
  {\bibinfo  {journal} {Phys. Lett. B}\ }\textbf {\bibinfo {volume} {428}},\
  \bibinfo {pages} {105} (\bibinfo {year} {1998})},\ \Eprint
  {https://arxiv.org/abs/hep-th/9802109} {arXiv:hep-th/9802109} \BibitemShut
  {NoStop}%
\bibitem [{\citenamefont {Witten}(1998)}]{Witten:1998qj}%
  \BibitemOpen
  \bibfield  {author} {\bibinfo {author} {\bibfnamefont {E.}~\bibnamefont
  {Witten}},\ }\bibfield  {title} {\bibinfo {title} {{Anti-de Sitter space and
  holography}},\ }\href {https://doi.org/10.4310/ATMP.1998.v2.n2.a2} {\bibfield
   {journal} {\bibinfo  {journal} {Adv. Theor. Math. Phys.}\ }\textbf {\bibinfo
  {volume} {2}},\ \bibinfo {pages} {253} (\bibinfo {year} {1998})},\ \Eprint
  {https://arxiv.org/abs/hep-th/9802150} {arXiv:hep-th/9802150} \BibitemShut
  {NoStop}%
\bibitem [{\citenamefont {Arkani-Hamed}\ \emph {et~al.}(2001)\citenamefont
  {Arkani-Hamed}, \citenamefont {Porrati},\ and\ \citenamefont
  {Randall}}]{Arkani-Hamed:2000ijo}%
  \BibitemOpen
  \bibfield  {author} {\bibinfo {author} {\bibfnamefont {N.}~\bibnamefont
  {Arkani-Hamed}}, \bibinfo {author} {\bibfnamefont {M.}~\bibnamefont
  {Porrati}},\ and\ \bibinfo {author} {\bibfnamefont {L.}~\bibnamefont
  {Randall}},\ }\bibfield  {title} {\bibinfo {title} {{Holography and
  phenomenology}},\ }\href {https://doi.org/10.1088/1126-6708/2001/08/017}
  {\bibfield  {journal} {\bibinfo  {journal} {JHEP}\ }\textbf {\bibinfo
  {volume} {08}},\ \bibinfo {pages} {017}},\ \Eprint
  {https://arxiv.org/abs/hep-th/0012148} {arXiv:hep-th/0012148} \BibitemShut
  {NoStop}%
\bibitem [{\citenamefont {Rattazzi}\ and\ \citenamefont
  {Zaffaroni}(2001)}]{Rattazzi:2000hs}%
  \BibitemOpen
  \bibfield  {author} {\bibinfo {author} {\bibfnamefont {R.}~\bibnamefont
  {Rattazzi}}\ and\ \bibinfo {author} {\bibfnamefont {A.}~\bibnamefont
  {Zaffaroni}},\ }\bibfield  {title} {\bibinfo {title} {{Comments on the
  holographic picture of the Randall-Sundrum model}},\ }\href
  {https://doi.org/10.1088/1126-6708/2001/04/021} {\bibfield  {journal}
  {\bibinfo  {journal} {JHEP}\ }\textbf {\bibinfo {volume} {04}},\ \bibinfo
  {pages} {021}},\ \Eprint {https://arxiv.org/abs/hep-th/0012248}
  {arXiv:hep-th/0012248} \BibitemShut {NoStop}%
\bibitem [{\citenamefont {Giudice}\ \emph {et~al.}(1999)\citenamefont
  {Giudice}, \citenamefont {Peloso}, \citenamefont {Riotto},\ and\
  \citenamefont {Tkachev}}]{Giudice:1999fb}%
  \BibitemOpen
  \bibfield  {author} {\bibinfo {author} {\bibfnamefont {G.~F.}\ \bibnamefont
  {Giudice}}, \bibinfo {author} {\bibfnamefont {M.}~\bibnamefont {Peloso}},
  \bibinfo {author} {\bibfnamefont {A.}~\bibnamefont {Riotto}},\ and\ \bibinfo
  {author} {\bibfnamefont {I.}~\bibnamefont {Tkachev}},\ }\bibfield  {title}
  {\bibinfo {title} {{Production of massive fermions at preheating and
  leptogenesis}},\ }\href {https://doi.org/10.1088/1126-6708/1999/08/014}
  {\bibfield  {journal} {\bibinfo  {journal} {JHEP}\ }\textbf {\bibinfo
  {volume} {08}},\ \bibinfo {pages} {014}},\ \Eprint
  {https://arxiv.org/abs/hep-ph/9905242} {arXiv:hep-ph/9905242} \BibitemShut
  {NoStop}%
\bibitem [{\citenamefont {Chacko}\ and\ \citenamefont
  {Mishra}(2013)}]{Chacko:2012sy}%
  \BibitemOpen
  \bibfield  {author} {\bibinfo {author} {\bibfnamefont {Z.}~\bibnamefont
  {Chacko}}\ and\ \bibinfo {author} {\bibfnamefont {R.~K.}\ \bibnamefont
  {Mishra}},\ }\bibfield  {title} {\bibinfo {title} {{Effective Theory of a
  Light Dilaton}},\ }\href {https://doi.org/10.1103/PhysRevD.87.115006}
  {\bibfield  {journal} {\bibinfo  {journal} {Phys. Rev. D}\ }\textbf {\bibinfo
  {volume} {87}},\ \bibinfo {pages} {115006} (\bibinfo {year} {2013})},\
  \Eprint {https://arxiv.org/abs/1209.3022} {arXiv:1209.3022 [hep-ph]}
  \BibitemShut {NoStop}%
\bibitem [{\citenamefont {Girmohanta}\ \emph {et~al.}(2024)\citenamefont
  {Girmohanta}, \citenamefont {Nakai}, \citenamefont {Shigekami},\ and\
  \citenamefont {Tobioka}}]{Girmohanta:2023tdr}%
  \BibitemOpen
  \bibfield  {author} {\bibinfo {author} {\bibfnamefont {S.}~\bibnamefont
  {Girmohanta}}, \bibinfo {author} {\bibfnamefont {Y.}~\bibnamefont {Nakai}},
  \bibinfo {author} {\bibfnamefont {Y.}~\bibnamefont {Shigekami}},\ and\
  \bibinfo {author} {\bibfnamefont {K.}~\bibnamefont {Tobioka}},\ }\bibfield
  {title} {\bibinfo {title} {{Light dilaton in rare meson decays and extraction
  of its CP property}},\ }\href {https://doi.org/10.1007/JHEP01(2024)153}
  {\bibfield  {journal} {\bibinfo  {journal} {JHEP}\ }\textbf {\bibinfo
  {volume} {01}},\ \bibinfo {pages} {153}},\ \Eprint
  {https://arxiv.org/abs/2310.16882} {arXiv:2310.16882 [hep-ph]} \BibitemShut
  {NoStop}%
\bibitem [{\citenamefont {Csaki}\ \emph {et~al.}(2001)\citenamefont {Csaki},
  \citenamefont {Graesser},\ and\ \citenamefont {Kribs}}]{Csaki:2000zn}%
  \BibitemOpen
  \bibfield  {author} {\bibinfo {author} {\bibfnamefont {C.}~\bibnamefont
  {Csaki}}, \bibinfo {author} {\bibfnamefont {M.~L.}\ \bibnamefont
  {Graesser}},\ and\ \bibinfo {author} {\bibfnamefont {G.~D.}\ \bibnamefont
  {Kribs}},\ }\bibfield  {title} {\bibinfo {title} {{Radion dynamics and
  electroweak physics}},\ }\href {https://doi.org/10.1103/PhysRevD.63.065002}
  {\bibfield  {journal} {\bibinfo  {journal} {Phys. Rev. D}\ }\textbf {\bibinfo
  {volume} {63}},\ \bibinfo {pages} {065002} (\bibinfo {year} {2001})},\
  \Eprint {https://arxiv.org/abs/hep-th/0008151} {arXiv:hep-th/0008151}
  \BibitemShut {NoStop}%
\bibitem [{\citenamefont {Goldberger}\ and\ \citenamefont
  {Wise}(1999)}]{Goldberger:1999uk}%
  \BibitemOpen
  \bibfield  {author} {\bibinfo {author} {\bibfnamefont {W.~D.}\ \bibnamefont
  {Goldberger}}\ and\ \bibinfo {author} {\bibfnamefont {M.~B.}\ \bibnamefont
  {Wise}},\ }\bibfield  {title} {\bibinfo {title} {{Modulus stabilization with
  bulk fields}},\ }\href {https://doi.org/10.1103/PhysRevLett.83.4922}
  {\bibfield  {journal} {\bibinfo  {journal} {Phys. Rev. Lett.}\ }\textbf
  {\bibinfo {volume} {83}},\ \bibinfo {pages} {4922} (\bibinfo {year}
  {1999})},\ \Eprint {https://arxiv.org/abs/hep-ph/9907447}
  {arXiv:hep-ph/9907447} \BibitemShut {NoStop}%
\bibitem [{\citenamefont {Bellazzini}\ \emph {et~al.}(2013)\citenamefont
  {Bellazzini}, \citenamefont {Csaki}, \citenamefont {Hubisz}, \citenamefont
  {Serra},\ and\ \citenamefont {Terning}}]{Bellazzini:2012vz}%
  \BibitemOpen
  \bibfield  {author} {\bibinfo {author} {\bibfnamefont {B.}~\bibnamefont
  {Bellazzini}}, \bibinfo {author} {\bibfnamefont {C.}~\bibnamefont {Csaki}},
  \bibinfo {author} {\bibfnamefont {J.}~\bibnamefont {Hubisz}}, \bibinfo
  {author} {\bibfnamefont {J.}~\bibnamefont {Serra}},\ and\ \bibinfo {author}
  {\bibfnamefont {J.}~\bibnamefont {Terning}},\ }\bibfield  {title} {\bibinfo
  {title} {{A Higgslike Dilaton}},\ }\href
  {https://doi.org/10.1140/epjc/s10052-013-2333-x} {\bibfield  {journal}
  {\bibinfo  {journal} {Eur. Phys. J. C}\ }\textbf {\bibinfo {volume} {73}},\
  \bibinfo {pages} {2333} (\bibinfo {year} {2013})},\ \Eprint
  {https://arxiv.org/abs/1209.3299} {arXiv:1209.3299 [hep-ph]} \BibitemShut
  {NoStop}%
\bibitem [{\citenamefont {Fujikura}\ \emph {et~al.}(2020)\citenamefont
  {Fujikura}, \citenamefont {Nakai},\ and\ \citenamefont
  {Yamada}}]{Fujikura:2019oyi}%
  \BibitemOpen
  \bibfield  {author} {\bibinfo {author} {\bibfnamefont {K.}~\bibnamefont
  {Fujikura}}, \bibinfo {author} {\bibfnamefont {Y.}~\bibnamefont {Nakai}},\
  and\ \bibinfo {author} {\bibfnamefont {M.}~\bibnamefont {Yamada}},\
  }\bibfield  {title} {\bibinfo {title} {{A more attractive scheme for radion
  stabilization and supercooled phase transition}},\ }\href
  {https://doi.org/10.1007/JHEP02(2020)111} {\bibfield  {journal} {\bibinfo
  {journal} {JHEP}\ }\textbf {\bibinfo {volume} {02}},\ \bibinfo {pages}
  {111}},\ \Eprint {https://arxiv.org/abs/1910.07546} {arXiv:1910.07546
  [hep-ph]} \BibitemShut {NoStop}%
\bibitem [{\citenamefont {Bringmann}\ \emph {et~al.}(2023)\citenamefont
  {Bringmann}, \citenamefont {Depta}, \citenamefont {Konstandin}, \citenamefont
  {Schmidt-Hoberg},\ and\ \citenamefont {Tasillo}}]{Bringmann:2023opz}%
  \BibitemOpen
  \bibfield  {author} {\bibinfo {author} {\bibfnamefont {T.}~\bibnamefont
  {Bringmann}}, \bibinfo {author} {\bibfnamefont {P.~F.}\ \bibnamefont
  {Depta}}, \bibinfo {author} {\bibfnamefont {T.}~\bibnamefont {Konstandin}},
  \bibinfo {author} {\bibfnamefont {K.}~\bibnamefont {Schmidt-Hoberg}},\ and\
  \bibinfo {author} {\bibfnamefont {C.}~\bibnamefont {Tasillo}},\ }\bibfield
  {title} {\bibinfo {title} {{Does NANOGrav observe a dark sector phase
  transition?}},\ }\href {https://doi.org/10.1088/1475-7516/2023/11/053}
  {\bibfield  {journal} {\bibinfo  {journal} {JCAP}\ }\textbf {\bibinfo
  {volume} {11}},\ \bibinfo {pages} {053}},\ \Eprint
  {https://arxiv.org/abs/2306.09411} {arXiv:2306.09411 [astro-ph.CO]}
  \BibitemShut {NoStop}%
\bibitem [{\citenamefont {Greene}\ and\ \citenamefont
  {Kofman}(1999)}]{Greene:1998nh}%
  \BibitemOpen
  \bibfield  {author} {\bibinfo {author} {\bibfnamefont {P.~B.}\ \bibnamefont
  {Greene}}\ and\ \bibinfo {author} {\bibfnamefont {L.}~\bibnamefont
  {Kofman}},\ }\bibfield  {title} {\bibinfo {title} {{Preheating of
  fermions}},\ }\href {https://doi.org/10.1016/S0370-2693(99)00020-9}
  {\bibfield  {journal} {\bibinfo  {journal} {Phys. Lett. B}\ }\textbf
  {\bibinfo {volume} {448}},\ \bibinfo {pages} {6} (\bibinfo {year} {1999})},\
  \Eprint {https://arxiv.org/abs/hep-ph/9807339} {arXiv:hep-ph/9807339}
  \BibitemShut {NoStop}%
\bibitem [{\citenamefont {Aad}\ \emph {et~al.}(2021)\citenamefont {Aad} \emph
  {et~al.}}]{ATLAS:2021kxv}%
  \BibitemOpen
  \bibfield  {author} {\bibinfo {author} {\bibfnamefont {G.}~\bibnamefont
  {Aad}} \emph {et~al.} (\bibinfo {collaboration} {ATLAS}),\ }\bibfield
  {title} {\bibinfo {title} {{Search for new phenomena in events with an
  energetic jet and missing transverse momentum in $pp$ collisions at $\sqrt
  {s}$ =13 TeV with the ATLAS detector}},\ }\href
  {https://doi.org/10.1103/PhysRevD.103.112006} {\bibfield  {journal} {\bibinfo
   {journal} {Phys. Rev. D}\ }\textbf {\bibinfo {volume} {103}},\ \bibinfo
  {pages} {112006} (\bibinfo {year} {2021})},\ \Eprint
  {https://arxiv.org/abs/2102.10874} {arXiv:2102.10874 [hep-ex]} \BibitemShut
  {NoStop}%
\bibitem [{\citenamefont {C\'\i{}scar-Monsalvatje}\ \emph
  {et~al.}(2024)\citenamefont {C\'\i{}scar-Monsalvatje}, \citenamefont
  {Ibarra},\ and\ \citenamefont {Vandecasteele}}]{Ciscar-Monsalvatje:2023zkk}%
  \BibitemOpen
  \bibfield  {author} {\bibinfo {author} {\bibfnamefont {M.}~\bibnamefont
  {C\'\i{}scar-Monsalvatje}}, \bibinfo {author} {\bibfnamefont
  {A.}~\bibnamefont {Ibarra}},\ and\ \bibinfo {author} {\bibfnamefont
  {J.}~\bibnamefont {Vandecasteele}},\ }\bibfield  {title} {\bibinfo {title}
  {{Matter-antimatter asymmetry and dark matter stability from baryon number
  conservation}},\ }\href {https://doi.org/10.1088/1475-7516/2024/01/028}
  {\bibfield  {journal} {\bibinfo  {journal} {JCAP}\ }\textbf {\bibinfo
  {volume} {01}},\ \bibinfo {pages} {028}},\ \Eprint
  {https://arxiv.org/abs/2307.02592} {arXiv:2307.02592 [hep-ph]} \BibitemShut
  {NoStop}%
\bibitem [{\citenamefont {Winkler}(2019)}]{Winkler:2018qyg}%
  \BibitemOpen
  \bibfield  {author} {\bibinfo {author} {\bibfnamefont {M.~W.}\ \bibnamefont
  {Winkler}},\ }\bibfield  {title} {\bibinfo {title} {{Decay and detection of a
  light scalar boson mixing with the Higgs boson}},\ }\href
  {https://doi.org/10.1103/PhysRevD.99.015018} {\bibfield  {journal} {\bibinfo
  {journal} {Phys. Rev. D}\ }\textbf {\bibinfo {volume} {99}},\ \bibinfo
  {pages} {015018} (\bibinfo {year} {2019})},\ \Eprint
  {https://arxiv.org/abs/1809.01876} {arXiv:1809.01876 [hep-ph]} \BibitemShut
  {NoStop}%
\bibitem [{\citenamefont {Albanese}\ \emph {et~al.}(2025)\citenamefont
  {Albanese} \emph {et~al.}}]{SHiP:2025ows}%
  \BibitemOpen
  \bibfield  {author} {\bibinfo {author} {\bibfnamefont {R.}~\bibnamefont
  {Albanese}} \emph {et~al.} (\bibinfo {collaboration} {SHiP, HI-ECN3 Project
  Team}),\ }\bibfield  {title} {\bibinfo {title} {{SHiP experiment at the SPS
  Beam Dump Facility}},\ }\href@noop {} {\  (\bibinfo {year} {2025})},\ \Eprint
  {https://arxiv.org/abs/2504.06692} {arXiv:2504.06692 [hep-ex]} \BibitemShut
  {NoStop}%
\bibitem [{\citenamefont {Tulin}\ and\ \citenamefont
  {Yu}(2018)}]{Tulin:2017ara}%
  \BibitemOpen
  \bibfield  {author} {\bibinfo {author} {\bibfnamefont {S.}~\bibnamefont
  {Tulin}}\ and\ \bibinfo {author} {\bibfnamefont {H.-B.}\ \bibnamefont {Yu}},\
  }\bibfield  {title} {\bibinfo {title} {{Dark Matter Self-interactions and
  Small Scale Structure}},\ }\href
  {https://doi.org/10.1016/j.physrep.2017.11.004} {\bibfield  {journal}
  {\bibinfo  {journal} {Phys. Rept.}\ }\textbf {\bibinfo {volume} {730}},\
  \bibinfo {pages} {1} (\bibinfo {year} {2018})},\ \Eprint
  {https://arxiv.org/abs/1705.02358} {arXiv:1705.02358 [hep-ph]} \BibitemShut
  {NoStop}%
\bibitem [{\citenamefont {Jungman}\ \emph {et~al.}(1996)\citenamefont
  {Jungman}, \citenamefont {Kamionkowski},\ and\ \citenamefont
  {Griest}}]{Jungman:1995df}%
  \BibitemOpen
  \bibfield  {author} {\bibinfo {author} {\bibfnamefont {G.}~\bibnamefont
  {Jungman}}, \bibinfo {author} {\bibfnamefont {M.}~\bibnamefont
  {Kamionkowski}},\ and\ \bibinfo {author} {\bibfnamefont {K.}~\bibnamefont
  {Griest}},\ }\bibfield  {title} {\bibinfo {title} {{Supersymmetric dark
  matter}},\ }\href {https://doi.org/10.1016/0370-1573(95)00058-5} {\bibfield
  {journal} {\bibinfo  {journal} {Phys. Rept.}\ }\textbf {\bibinfo {volume}
  {267}},\ \bibinfo {pages} {195} (\bibinfo {year} {1996})},\ \Eprint
  {https://arxiv.org/abs/hep-ph/9506380} {arXiv:hep-ph/9506380} \BibitemShut
  {NoStop}%
\bibitem [{\citenamefont {Ma}\ \emph {et~al.}(2023)\citenamefont {Ma} \emph
  {et~al.}}]{PandaX:2022aac}%
  \BibitemOpen
  \bibfield  {author} {\bibinfo {author} {\bibfnamefont {W.}~\bibnamefont {Ma}}
  \emph {et~al.} (\bibinfo {collaboration} {PandaX}),\ }\bibfield  {title}
  {\bibinfo {title} {{Search for Solar B8 Neutrinos in the PandaX-4T Experiment
  Using Neutrino-Nucleus Coherent Scattering}},\ }\href
  {https://doi.org/10.1103/PhysRevLett.130.021802} {\bibfield  {journal}
  {\bibinfo  {journal} {Phys. Rev. Lett.}\ }\textbf {\bibinfo {volume} {130}},\
  \bibinfo {pages} {021802} (\bibinfo {year} {2023})},\ \Eprint
  {https://arxiv.org/abs/2207.04883} {arXiv:2207.04883 [hep-ex]} \BibitemShut
  {NoStop}%
\bibitem [{\citenamefont {Abdukerim}\ \emph {et~al.}(2025)\citenamefont
  {Abdukerim} \emph {et~al.}}]{PandaX:2024oxq}%
  \BibitemOpen
  \bibfield  {author} {\bibinfo {author} {\bibfnamefont {A.}~\bibnamefont
  {Abdukerim}} \emph {et~al.} (\bibinfo {collaboration} {PANDA-X, PandaX}),\
  }\bibfield  {title} {\bibinfo {title} {{PandaX-xT{\textemdash}A deep
  underground multi-ten-tonne liquid xenon observatory}},\ }\href
  {https://doi.org/10.1007/s11433-024-2539-y} {\bibfield  {journal} {\bibinfo
  {journal} {Sci. China Phys. Mech. Astron.}\ }\textbf {\bibinfo {volume}
  {68}},\ \bibinfo {pages} {221011} (\bibinfo {year} {2025})},\ \Eprint
  {https://arxiv.org/abs/2402.03596} {arXiv:2402.03596 [hep-ex]} \BibitemShut
  {NoStop}%
\bibitem [{\citenamefont {Abe}\ \emph {et~al.}(2021)\citenamefont {Abe} \emph
  {et~al.}}]{Super-Kamiokande:2020bov}%
  \BibitemOpen
  \bibfield  {author} {\bibinfo {author} {\bibfnamefont {K.}~\bibnamefont
  {Abe}} \emph {et~al.} (\bibinfo {collaboration} {Super-Kamiokande}),\
  }\bibfield  {title} {\bibinfo {title} {{Neutron-antineutron oscillation
  search using a 0.37 megaton-years exposure of Super-Kamiokande}},\ }\href
  {https://doi.org/10.1103/PhysRevD.103.012008} {\bibfield  {journal} {\bibinfo
   {journal} {Phys. Rev. D}\ }\textbf {\bibinfo {volume} {103}},\ \bibinfo
  {pages} {012008} (\bibinfo {year} {2021})},\ \Eprint
  {https://arxiv.org/abs/2012.02607} {arXiv:2012.02607 [hep-ex]} \BibitemShut
  {NoStop}%
\bibitem [{\citenamefont {Santoro}\ \emph {et~al.}(2024)\citenamefont {Santoro}
  \emph {et~al.}}]{Santoro:2024lvc}%
  \BibitemOpen
  \bibfield  {author} {\bibinfo {author} {\bibfnamefont {V.}~\bibnamefont
  {Santoro}} \emph {et~al.},\ }\bibfield  {title} {\bibinfo {title} {{HighNESS
  conceptual design report: Volume II. The NNBAR experiment.}},\ }\href
  {https://doi.org/10.3233/jnr-230951} {\bibfield  {journal} {\bibinfo
  {journal} {J. Neutron Res.}\ }\textbf {\bibinfo {volume} {25}},\ \bibinfo
  {pages} {315} (\bibinfo {year} {2024})}\BibitemShut {NoStop}%
\bibitem [{\citenamefont {Abed~Abud}\ \emph {et~al.}(2024)\citenamefont
  {Abed~Abud} \emph {et~al.}}]{DUNE:2024wvj}%
  \BibitemOpen
  \bibfield  {author} {\bibinfo {author} {\bibfnamefont {A.}~\bibnamefont
  {Abed~Abud}} \emph {et~al.} (\bibinfo {collaboration} {DUNE}),\ }\bibfield
  {title} {\bibinfo {title} {{DUNE Phase~II: scientific opportunities, detector
  concepts, technological solutions}},\ }\href
  {https://doi.org/10.1088/1748-0221/19/12/P12005} {\bibfield  {journal}
  {\bibinfo  {journal} {JINST}\ }\textbf {\bibinfo {volume} {19}}\bibfield
  {number} {\bibinfo  {number} { (12)},\ \bibinfo {pages} {P12005}},\ }\Eprint
  {https://arxiv.org/abs/2408.12725} {arXiv:2408.12725 [physics.ins-det]}
  \BibitemShut {NoStop}%
\bibitem [{\citenamefont {Abe}\ \emph {et~al.}(2018)\citenamefont {Abe} \emph
  {et~al.}}]{Hyper-Kamiokande:2018ofw}%
  \BibitemOpen
  \bibfield  {author} {\bibinfo {author} {\bibfnamefont {K.}~\bibnamefont
  {Abe}} \emph {et~al.} (\bibinfo {collaboration} {Hyper-Kamiokande}),\
  }\bibfield  {title} {\bibinfo {title} {{Hyper-Kamiokande Design Report}},\
  }\href@noop {} {\  (\bibinfo {year} {2018})},\ \Eprint
  {https://arxiv.org/abs/1805.04163} {arXiv:1805.04163 [physics.ins-det]}
  \BibitemShut {NoStop}%
\bibitem [{\citenamefont {Abusleme}\ \emph {et~al.}(2022)\citenamefont
  {Abusleme} \emph {et~al.}}]{JUNO:2021vlw}%
  \BibitemOpen
  \bibfield  {author} {\bibinfo {author} {\bibfnamefont {A.}~\bibnamefont
  {Abusleme}} \emph {et~al.} (\bibinfo {collaboration} {JUNO}),\ }\bibfield
  {title} {\bibinfo {title} {{JUNO physics and detector}},\ }\href
  {https://doi.org/10.1016/j.ppnp.2021.103927} {\bibfield  {journal} {\bibinfo
  {journal} {Prog. Part. Nucl. Phys.}\ }\textbf {\bibinfo {volume} {123}},\
  \bibinfo {pages} {103927} (\bibinfo {year} {2022})},\ \Eprint
  {https://arxiv.org/abs/2104.02565} {arXiv:2104.02565 [hep-ex]} \BibitemShut
  {NoStop}%
\bibitem [{\citenamefont {Garcia-Bellido}\ \emph {et~al.}(2021)\citenamefont
  {Garcia-Bellido}, \citenamefont {Murayama},\ and\ \citenamefont
  {White}}]{Garcia-Bellido:2021zgu}%
  \BibitemOpen
  \bibfield  {author} {\bibinfo {author} {\bibfnamefont {J.}~\bibnamefont
  {Garcia-Bellido}}, \bibinfo {author} {\bibfnamefont {H.}~\bibnamefont
  {Murayama}},\ and\ \bibinfo {author} {\bibfnamefont {G.}~\bibnamefont
  {White}},\ }\bibfield  {title} {\bibinfo {title} {{Exploring the early
  Universe with Gaia and Theia}},\ }\href
  {https://doi.org/10.1088/1475-7516/2021/12/023} {\bibfield  {journal}
  {\bibinfo  {journal} {JCAP}\ }\textbf {\bibinfo {volume} {12}}\bibfield
  {number} {\bibinfo  {number} { (12)},\ \bibinfo {pages} {023}},\ }\Eprint
  {https://arxiv.org/abs/2104.04778} {arXiv:2104.04778 [hep-ph]} \BibitemShut
  {NoStop}%
\bibitem [{\citenamefont {Randall}\ and\ \citenamefont
  {Sundrum}(1999)}]{Randall:1999ee}%
  \BibitemOpen
  \bibfield  {author} {\bibinfo {author} {\bibfnamefont {L.}~\bibnamefont
  {Randall}}\ and\ \bibinfo {author} {\bibfnamefont {R.}~\bibnamefont
  {Sundrum}},\ }\bibfield  {title} {\bibinfo {title} {{A Large mass hierarchy
  from a small extra dimension}},\ }\href
  {https://doi.org/10.1103/PhysRevLett.83.3370} {\bibfield  {journal} {\bibinfo
   {journal} {Phys. Rev. Lett.}\ }\textbf {\bibinfo {volume} {83}},\ \bibinfo
  {pages} {3370} (\bibinfo {year} {1999})},\ \Eprint
  {https://arxiv.org/abs/hep-ph/9905221} {arXiv:hep-ph/9905221} \BibitemShut
  {NoStop}%
\bibitem [{\citenamefont {Goldberger}\ and\ \citenamefont
  {Wise}(2000)}]{Goldberger:1999un}%
  \BibitemOpen
  \bibfield  {author} {\bibinfo {author} {\bibfnamefont {W.~D.}\ \bibnamefont
  {Goldberger}}\ and\ \bibinfo {author} {\bibfnamefont {M.~B.}\ \bibnamefont
  {Wise}},\ }\bibfield  {title} {\bibinfo {title} {{Phenomenology of a
  stabilized modulus}},\ }\href {https://doi.org/10.1016/S0370-2693(00)00099-X}
  {\bibfield  {journal} {\bibinfo  {journal} {Phys. Lett. B}\ }\textbf
  {\bibinfo {volume} {475}},\ \bibinfo {pages} {275} (\bibinfo {year}
  {2000})},\ \Eprint {https://arxiv.org/abs/hep-ph/9911457}
  {arXiv:hep-ph/9911457} \BibitemShut {NoStop}%
\bibitem [{\citenamefont {Creminelli}\ \emph {et~al.}(2002)\citenamefont
  {Creminelli}, \citenamefont {Nicolis},\ and\ \citenamefont
  {Rattazzi}}]{Creminelli:2001th}%
  \BibitemOpen
  \bibfield  {author} {\bibinfo {author} {\bibfnamefont {P.}~\bibnamefont
  {Creminelli}}, \bibinfo {author} {\bibfnamefont {A.}~\bibnamefont
  {Nicolis}},\ and\ \bibinfo {author} {\bibfnamefont {R.}~\bibnamefont
  {Rattazzi}},\ }\bibfield  {title} {\bibinfo {title} {{Holography and the
  electroweak phase transition}},\ }\href
  {https://doi.org/10.1088/1126-6708/2002/03/051} {\bibfield  {journal}
  {\bibinfo  {journal} {JHEP}\ }\textbf {\bibinfo {volume} {03}},\ \bibinfo
  {pages} {051}},\ \Eprint {https://arxiv.org/abs/hep-th/0107141}
  {arXiv:hep-th/0107141} \BibitemShut {NoStop}%
\bibitem [{\citenamefont {Goldberger}\ \emph {et~al.}(2008)\citenamefont
  {Goldberger}, \citenamefont {Grinstein},\ and\ \citenamefont
  {Skiba}}]{Goldberger:2007zk}%
  \BibitemOpen
  \bibfield  {author} {\bibinfo {author} {\bibfnamefont {W.~D.}\ \bibnamefont
  {Goldberger}}, \bibinfo {author} {\bibfnamefont {B.}~\bibnamefont
  {Grinstein}},\ and\ \bibinfo {author} {\bibfnamefont {W.}~\bibnamefont
  {Skiba}},\ }\bibfield  {title} {\bibinfo {title} {{Distinguishing the Higgs
  boson from the dilaton at the Large Hadron Collider}},\ }\href
  {https://doi.org/10.1103/PhysRevLett.100.111802} {\bibfield  {journal}
  {\bibinfo  {journal} {Phys. Rev. Lett.}\ }\textbf {\bibinfo {volume} {100}},\
  \bibinfo {pages} {111802} (\bibinfo {year} {2008})},\ \Eprint
  {https://arxiv.org/abs/0708.1463} {arXiv:0708.1463 [hep-ph]} \BibitemShut
  {NoStop}%
\bibitem [{\citenamefont {Pilaftsis}(1999)}]{Pilaftsis:1998pd}%
  \BibitemOpen
  \bibfield  {author} {\bibinfo {author} {\bibfnamefont {A.}~\bibnamefont
  {Pilaftsis}},\ }\bibfield  {title} {\bibinfo {title} {{Heavy Majorana
  neutrinos and baryogenesis}},\ }\href
  {https://doi.org/10.1142/S0217751X99000932} {\bibfield  {journal} {\bibinfo
  {journal} {Int. J. Mod. Phys. A}\ }\textbf {\bibinfo {volume} {14}},\
  \bibinfo {pages} {1811} (\bibinfo {year} {1999})},\ \Eprint
  {https://arxiv.org/abs/hep-ph/9812256} {arXiv:hep-ph/9812256} \BibitemShut
  {NoStop}%
\bibitem [{\citenamefont {Girmohanta}\ and\ \citenamefont
  {Qiu}(2025)}]{Girmohanta:2025uue}%
  \BibitemOpen
  \bibfield  {author} {\bibinfo {author} {\bibfnamefont {S.}~\bibnamefont
  {Girmohanta}}\ and\ \bibinfo {author} {\bibfnamefont {Y.-C.}\ \bibnamefont
  {Qiu}},\ }\bibfield  {title} {\bibinfo {title} {{Quest for a
  phenomenologically consistent low cutoff theory}},\ }\href@noop {} {\
  (\bibinfo {year} {2025})},\ \Eprint {https://arxiv.org/abs/2505.10233}
  {arXiv:2505.10233 [hep-ph]} \BibitemShut {NoStop}%
\bibitem [{\citenamefont {Lee}\ \emph {et~al.}(2022)\citenamefont {Lee},
  \citenamefont {Nakai},\ and\ \citenamefont {Suzuki}}]{Lee:2021wau}%
  \BibitemOpen
  \bibfield  {author} {\bibinfo {author} {\bibfnamefont {S.~J.}\ \bibnamefont
  {Lee}}, \bibinfo {author} {\bibfnamefont {Y.}~\bibnamefont {Nakai}},\ and\
  \bibinfo {author} {\bibfnamefont {M.}~\bibnamefont {Suzuki}},\ }\bibfield
  {title} {\bibinfo {title} {{Multiple hierarchies from a warped extra
  dimension}},\ }\href {https://doi.org/10.1007/JHEP02(2022)050} {\bibfield
  {journal} {\bibinfo  {journal} {JHEP}\ }\textbf {\bibinfo {volume} {02}},\
  \bibinfo {pages} {050}},\ \Eprint {https://arxiv.org/abs/2109.10938}
  {arXiv:2109.10938 [hep-ph]} \BibitemShut {NoStop}%
\bibitem [{\citenamefont {Girmohanta}\ \emph {et~al.}(2023)\citenamefont
  {Girmohanta}, \citenamefont {Lee}, \citenamefont {Nakai},\ and\ \citenamefont
  {Suzuki}}]{Girmohanta:2023sjv}%
  \BibitemOpen
  \bibfield  {author} {\bibinfo {author} {\bibfnamefont {S.}~\bibnamefont
  {Girmohanta}}, \bibinfo {author} {\bibfnamefont {S.~J.}\ \bibnamefont {Lee}},
  \bibinfo {author} {\bibfnamefont {Y.}~\bibnamefont {Nakai}},\ and\ \bibinfo
  {author} {\bibfnamefont {M.}~\bibnamefont {Suzuki}},\ }\bibfield  {title}
  {\bibinfo {title} {{Multi-brane cosmology}},\ }\href
  {https://doi.org/10.1007/JHEP07(2023)182} {\bibfield  {journal} {\bibinfo
  {journal} {JHEP}\ }\textbf {\bibinfo {volume} {07}},\ \bibinfo {pages}
  {182}},\ \Eprint {https://arxiv.org/abs/2304.05586} {arXiv:2304.05586
  [hep-ph]} \BibitemShut {NoStop}%
\end{thebibliography}%

\end{document}